\documentclass[11pt,a4paper]{article}
\usepackage[T1]{fontenc}
\usepackage[utf8]{inputenc}
\usepackage{lmodern}
\usepackage{amsmath,amssymb}
\usepackage{graphicx}
\usepackage{booktabs}
\usepackage{array}
\usepackage{tabularx}
\usepackage{longtable}
\usepackage{multirow}
\usepackage{caption}
\usepackage{xcolor}
\usepackage{microtype}
\usepackage{tikz}
\usetikzlibrary{positioning,fit,backgrounds,arrows.meta,calc}
\usepackage{geometry}
\usepackage[hidelinks,bookmarks=true]{hyperref}
\usepackage[backend=biber,style=numeric,sorting=none]{biblatex}
\AtBeginBibliography{\raggedright\Urlmuskip=0mu}

\newcolumntype{Y}{>{\raggedright\arraybackslash}X}
\hypersetup{
  pdftitle={Beyond Policy Engines: Binding-Complete Execution Admission},
  pdfsubject={Toward an Engine-Agnostic Execution Boundary Architecture},
  pdfauthor={Lin Wang, Mengting Wu, Yong Zhang, Jiang Deng},
  pdfkeywords={Execution Admission; Binding Completeness; Policy Engines; Governance; Authorization; Engine-Agnostic Architecture; Execution Control}
}

\title{Beyond Policy Engines: Binding-Complete Execution Admission\\[6pt]
\large Toward an Engine-Agnostic Execution Boundary Architecture}
\author{Lin Wang\thanks{Corresponding Author: Lin Wang. Email: \texttt{charles@havenlon.com}}\quad
Mengting Wu\quad
Yong Zhang\quad
Jiang Deng\\[4pt]
Chengdu Havenlon Security Technology Co., Ltd.}
\date{}

\begin{document}
\maketitle

\begin{abstract}
Policy engines determine whether evaluated policies permit a request, but a correct \texttt{ALLOW} decision does not establish that every required governing source participated, that the designated policy objects were bound to the intended action and state, or that the decision is sufficient to authorize execution. This paper examines the gap between policy evaluation and execution admission.

We introduce the \textbf{Binding-Complete Execution Admission Contract (BCEAC)}, a governance-relative contract that defines independently determined source obligations, exact policy and contextual binding, and typed admission outcomes under explicit trust assumptions. We propose a conditionally engine-agnostic execution-admission architecture that requires a canonical semantic and governance-binding contract for participating backends, while an Execution Boundary Language (EBL) authoring syntax remains optional. The architecture separates governance resolution, policy evaluation, independent admission verification, and downstream execution mediation. Backends such as a future EBL Native runtime, Open Policy Agent (OPA), or Cedar may participate only if they satisfy declared bounded semantic profiles, adapter contracts, and contextual correspondence obligations; their decisions do not independently establish admission eligibility.

We investigate the contract through two distinct experimental tracks: a finite Governance Source Completeness study and an OPA/Cedar experiment with archived native-engine records subsequently subjected to independent offline review. The archived OPA/Cedar track comprises 72 scenarios and 144 strict case--engine records (120 recorded as native-invoked and 24 pre-native), with eight additional permissive controls outside the strict denominator. The subsequent independent offline audit examined archived evidence without rerunning the original native OPA/Cedar executables. The results provide bounded evidence for the proposed distinctions, while preserving historical failures, unresolved implementation defects, and limitations of the archived experimental record.

The work does not establish constitutional adequacy, general semantic refinement, authenticated currentness, machine-checked execution-safety theorems, or protected external effects. Its contribution is an explicit, falsifiable framework for determining what must be established between a policy decision and an execution-admission judgment.

\end{abstract}

\section{Introduction}

\subsection{A correct policy answer is not yet an admissible action}

Policy evaluation is often represented as a compact decision: a request and contextual information are submitted to a policy evaluator, which returns a permit, a denial, or an indeterminate result. This interface is useful precisely because it makes the evaluation of a supplied policy separable from the systems that prepare requests and act upon decisions. Yet that separation creates a question that the evaluator's answer cannot settle on its own. \textbf{Who established that the evaluator received the governance sources, versions, and context that were actually required for this particular action?} A correct answer to the policies presented is not necessarily a justified answer to the larger question of whether the action may proceed toward execution.

The distinction becomes important when authorization is distributed rather than authored by a single policy owner. A proposed operation may be governed by a constitution or source registry, several independently designated policy issuers, a joint-governance rule, an authorized choice between policy paths, and constraints tied to an earlier governance state. Policy sources may be selected and materialized by components other than the evaluator. Candidate identity and business state may be transformed by adapters. A later component may consume a favorable policy result after the governing epoch or the exact execution target has changed. Any one of these transitions can preserve the appearance of a familiar \texttt{ALLOW} response while changing what that response actually supports.

There is no need to suppose that a policy evaluator is defective in such a case. An evaluator that faithfully executes a supplied program can still be evaluating an incomplete \emph{selection} of governing policy bodies, or a correct policy under the wrong candidate or state. Conversely, a fully identified set of governing sources can be supplied to an evaluator that computes the wrong business predicate. These are different failures, require different assumptions, and cannot be resolved by counting how often one implementation agrees with another on a finite set of favorable examples. The problem is therefore not simply how to obtain a trustworthy policy decision, but how to establish the independently justified conditions under which that decision is eligible to contribute to an execution-admission judgment.

We study this intermediate judgment as \textbf{binding-complete execution admission}. The proposed Binding-Complete Execution Admission Contract (BCEAC) makes the governing basis, required source obligations, exact subject of evaluation, authorization context, policy-body provenance, and backend-result correlation explicit. Its goal is not to turn the evaluator into a universal authority over governance or physical execution. Rather, it asks which finite, checkable obligations must be satisfied before a positive policy decision may be considered by a separately governed execution boundary. The answer is deliberately conditional on an independently accepted constitutional and semantic environment; the paper neither derives the rightful set of all real-world authorities from untrusted input nor proves that an admitted operation has caused an external effect.

This question extends the conceptual trajectory of prior work distinguishing authorization from the causal ability to make an action happen \cite{Wu2026Authority}, defining execution-control and exact-version authorization stages \cite{Wu2026EBLCore}, and treating evidence coverage under explicitly scoped source assumptions \cite{Wu2026Evidence}. The present problem is narrower and temporally different from retrospective evidence coverage: it concerns the sources and context that must govern a \emph{candidate before execution admission}. It also does not revise the established definitions of an Execution Grant or Redemption; these belong to later control phases.

\subsection{A minimal counterexample: one required policy never reaches the evaluator}

Consider a precisely identified candidate action $c$ under governance boundary $b$ and epoch $e$. An independently accepted constitutional basis $C_e$, together with applicable authorized facts, establishes two required source seats, $s_A$ and $s_B$. Both must participate in the fixed governing path for $c$. The designated current body for $s_A$, denoted $p_A$, permits the action, while the designated body $p_B$ for $s_B$ denies it. Suppose a selector supplies only $S=\{p_A\}$ to an otherwise correct policy evaluator. For the supplied input, the resulting \texttt{ALLOW} may be entirely faithful to the evaluator's semantics. Nevertheless, the constitutionally required seat $s_B$ remains undischarged.

The critical distinction can be written without assuming a new general theorem:

\begin{equation*}
\operatorname{Eval}_{\pi}(c,\{p_A\},x)=\mathrm{ALLOW}
\quad\not\Rightarrow\quad
\operatorname{AdmissionEligible}_{\rho}(c).
\end{equation*}

Here $\rho$ names a declared admission profile, and the non-implication is witnessed by the omitted mandatory source. The missing object is not a missing \emph{permission} to be inferred from the other body. A constitutionally required \textbf{SourceSlot} exists independently of the availability, status, or favorable output of any policy body intended to discharge it. An expired, revoked, missing, or incorrectly sourced body cannot erase that requirement. If the relevant inventory and selection boundary are independently trusted and complete, an omitted required body can support a rejection. If necessary authority or materialization evidence is unavailable, the correct normative status may instead be unresolved. Neither condition supports positive admission.

This counterexample has a precise epistemic limit. It presupposes that $C_e$ adequately declares both required source seats for this boundary and candidate. If the accepted constitution itself lists only $s_A$ while a rightful but invisible external authority requires $s_B$, the verifier cannot deduce that hidden authority from an unchanged accepted view alone. That is the separate \textbf{A0 constitutional-adequacy assumption}, not a selection-omission defect that a local hashing or completeness algorithm necessarily detects. Binding completeness in this paper is therefore \emph{constitution-relative}, never an unconditional claim of discovering every legitimately applicable authority.

Other small changes expose different obligations. Substituting an object with the expected source label but a wrong issuer, revision, or digest may leave the counted number of bodies unchanged while invalidating source discharge. Evaluating a candidate that removes principal $C$ cannot authorize a proposal and approvals bound to removing principal $B$, even if the evaluator is correct for the former candidate. If an authorized governance path chooses branch $A$, a favorable answer on an unchosen branch $B$ cannot retroactively authorize that path. If governance membership itself is being changed, using the proposed \emph{new} membership to approve the transition may replace the frozen pre-state authorization rule with a self-serving one. These failures concern provenance, candidate identity, authorized path, and governance state respectively; a raw \texttt{ALLOW} does not encode their independent validation.

The argument does not depend on a claim that conventional policy-decision architectures are intrinsically incapable of performing such validation. A policy decision point equipped with a correctly designed, independently trusted source inventory and wrapper could reject the same omission. The problem we isolate is how to state these surrounding admission obligations as an explicit, falsifiable contract and evaluate selected aspects of that contract, without inferring completeness from the evaluator's local success. The contribution sought here is a formalized decomposition and a bounded adversarial evaluation of that contract, not exclusive ownership of the architectural pattern.

\subsection{Research problem, assumptions, and questions}

The central research question is: \textbf{Given an exactly identified candidate and an independently trusted, declared governance scope, what checkable conditions permit a policy evaluator's result to contribute to execution admission without silently omitting or substituting mandatory governance sources, identities, versions, evidence, or authorization context?} The phrase \emph{contribute to} is essential. Even after all relevant pre-execution admission checks, a favorable result does not itself confer a later Execution Grant, authorize Redemption, or demonstrate that an external action occurred.

The reference setting fixes a canonical candidate $c$, its identity commitment, a governance boundary $b$, and an applicable epoch $e$. A trusted verification environment includes the constitution, relevant source inventory and designated heads, issuer rights, application facts, any required checkpoint and time information, and rules for authorized policy alternatives. These are not authenticated merely because they appear in a request or share a digest with another self-reported object. The environment must be established through the trust roots and evidence required by the declared profile. Its actual completeness with respect to rightful real-world authority remains an external adequacy obligation. Unavailable or conflicting required information must remain visible as a nonpositive condition instead of being silently removed from the required inventory.

Within this setting, the study addresses three research questions and identifies one downstream open question. \textbf{RQ1 --- Governance-derived completeness:} How can applicable required source obligations be resolved from an independently accepted constitutional basis without allowing the currently selected or loaded policy bodies to define what was required? \textbf{RQ2 --- Binding-complete admission:} Which exact source, candidate, target, issuer, revision, digest, authorized-path, frozen-governance, and evaluator-invocation relationships must hold before a positive admission judgment is eligible? \textbf{RQ3 --- Conditional backend participation:} Can distinct policy evaluation technologies serve the bounded business-evaluation role under a declared common semantic profile while governance and admission remain independently specified responsibilities? \textbf{OQ1 --- Protected downstream execution:} Can that eligibility be connected to genuine, current, non-bypassable release, Grant, Redemption, and external effects? OQ1 is explicitly a future integration and verification problem; the present experiments do not answer it.

Two distinctions constrain the possible answers. First, \emph{structural binding} and \emph{business-policy correctness} are separate. Correctly identifying all mandatory sources does not prove that an adapter or native evaluator implements the complete intended business relation. In a qualified profile, correct business evaluation and, where claimed, preservation of the independently fixed authorized path are additional premises. Second, \emph{fail-secure} does not mean collapsing every nonpositive result into a single undifferentiated denial. A conclusively invalid requirement and a necessary but unavailable fact justify different epistemic outcomes, even though neither permits positive admission. This difference is important both for security analysis and for distinguishing a system that safely refuses from one that faithfully provides authorized service.

We do not assume that policies with identical names or syntactic structure are semantically interchangeable. Nor does the research posit that two implementations agreeing on a finite set of expected outcomes establish an independent, universally correct oracle. The trust and domain premises are part of the evaluated proposition. Where an independently necessary input is absent, a favorable evaluator decision cannot make it appear; where a backend itself miscomputes the business predicate, a Gate that deliberately does not re-evaluate that predicate cannot generally repair the semantic error. These limits are tested or disclosed rather than concealed.

\subsection{BCEAC and the proposed execution-boundary architecture}

To answer RQ1 and RQ2, we formulate BCEAC as a profile-indexed contract for the \emph{eligibility of a policy decision}, rather than as another mandatory policy language. Its independently resolvable obligations include: a trusted constitutional basis and applicable governance scope; a complete finite set of required SourceSlots for the fixed authorized path; exact designated policy-body discharge with issuer, revision, digest, validity and semantic identity; equality of candidate, proposal action and target, authorized frozen pre-state and evaluation context; trustworthy mapping and invocation association for a supported policy backend; and separately applicable host-level safety requirements. A positive admission result requires all applicable necessary obligations to be established, together with a favorable supported business-policy outcome. An authenticated contradiction can justify rejection; unresolved necessary evidence prevents a positive result without being mislabeled as a conclusive contradiction.

An important part of the specification is the distinction between an authorized source \emph{seat} and the mutable \emph{body} occupying it. Required source obligations follow the independently accepted constitution and applicability rules, not whatever happens to be selected or fetched. When a governance rule permits alternatives, a lawful choice witness $w$ fixes the path $\Psi=\Phi\mid_w$ for the attempt. The system must not replace an unfavorable chosen path with a more convenient branch after seeing the native result. Equally, a compiled EffectiveArtifact or a binding record $\beta$ can provide commitments about the checked inputs and decisions without becoming an execution-capable authority itself.

The proposed \textbf{Engine-Agnostic EBL Architecture} separates three conceptual layers. Its first layer defines canonical input, policy and evidence semantics; an EBL Language and Native Runtime may realize those semantics, but an original EBL domain-specific language is not a prerequisite for the admission contract. The second layer resolves independently trusted authority and governance, binds exact source bodies and artifacts, invokes a qualified policy evaluator through a declared adapter, and verifies the required admission relationships. In this layer, independently expected resolution and bound-context information must be compared against the actual evaluated candidate, materialized inputs and result for the \emph{same attempt}. The third layer is a separately mediated execution-control system involving Hub arbitration, a Security Module, and an Executor, with later authorization, freshness and effect-observation duties.

This decomposition is \textbf{logical, not a proof that physically separated processes have been implemented or that one architecture has a monopoly on the needed controls}. A conventional policy decision point and correctly implemented trusted enforcement wrapper can discharge comparable finite admission obligations under the same external trust and semantic assumptions. Conversely, attaching OPA or Cedar to an untrusted wrapper would not make governance complete or execution non-bypassable. \emph{Engine agnosticism} here means that the evaluator role may be populated by different \textbf{qualified engine--adapter pairs} while the independently specified admission obligations stay in force. It does not imply that arbitrary policy languages, error behavior, evaluation domains or costs are equivalent.

This separation also constrains any security theorem one might wish to establish. To justify preservation of an authorized path, business permission must imply permission under the fixed path on a separately justified, guarded subdomain; this is distinct from the backend's correctly computing the complete business predicate. To justify a positive admission judgment, implementation-level resolution and evaluation matching must faithfully refine the typed contract. Neither requirement becomes true merely because the corresponding interface has been named. The broad T1 theorem target, source-to-program derivation proof $\delta$, authentic currentness guarantees and complete implementation refinement remain open. The execution layer adds further obligations concerning protected Grant issuance, Redemption, non-bypassable mediation and externally observable effects, none of which is settled by pre-execution eligibility alone.

\subsection{Empirical approach and what its records establish}

The experimental program uses \textbf{two distinct research tracks}, not a single deployed implementation of the entire proposed architecture. The first, GSC-1, is a finite Go-based study of constitution-relative source closure, chosen-path binding, typed uncertainty and independently stipulated source/program comparisons. Its archived final run contains \textbf{44 named cases} (21 P0 and 23 P1), \textbf{48 recorded variants}, \textbf{four additional falsification controls}, and \textbf{360 prototype timing observations}. The associated \textbf{452 measurement CSV data rows} combine 44 named-case rows, 48 variant rows and 360 timing rows; they are not 452 statistically independent security tests or external actions. The named and variant records do not demonstrate release or external execution: their recorded Release and Execution states are \texttt{INELIGIBLE} and \texttt{NOT\_ATTEMPTED}. The study does not implement the general derivation checker $\delta$ or prove T1. Historical archive-integrity review also identified four P0 CSV line-ending drifts and retained an original C5 \texttt{NO-GO/BLOCKED} finding. Later derived CRLF reconstructions permit some bounded analysis but do not become independent copies of the original historical acquisitions.

The second track exercises the limited \texttt{e1lite.amount-limit.v1} profile using recorded native Open Policy Agent (OPA) and Cedar evaluations, declared adapters, a separately implemented admission Gate and Reference, and a finite family of adversarial inputs. Its R3 Final archive covers \textbf{72 scenarios}, giving \textbf{144 strict case--engine records}: \textbf{120 marked native-invoked} and \textbf{24 pre-native early stops}, with \textbf{eight additional permissive native controls} outside the strict denominator. The 144 strict records are classified as \textbf{111 \texttt{PASS}}, \textbf{24 \texttt{PASS\_PRE\_NATIVE}}, \textbf{five \texttt{UNSUPPORTED}}, \textbf{two \texttt{NOT\_CLAIMED}}, and \textbf{two \texttt{A0\_LIMITATION}}. A \texttt{PASS} here means that a record met its fixture-specific expectation; it is not synonymous with an admitted action. In the archived strict table, only \textbf{six} of the 111 \texttt{PASS} records carry Gate \texttt{ADMIT}, while the remainder include expected rejections and indeterminate results. The two backends' outputs and adapters are examined within the supported profile, not as demonstrations of general language equivalence or an independently completed three-engine comparison.

The negative history is part of the evidence. Independent adversarial reviews returned \texttt{REJECT} for R1 and R2 after finding missing validation, admission/reference conflicts and weaknesses not exposed by the original expected-case suites. R3 improved the contract but received a \texttt{CONDITIONAL} audit judgment. R3 Final subsequently received a \textbf{bounded offline \texttt{ACCEPT}} following additional input checks and strengthened archival correlation. These are versioned and scoped judgments; the final status does not rewrite earlier failures. A simulated faulty-backend \texttt{ALLOW} ablation exhibits the admitted dependency on semantic correctness: when a backend deliberately allows a disallowed business action, a Gate that checks only the separately assigned structural and contextual obligations need not correct the decision. The distinction is between a violated semantic premise and a defect in a correct-backend binding check.

The evidence has a second, independent limit. Preserved logs, raw output, source code, input commitments and checksum records support reproducible \textbf{offline consistency checks and analytical replay}. They do not, without additional provenance, constitute fresh independent observation of the historical Mac native processes. The later review re-evaluated the archived Gate/Reference relationships without rerunning native OPA/Cedar executables, and some binary and historical audit-supplement artifacts are absent from the accessible archive. Furthermore, a later adversarial review found a \textbf{required-but-nullable field-presence defect shared by the archived Gate and Reference}, outside the frozen 72-case cohort: removing \texttt{proposal.target\_member} from an otherwise valid payment input could still produce \texttt{ADMIT} and \texttt{PERMIT}. Agreement between the two implementations on the frozen records is therefore not proof of full input-schema soundness. Related legacy diagnostic-classification differences and an incorrect Claim Ledger ablation citation also remain tracked as F-04 and F-06/C11. These findings are not silently counted as failures or successful fixes within the original cohort.

Together, the tracks provide \textbf{bounded empirical support for separating governance-source and contextual admission checks from business-policy evaluation}. They do not measure the probability of resisting a future attack, certify a general checker, establish cryptographic independence of real governing parties, or demonstrate protected external execution. They also show why transparent adverse observations, test denominators and archive provenance are part of an accurate scientific claim rather than inconvenient implementation details.

\subsection{Contributions and non-contributions}

This paper makes four connected contributions, each bounded by its stated evidence and assumptions.

\textbf{First, a governance-relative admission problem and explicit contract.} We define the missing judgment between a local policy outcome and the eligibility of a precisely identified action for subsequent execution control. The BCEAC distinguishes an independently constituted Required set from the Selected or Loaded bodies evaluated by a backend, and identifies the necessary candidate, source, authorized-path and frozen-governance commitments. The formulation includes a typed distinction between disqualifying facts and unresolved necessary evidence. It is a specification of verifiable obligations under accepted authority, not proof that the authority is globally adequate.

\textbf{Second, a compositional architecture for conditionally qualified evaluators.} We describe an execution-admission architecture in which canonical semantics, constitution and source resolution, exact materialization, backend evaluation, independent result matching and admission verification have distinct responsibilities. A supported engine--adapter pair must meet an entire declared business and context contract; changing the evaluator does not remove governance or downstream enforcement obligations. The design accommodates a possible EBL Language and Native Runtime but does not require their completed implementation or assert unconditional backend substitution.

\textbf{Third, a bounded, falsifiable experimental protocol across two separate study profiles.} We preserve GSC-1's finite closure and binding evidence separately from the OPA/Cedar \texttt{e1lite} records. The evaluation exercises omissions, malformed inputs, policy provenance substitutions, authorization-state mismatches, backend-output mapping, controls and evidence correlations, while keeping actual native records distinct from pre-native handling, simulated backend faults and analytical replay. The result is a finite conformance and failure-analysis study, not a general proof of engine or Gate correctness.

\textbf{Fourth, an auditable counterexample-to-repair record.} By retaining R1/R2 rejection, R3 conditional acceptance, R3 Final's bounded offline acceptance, the GSC-1 C5 integrity history and subsequent open defects, the study exposes where a design's stated guarantees depend on incomplete schemas, shared oracles, archive provenance and external trust. This historical discipline makes the claimed boundary inspectable without relabeling repairs as proof of all future behavior.

Equally important are the \textbf{non-contributions}. We do not claim to have invented policy discovery, complete mediation, a new universal authorization language, or a capability that no trusted PDP/PEP wrapper could implement. The work does not establish globally complete governing authority, arbitrary-policy-language equivalence, a verified compilation derivation, production-grade currentness, fully authentic multi-party governance, general implementation soundness, a fair three-engine performance ranking, or actual non-bypassable external execution. These are either independent premises, future proof obligations or distinct engineering studies. Stating them as such is necessary to interpret the four positive contributions precisely.

\subsection{Organization of the paper}

Section 2 situates the admission problem relative to earlier Execution Gap and Execution Control work, evidence-coverage models, authorization and policy-decision architectures, and systems such as OPA, Cedar and conventional PDP/PEP with trusted wrappers. It asks which obligations have already been studied and what is distinctive about making this particular admission contract and its falsification criteria explicit; it does not assume that predecessor systems lack the ability to realize comparable checks.

Section 3 defines the exact candidate and governance trust model, separates Required from Selected sources, and presents omission, substitution and authorized-path counterexamples. Section 4 gives the BCEAC: independently resolved SourceSlots, source and context discharge, frozen governance, qualified business evaluation, and typed admission obligations. Section 5 develops the corresponding engine-agnostic EBL architecture, distinguishing optional authoring semantics, admission interfaces and the still-separate execution-control layer. Section 6 describes the two independent implementation/evaluation profiles, versioned adversarial history, finite experimental results and archival boundaries. Section 7 analyzes constitutional adequacy, semantic substitution, functional false denials, research validity, known implementation defects and the limits of extending admission into execution. Section 8 concludes with the resulting bounded contribution and outstanding obligations.

The organizing proposition is therefore one of \textbf{non-equivalence without architectural exclusivity}: a correct policy evaluation and an admissible execution candidate are different judgments, but the mechanisms that discharge their connecting obligations need not belong to a proprietary policy engine. Under a justified governance basis and semantic profile, the evaluator may be conditionally replaceable; the obligation to establish required-source completeness, exact binding and independent admission remains. Whether a later, protected execution system faithfully enforces that eligibility and causes the intended effect is a further question, not a conclusion manufactured from the evaluator's \texttt{ALLOW}.

\section{Background and Related Work}

\subsection{The comparison problem: evaluation, obligation selection, and execution admission}

Authorization systems answer questions about whether an identified subject may perform a specified operation on a resource under an associated context. Such answers may depend on applicable policies, hierarchical permissions, attributes, relationships, delegated authorities, and the integrity of the information consulted. The abstract distinction between a policy decision point (PDP) and a policy enforcement point (PEP) remains useful, but neither the existence of separate components nor a favorable decision label, by itself, determines the external conditions under which a consequential action may acquire execution authority \cite{OASIS2017XACML3,Saltzer1975Protection}.

For the question studied here, several familiar properties must be kept distinct. \emph{Applicability} determines which policies in an identified universe match a request. \emph{Combination} specifies how their results interact. \emph{Issuer authority and delegation} establish whether a source can authoritatively assert a rule. \emph{Discovery} obtains relevant policies or credentials. \emph{Consistency} associates a decision with an appropriate version of governing data. \emph{Enforcement} connects decisions with protected effects. A further question is whether an independently justified, action-relative governance basis identifies all the \textbf{required} source obligations before the selected materials are allowed to define the evaluation problem. A successful search for a permission proof, a complete evaluation of a supplied policy set, and a cryptographically valid policy bundle answer different questions from that last one.

The distinction is relative to declared, trusted premises. Let a canonical Candidate $c$ and a declared boundary, epoch, and admissible contextual facts determine an independently accepted constitutional basis $C_e$. The \emph{SourceSlots} required for $c$ are derived from this basis and its legally authorized path rather than from the policy bodies that happened to be retrieved. The precise formulation is given in Sections 3 and 4; this section uses the distinction to compare established approaches, not to assert that existing authorization systems are incapable of expressing it. In particular, no internal calculation can establish that $C_e$ includes every rightful authority in the external world without an additional adequacy premise. We call this the A0 boundary, consistently with the rest of the paper.

This comparison deliberately gives existing systems their strongest reasonable deployment assumptions. A conventional PDP combined with a correctly implemented, appropriately trusted wrapper may independently establish the expected policy inventory, check exact sources and versions, compare actual evaluation inputs with the expected Candidate and context, and separately enforce an admission result. Such a design may discharge the same bounded obligations. Our question is therefore not whether a particular product owns an exclusive mechanism, but how to make the necessary contract, its trust dependencies, its typed failures, and its empirical conformance criteria explicit.

\subsection{XACML, policy combination, and issuer-authority control}

The XACML 3.0 standard is a direct antecedent to several mechanisms used here. It distinguishes policy evaluation and enforcement responsibilities and specifies Targets, Rules, Policies, PolicySets, policy references, combining algorithms, decision types including \texttt{Permit}, \texttt{Deny}, \texttt{NotApplicable}, and \texttt{Indeterminate}, and obligations and advice \cite{OASIS2017XACML3}. These are substantive policy-applicability and combination semantics, not merely a Boolean function for a single isolated rule. Hierarchical PolicySets and version-aware references can also organize substantial governance logic. A claim that XACML intrinsically cannot express multiple applicable constraints would therefore be inaccurate.

An especially relevant antecedent is the XACML Administration and Delegation Profile, which specifies policy-issuer authority and delegation-related reduction procedures \cite{OASIS2014XACMLDelegation}. This source is a \textbf{2014 Committee Specification Draft/Public Review Draft}, not a mandatory feature of the XACML 3.0 Core OASIS Standard and not evidence that every compliant PDP implements the profile. Its presence nevertheless prevents us from presenting issuer authentication or delegated policy authority as concepts introduced by this paper.

These facilities clarify the remaining distinction. A valid issuer chain can justify \emph{who may issue} an encountered policy. Target matching and combining can justify \emph{how an identified PolicySet is evaluated}. Neither fact alone establishes that the accepted top-level PolicySet, its externally provisioned references, or the repository from which it was assembled constitutes an adequate inventory of every independently mandatory governor for an action. Conversely, a deployment can provide that inventory separately and bind it to the XACML input: an authorized top-level PolicySet, an exact reference/version map, and a trusted resolver are plausible components of a BCEAC-conforming implementation. Their correctness and consistency would still need verification. This is a separation of proof obligations, not an impossibility result about XACML.

Typed outcomes are particularly important for a fair comparison. \texttt{NotApplicable}, \texttt{Indeterminate}, missing required policy material, and positively established invalidity cannot be silently treated as the same Boolean \texttt{Deny} without specifying an adapter mapping and the surrounding admission profile. Similarly, a deny-overrides algorithm does not itself prove that an absent mandatory PolicyObject was included and evaluated. An admission contract must justify the expected source obligation \emph{before} concluding that the policies present in the evaluator were sufficient for that admission attempt.

\subsection{Contemporary evaluation engines and managed policy stores}

Open Policy Agent (OPA) makes policy evaluation configurable through Rego, policy/data bundles, discovery, and operational interfaces. Its bundle format supports a manifest with revisions and namespace roots, and it provides optional cryptographic signing and signature verification under separately configured trust keys \cite{OPA2026Bundles}. Discovery can supply configuration for subsequent bundle retrieval and verification \cite{OPA2026Discovery}. These are meaningful source-identity and distribution mechanisms. The official documentation also distinguishes verified bundle activation from pre-production commands that do not perform the same signature verification, and warns about ordering and conflicts among multiple bundle sources. The precise consequences depend on the loader, configuration, and selected version, not merely on the presence of \texttt{.manifest} or a digest.

A signed OPA bundle establishes integrity relative to the key and files covered by the signature; its \texttt{roots} describe a namespace ownership boundary. Neither property, by itself, proves that an external governance constitution required \emph{only} the bundles named in a supplied discovery configuration. If two independently mandatory source slots are $A$ and $B$, a correctly evaluated signed bundle for $A$ does not justify omission of $B$. An independent signed required-source map, exact version commitments, and an appropriately stabilized evaluator view could provide a compatible remedy. Whether a concrete multi-bundle deployment actually establishes the needed common version view remains an implementation-specific obligation, not a deficiency universally attributed to OPA. No such alternative OPA-plus-wrapper deployment is implemented or benchmarked in this paper.

Cedar provides an additional and in some respects stronger comparison than a minimal PDP abstraction. It defines authorization over principal, action, resource, and context (PARC); supports scoped policies, permit/forbid statements, policy validation, and analysis; and has associated formal-language and verification-guided research \cite{Cutler2024CedarExtended,Disselkoen2024CedarVGD}. Cedar's published authorization semantics employ default denial and forbid-overrides-permit. Its official authorizer documentation also describes \textbf{skip-on-error}: an erroneous policy is omitted from the combination while error diagnostics are returned. Consequently, an application's fail-secure requirements cannot be inferred solely from the final native \texttt{Allow}/\texttt{Deny} label; the adapter and admission boundary must specify the treatment of errors that are material to a required obligation \cite{CedarDocsAuthorization2026}.

Cedar explicitly recognizes the cost of supplying an entire policy and entity universe to each request and provides structured means for determining relevant data, including level-based validation and slicing \cite{CedarDocsAuthorization2026,CedarDocsLevelValidation2026}. Such selection is not presumed unsound. The distinction is between preserving the decision relative to an already justified \emph{parent set} and establishing the adequacy of that parent set against independent governance obligations. A valid slice of a constitutionally incomplete universe remains incomplete in that different, external sense; however, a trusted, complete parent inventory together with sound selection and faithful evaluation can satisfy the corresponding source obligations. We do not claim that every Cedar version, extension, or deployed Rust interface has an end-to-end mechanized proof, or that arbitrary Cedar policies are equivalent to the EBL path expression.

Managed authorization services further narrow any plausible novelty claim. Amazon Verified Permissions names the \emph{considered policies} selected for an authorization request and uses policy-store scope to select relevant policies from a named store \cite{AWS2026AVPTerms,AWS2026AVPStores}. A requesting application identifies the store; the service evaluates policies relevant to that request. A system may therefore perform genuine policy selection without implementing BCEAC as a distinct research contract. The question for an action spanning several independently governed sources is who establishes that the selected store or stores constitute the required governance universe and how their exact material, state, and version are bound to the admission attempt. That question may be answerable by application integration and trusted surrounding controls; this paper does not establish that a managed service cannot answer it.

AWS Identity and Access Management (IAM) also demonstrates that mature authorization evaluation is not reducible to a flat conjunction of equally situated rules. Identity- and resource-based permissions, permissions boundaries, organization-level service or resource controls, and explicit denies interact according to policy-type-specific combination rules \cite{AWS2026IAMPolicyEvaluation}. A BCEAC profile cannot replace those semantics with an indiscriminate global \texttt{AND} without changing the authorization relation. Its required obligations must reflect the \emph{declared and legitimately authorized} combination and alternative-path rules of the supported profile. In our bounded real-engine evaluation, the native business relation itself remains separate from the independent source-binding checks; neither the presence of an admission Gate nor a favorable gate outcome proves the business relation correct for arbitrary inputs.

\subsection{Distributed credentials, delegated authorization, and proofs}

Distributed policy discovery and verifiable authorization have a substantial research history. Proof-carrying authorization systems can acquire policy fragments across administrative boundaries and construct authorization proofs from fetched material. Bauer, Schneider, and Felten demonstrate a web access-control architecture with iterative proof construction and policy retrieval across hosts \cite{Bauer2002FlexibleWebAccess}. SPKI/SDSI certificate-chain discovery supports delegated authorization and includes threshold subjects requiring cooperation among principals \cite{Clarke2001SPKISDSI}. SecPAL supplies a language for decentralized authorization, controlled delegation, constraints, and an execution method whose soundness, completeness, and termination are established under stated syntactic restrictions \cite{Becker2007SecPAL}. These works preclude any claim that the present paper first combines distributed policy acquisition, multi-principal authority, and machine-checkable authorization reasoning.

The remaining comparison concerns the \emph{statement being proved}. A proof that a particular requester has sufficient credentials for an authorization goal need not, by that fact alone, demonstrate that all independently mandatory restrictions on this exact Candidate were identified and checked. The difference can be expressed using a trusted, declared required-source inventory: if a source slot must be discharged, discovering one successful permission chain cannot make that unaddressed source disappear. Yet the same older proof systems may be combined with a trusted inventory and required queries so that the more demanding statement becomes part of the proof goal. Accordingly, we do not infer that SecPAL, SPKI/SDSI, or proof-carrying authorization are incapable of representing an equivalent obligation in an appropriately constrained setting.

Three distinct notions of completeness must be maintained. Logical completeness concerns whether an inference mechanism can derive the consequences specified by its language and premises. Discovery completeness concerns a suitably characterized set of available policy or credential material. \emph{Constitution-relative admission completeness} concerns whether all independently required obligations for the Candidate have been discharged under a justified governance basis. None implies adequacy of a hidden real-world authority universe merely from a syntactically correct policy, complete derivation, valid credential chain, or signed manifest. The additional accepted-governance premise remains external; describing it precisely is not equivalent to discharging it cryptographically.

\subsection{Temporal authorization, usage control, and protected effects}

Version and time are not novel concerns in authorization research. Zanzibar presents a globally distributed authorization architecture with consistency properties designed to preserve causal ordering amid changes to access-control lists and object contents \cite{Pang2019Zanzibar}. The UCON$_{ABC}$ model explicitly addresses authorizations, obligations, conditions, ongoing controls, and mutable attributes \cite{Park2004UCONABC}. These are substantial precedents for treating authorization as dependent on state and for re-evaluating a decision as relevant facts change. The temporal questions in this paper must therefore be read as precise obligations for a \emph{particular} declared governance snapshot, policy-source head, Candidate, and admission attempt, rather than as an assertion that previous authorization systems ignored freshness or revocation.

Even a correctly formed binding record does not prove that its referenced state is current at an eventual release. Verification time, release time, Redemption time, and external-effect time may differ, and no cryptographic digest alone solves all intervening revocation or concurrency questions. Zanzibar-like consistency machinery could inform a future deployment; this paper does not establish that the archived prototypes implement Zanzibar's consistency contract or a substitute for it. The necessary checkpoint, epoch, anti-rollback, and cross-phase mediation guarantees remain independently stated research and engineering obligations.

The protection principle of \emph{complete mediation} likewise long predates this study \cite{Saltzer1975Protection}. A PDP, trusted wrapper, or policy verifier can reject an unsafe request while an unmediated administrative, recovery, or direct device path still causes an effect. Conversely, a legitimate admission decision may fail to produce an effect because an Executor, credential, device, or external service refuses or fails. Thus the BCEAC concerns the conditions under which an evaluator's answer is eligible to contribute to \textbf{pre-execution admission}. It does not establish the separate Hub, Security Module, Grant, Redemption, Executor, or real-world-effect properties proposed for a later protected execution boundary.

\subsection{Relationship to the preceding execution-control research}

The present question also builds on three earlier HAVENLON preprints, whose different objects of study must not be collapsed into a single previously proved theory. The authority-decomposition work asks which trust-domain coalitions can actually cause protected execution, including alternate control paths, and expressly does not certify global non-bypassability \cite{Wu2026Authority}. EBL-Core specifies a fully materialized Candidate, evidence and policy obligations, a decision-binding Execution Release Contract (ERC), and a separate Execution Grant with version-constrained Redemption; the ERC is not itself the capability to execute \cite{Wu2026EBLCore}. The evidence-coverage study uses declared scope, branch, trigger, deadline, verifier view, and cutoff to assess \textbf{retrospective} coverage of records due for an execution account \cite{Wu2026Evidence}.

Paper IV addresses a different, \textbf{prospective} question. Before an evaluator's answer can contribute to admission, which independently governing sources had to participate, which exact PolicyObjects could discharge their SourceSlots, which authorized path was selected, and was the exact Candidate, governance snapshot, context, and backend invocation preserved? Retrospective evidence completeness and prospective policy-source binding may share the methodological discipline of defining an independently justified required set and reporting unknowns explicitly, but they are not interchangeable predicates. Likewise, EBL-Core's release-and-redemption profile does not already prove the source-to-program derivation, governance inventory adequacy, or concrete interface refinement sought here. The earlier preprints are related conceptual and normative work, not independent evidence that Paper IV's currently proposed checks are deployed, generally sound, or accepted by peer review.

\subsection{Equal-assumption comparison and the actual research increment}

A fair comparison places BCEAC and conventional authorization compositions under the \textbf{same} external trust, completeness, freshness, and business-semantic assumptions. Consider a conventional evaluator wrapped by a trusted admission component. Before evaluation, the wrapper obtains the independently authorized governance constitution; resolves an action-relative, finite source obligation expression; fixes a legally authorized path; checks the designated issuers, revisions, digests, state and selected/loaded material; and records the exact evaluator invocation. It then interprets the qualified backend result and applies a typed admission rule. If all these operations are correct and non-bypassably connected to the relevant admission interface, such a composition can meet the same scoped BCEAC obligations. Describing its architecture as a PDP with a trusted wrapper rather than an EBL Native Runtime changes no relevant safety premise by itself.

This possibility is a deliberate \emph{falsification challenge} to the research framing, not a result of a performed comparison. The source-version ledger identifies candidate OPA, Cedar, and XACML/AuthzForce artifacts for a future comparable profile, but a complete equal-assumption three-way implementation and whole-system-cost study was not completed. The archived real-engine evaluation covers a specified OPA/Cedar profile with its adapters; it does not show that competing wrapped systems cannot reproduce the same finite properties or that a particular engine has lower cost or weaker security. In particular, the earlier XACML comparison work was blocked by local preparation and artifact-readiness issues and supplies no fair three-engine performance ranking.

The paper's supported increment is consequently \emph{contractual and methodological}: (i) state independently determined, constitution-relative required-source obligations rather than deriving them from the selected set; (ii) bind each designated source, exact Candidate, frozen governance, authorized path, context, and evaluator input/output under typed fail-secure admission conditions; (iii) distinguish complete business semantics and any separately guarded path-preservation obligation from structural binding; and (iv) make the claims testable through explicitly bounded falsification cases, archival provenance, adverse review, and recorded counterexample-to-repair history. These are specifications, architecture constraints, and finite empirical observations under stated assumptions, not a claim of historical firstness or an unconditional theorem of secure execution.

The condition matters as much as the construction. An omitted rightful governor outside an inadequate accepted constitution remains the A0 limitation. A semantically faulty backend may return \texttt{ALLOW} despite correct structural binding; an independent Gate not designed to recompute the entire business predicate cannot generally repair that error. A one-way no-unsafe-positive property can hold for an always-denying implementation while functional equivalence fails. Missing authenticated source and context evidence cannot be transmuted into permission. Finally, successful admission does not imply a protected external effect. Section 3 makes these distinctions precise, Sections 4--5 specify the necessary contract and architectural interfaces, and Sections 6--8 describe their bounded evaluation, limitations, and conclusions.

\subsection{Summary of the relationship to prior art}

Existing authorization research already supplies expressive policy combination, delegated issuer authority, remote policy discovery, formal-language reasoning, policy and entity selection, consistency-aware authorization, ongoing usage control, and the principle of complete mediation. The distinction defended here is narrower: a high-risk action's \emph{admission} should carry a checkable justification relative to an independently accepted governance basis and the exact artifacts and evaluator context actually used. This formulation does not displace those earlier mechanisms. It exposes explicit proof obligations at their integration boundary and states which obligations remain unverified. The present evidence supports that bounded framing, not an exclusive new authorization primitive, complete constitutional discovery, arbitrary engine interchangeability, or already protected execution.

\section{Problem Definition and Trust Model}

\subsection{From a Policy Decision to an Admissible Execution Candidate}

Policy engines answer questions relative to the policies and contextual inputs presented to them. A correctly implemented evaluator may return \texttt{ALLOW} because every relevant expression \emph{in the supplied policy program} evaluates favorably. The logical scope of that result is important: correctness relative to a supplied program does not establish that the program includes every source entitled to constrain an action, that those sources represent their current authorized revisions, or that the object evaluated is identical to the object later proposed for release. A policy engine need not have made an evaluation error for the surrounding authorization workflow to be incomplete.

We call this gap the \emph{admission problem}: the transition from a policy result to a justified claim that a particular proposed action may enter a later execution-control phase. Admission is not synonymous with policy evaluation, approval, or execution. A useful admission judgment must identify the governing boundary; determine which sources of authority were required for the exact action; account for their authorized policy bodies and governance state; and bind the native evaluator's output to the same contextual commitments. These obligations can be independently specified even when their concrete implementation is integrated into an existing policy-decision point or trusted enforcement wrapper. We do \textbf{not} claim that conventional access-control architectures are incapable of satisfying them.

The distinction extends a sequence of execution-oriented research questions. Analysis of execution authority asks which principals can actually cause a protected state transition, rather than merely who possesses a workflow label or approval \cite{Wu2026Authority}. The EBL-Core account distinguishes a fully materialized Candidate and its bound Execution Release Contract (ERC) from a subsequent Execution Grant and version-constrained Redemption \cite{Wu2026EBLCore}. Work on evidence coverage asks whether an independently justified inventory of post-execution recording obligations has been covered, rather than treating observed records as their own proof of completeness \cite{Wu2026Evidence}. Here the relevant inventory is different: we examine the \emph{pre-execution governance sources and conditions} that must constrain a Candidate before a backend decision is eligible to contribute to admission. Retrospective evidence coverage and prospective governance-source binding are not the same predicate.

Our target is a \textbf{governance-relative execution-admission contract}. It is engine-agnostic only in the following conditional sense: a backend that implements a declared, supported semantic profile and a faithful adapter may provide the policy-evaluation component of the contract, while admission-specific verification remains separately accountable. An EBL Native evaluator, OPA, or Cedar can therefore be considered for the backend position without asserting that their native languages, feature sets, error conventions, or arbitrary programs are equivalent. The contract does not require all third-party policies to be authored in an EBL-specific textual language; it does require an explicit common semantic and binding interface for any particular interoperability claim.

The proposed architecture organizes this distinction into three responsibilities. An external EBL semantic interface describes Candidates, governance conditions, evidence, state, and execution boundaries. An admission architecture resolves required governance seats, validates their bindings and frozen approval context, invokes a qualified policy backend where appropriate, and produces a typed pre-execution result. A separate execution-control boundary---potentially Hub Arbitration, a Security Module, and an Executor---may later mediate release and effects. This third boundary is \textbf{proposed but not established by the experiments reported in this paper}. Admission eligibility does not confer a physical capability to force the protected effect.

\subsection{Candidate Identity and the Declared Governance Basis}

Let $c$ be a fully materialized, canonical Candidate in a declared encoding profile, and let $\operatorname{cid}(c)$ denote its commitment. The Candidate includes the action, the relevant object or target, and all parameters that can affect the decision. The commitment must be interpreted relative to its canonicalization and domain-separation rules: equal-looking descriptions are not necessarily equal committed Candidates, and any change to a committed decision-relevant field yields a different Candidate. Exact commitment is a property of the represented action; it does not, by itself, establish that the human or agent who supplied the request intended that action.

Governance is evaluated relative to a boundary $b$ and an applicable epoch $e$. We write $C_e$ for the independently authorized constitution at that epoch and $f$ for authenticated or explicitly trusted applicability facts. A constitution specifies which classes of actors or source seats may govern particular operations, what policy-source and governance-transition rules are in force, which issuers and revisions are authorized, and which alternatives are legally available. A \emph{governance constitution} in this model is an authority-bearing reference state; it is not simply another caller-supplied policy file. An evaluated set of policies cannot appoint its own missing or replacement governors without an independently authorized transition.

The central modeling distinction is between a \textbf{SourceSlot} and a \textbf{PolicyObject}. A SourceSlot $s$ is a constitutive position of required authority. A policy reference $r_s$ names an authorized policy body that may discharge the position at the relevant phase. The reference may carry a slot identifier, issuer, revision, semantic-profile identifier, scope, content digest, validity status, and applicable version constraints. Such attributes must be validated under the declared profile rather than accepted because an untrusted body describes itself as legitimate. A slot's normative existence is not contingent upon successfully fetching its body. Accordingly, a missing, expired, revoked, or conflicting policy version can be a reason for non-admission, but cannot by itself remove the slot that made the source necessary.

To avoid circular definitions of completeness, applicability is computed from the accepted constitution and permitted facts, \textbf{not} from the policies the caller happened to select or load. In the broader model, let

\begin{equation}
\Phi(c,b,e,f)=\operatorname{ResolveObligations}(C_e,c,f).
\label{eq:c3-1}
\end{equation}

The resolver is a \emph{partial-information, typed} operation, not an unconditionally defined Boolean oracle: its successful branch yields a finite governance-obligation expression and its other branches report conclusively invalid or unresolved evidence. The resolver must use an independently authorized Candidate/action-to-boundary mapping rather than treating a caller-provided boundary label as sufficient authority. The profile must specify which trusted facts and current heads are available at the verification phase, and whether each fact is verified, conclusively invalid, or unavailable. A representative grammar for the successful branch is

\begin{equation}
\Phi ::= \top \mid \operatorname{require}(s)
\mid \operatorname{all}(\Phi_1,\ldots,\Phi_n)
\mid \operatorname{anyAuthorized}(q,\Phi_1,\ldots,\Phi_m),
\label{eq:c3-2}
\end{equation}

where conjunction denotes jointly mandatory obligations and the last constructor represents \textbf{constitutionally authorized alternatives} named by a governance rule $q$. The bounded grammar is finite; $n\geq 1$ for conjunction and $m\geq 2$ for authorized alternatives, following the applicable formal-model profile. A path witness is keyed by unique reached alternative-node identity (e.g., a tree address), not merely a rule label that might appear more than once. The terminal $\top$ denotes an independently established absence of additional slot obligations; it must not be inferred from an empty query response. Mandatory Root or host constraints are not waived by an apparently empty policy catalog. In this bounded formal framing, the obligation resolver is required to terminate or return \texttt{UNRESOLVED}; an unknown guard is not silently interpreted as a false applicability condition.

When an authorized alternative exists, an authenticated witness $w$ fixes a permissible choice for the reached alternative nodes, bound to $\operatorname{cid}(c)$, $e$, relevant facts and the applicable constitutional rule. Instantiation produces the fixed legal path

\begin{equation}
\Psi=\operatorname{Instantiate}(\Phi,w)=\Phi\!\mid_w.
\label{eq:c3-3}
\end{equation}

An unselected sibling branch cannot later contribute favorable results to the branch that was selected. Witness absence, ambiguity, conflict, or unauthorized branch selection prevents a positive closure judgment. This path machinery comes from the broader governance-scoped formal model. The final OPA/Cedar amount-limit experiment does \textbf{not} implement or prove arbitrary nested authorized-alternative semantics; these are formal obligations for a broader admission profile.

The claim is explicitly relative to the declared authority basis. A valid digest, signature, or internally consistent manifest can establish identity and integrity relative to trusted keys and rules, but cannot demonstrate that an honest-looking authority omitted no source it was obligated to declare. In two otherwise indistinguishable worlds, the accepted constitution may list only source $s_A$, while an undisclosed rightful source $s_B$ also exists in one world. If the verifier has no independent observation distinguishing them, it cannot infer the hidden obligation from the same manifest bytes. We therefore assume \textbf{constitutional adequacy for the stated boundary}, not global knowledge of all possible authorities. Completeness below means closure \emph{within that independently justified boundary}, never discovery of every obligation in the world.

\subsection{Binding, Evaluation, and Admission Are Distinct Judgments}

Let $S$ be the set of selected, validated policy bodies or references supplied for policy evaluation. Define the predicate $\operatorname{Discharges}(\Psi,S)$ to mean that each mandatory slot along the fixed authorized path has been discharged by the unique appropriately designated active policy object, under the constitution's issuer, revision, digest, status, and semantic rules. This is not mere inclusion of strings naming policies. Two objects with the same visible slot label can differ in governing issuer; a valid object at the wrong revision can be stale; and a body belonging to an unselected authorized branch cannot automatically replace a mandatory source on the chosen branch.

The discharge predicate is deliberately \textbf{structural}. It can be true when a correctly included mandatory policy denies the Candidate. Conversely, it can be false when the only policy that was supplied correctly allows the Candidate. Extra properly authorized restrictions may coexist with required sources under explicit rules; they may not masquerade as required sources or introduce permission by replacing a mandatory constraint. This distinction also leaves ordinary local business semantics intact: a duly authorized Auth Key whitelist rule may change which internal expression a policy evaluates without deleting independently constituted Hub or Root source obligations.

For clarity, consider three logically distinct outcomes:

\begin{equation}
\begin{aligned}
\mathrm{BindingResult}&\in\{\mathrm{VERIFIED},\mathrm{REJECTED},\mathrm{UNDETERMINED}\},\\
\mathrm{PolicyDecision}&\in\{\mathrm{ALLOW},\mathrm{DENY},\mathrm{INDETERMINATE},\mathrm{ERROR}\},\\
\mathrm{AdmissionDecision}&\in\{\mathrm{ELIGIBLE},\mathrm{REJECT},\mathrm{UNRESOLVED}\}.
\end{aligned}
\label{eq:c3-4}
\end{equation}

These names are \textbf{paper-level abstract result types}, not a representation of the exact enumerations in any frozen implementation. In particular, the R3 Final Gate reports \texttt{ADMIT}, \texttt{REJECT}, or \texttt{INDETERMINATE}, while its independent Reference reports \texttt{PERMIT} or \texttt{FORBID}; the GSC-1 research specification uses its own typed outcomes. No proof follows from renaming those statuses to the abstract algebra above. What matters is the non-coercion principle: \texttt{VERIFIED} binding does not imply \texttt{ALLOW}; policy \texttt{ALLOW} does not imply \texttt{ELIGIBLE}; \texttt{ELIGIBLE} does not issue a Grant or demonstrate an external effect.

Write $\operatorname{Eval}_{\pi}(c,S,x)$ for a policy decision under backend-and-adapter profile $\pi$, where $x$ is the exact mapped contextual input. A positive admission judgment requires more than that native result. Schematically, for an admission context $\mathcal{A}$,

\begin{equation}
\begin{aligned}
\operatorname{Eligible}(c,\mathcal{A})\Longrightarrow{}&
\operatorname{TrustedGovernance}(b,e,C_e,\kappa,f)\;\land\\
&\operatorname{RequiredPathResolved}(C_e,c,f,w,\Psi)\;\land\\
&\operatorname{Discharges}(\Psi,S)\;\land\\
&\operatorname{ExactContextBinding}(c,b,e,w,S,x)\;\land\\
&\operatorname{ValidBackendInvocation}(\pi,c,S,x)\;\land\\
&\bigl(\operatorname{Eval}_{\pi}(c,S,x)=\mathrm{ALLOW}\bigr)\;\land\\
&\operatorname{AdmissionHostFloor}(c,\mathcal{A}).
\end{aligned}
\label{eq:c3-5}
\end{equation}

Here $\kappa$ denotes independently established checkpoint/currentness information where the chosen deployment profile requires it. Equation~\eqref{eq:c3-5} is a \textbf{necessary-condition specification}, not a proved biconditional and not an implementation-wide soundness theorem. The conjunction collects requirements that may be checked by several components or in different phases; it is not a claim that one sequential function implements every line. The Section 4 contract distinguishes the abstract, path-relative permission relation $\operatorname{Permit}_{\Psi}$ from a backend's interpreted business permission. \textbf{For any profile that claims semantic non-weakening, a separate domain-scoped obligation must connect the two:} the backend's correctly interpreted positive permission must imply permission under the authenticated fixed governance path. Equation~\eqref{eq:c3-5} does not silently assert that relationship is verified by R3 Final; it remains a separately named premise or proof obligation. The Section 4 contract will give profile-indexed input and outcome rules, including which errors are provable rejection and which are unresolved. It will not infer authenticity from the mere presence of fields or claim that all conjuncts were tested in R3 Final.

The backend and Gate occupy different proof responsibilities. A backend evaluates the policy's business predicate over a correctly mapped input. Admission-specific logic checks independently observable provenance, source completeness, candidate identity, governance commitments, supported-result mapping, and non-waivable host conditions. A Gate that does \textbf{not} reevaluate the backend's business predicate cannot necessarily detect a backend that falsely returns \texttt{ALLOW} for a prohibited business action. A correct business evaluator alone is also insufficient for a \emph{path-preservation} claim unless the intended business relation is normatively connected to the selected $\Psi$; otherwise a backend could correctly compute a permissive OR while a mandatory conjunct on the fixed path denies. Therefore, backend semantic correctness and adapter fidelity are explicit trust or verification obligations. Separating responsibilities makes this premise visible; it does not eliminate it.

There is also an epistemic distinction between \texttt{REJECT} and \texttt{UNRESOLVED}. If an independently trusted catalog conclusively states that a required version is revoked, binding fails on known evidence. If an authenticated checkpoint cannot be obtained, the verifier lacks grounds for currentness; calling the action safe would be unjustified, even if no specific revocation has been demonstrated. Both outcomes prohibit positive admission in a fail-secure design, but keeping their reasons separate matters for auditability, availability analysis, and future phase-specific recovery. The frozen R3 fixtures do not provide real checkpoint availability evidence and should not be presented as testing this complete distinction. Moreover, abstract $\mathrm{REJECT}$/$\mathrm{UNRESOLVED}$ is a \emph{normative classification}, not a literal description of every historical Gate branch: R3 Final classifies at least one known malformed quorum threshold as \texttt{INDETERMINATE}. The distinction is diagnostic; both statuses remain non-positive in the archived cases.

\subsection{Minimal Omission and Context-Substitution Counterexamples}

A small omission example shows why evaluation correctness and binding completeness are different propositions. Let $C_e$ independently require two slots, $s_A$ and $s_B$, for Candidate $c$. The instantiated path is $\Psi=\operatorname{all}(\operatorname{require}(s_A),\operatorname{require}(s_B))$. Suppose the current, authentic policy body for $s_A$ permits $c$, while the body for $s_B$ denies it. An upstream selector supplies only $S=\{p_A\}$. The policy evaluator may correctly compute

\begin{equation}
\operatorname{Eval}_{\pi}(c,\{p_A\},x)=\mathrm{ALLOW},
\qquad
\neg\operatorname{Discharges}(\Psi,\{p_A\}).
\label{eq:c3-6}
\end{equation}

No contradiction exists: the policy evaluator accurately evaluated what it received; the selector failed to present a required source. Consequently, \texttt{ALLOW} cannot be promoted to admission eligibility. If an independently mandated source remains absent, the correct structural outcome is non-positive irrespective of whether an evaluated policy permits. This is an adversarial \emph{selection omission}, not necessarily a defect in OPA, Cedar, or another evaluator. A complete policy-decision workflow with a trustworthy source inventory and wrapper could implement the same check.

The example also distinguishes \textbf{completeness from favorability}. Suppose both authentic bodies are supplied and the conjunction is evaluated faithfully. The source seats are discharged, so structural binding may be verified, but $p_B$ returns \texttt{DENY}. The correct admission outcome remains non-positive for a policy reason, not because the source inventory is incomplete. Treating any backend denial as evidence of missing binding would wrongly classify an intentionally enforced restriction as a structural defect.

A second variant replaces the policy body for $s_B$ with one bearing a familiar identifier but an unauthorized issuer or the wrong digest. Counting two objects is then insufficient. The designated $s_B$ is still not discharged because authority-bearing provenance is part of source identity. The same applies if an old valid revision was superseded or conclusively revoked. Conversely, lack of current status information is not proof of revocation; it supports an unresolved result. These alternatives motivate typed rather than Boolean completeness judgments.

Context substitution can survive an otherwise complete-looking source list. Imagine a proposal and its approvals commit to removing principal $B$, but the candidate presented to a backend targets principal $C$. The backend may correctly permit the latter action under its mapped business policy, while the approval context authorizes a different operation target. A pre-execution admission contract must bind the candidate action, target, proposal identity, and relevant governance version to the approved object, rather than treat the backend's result as transferable between Candidates. In the R3 Final fixtures, exact target binding is exercised by \texttt{candidate\_target\_not\_approved}; this is a bounded schema/governance test, not proof that actual human signers endorsed either object. Similarly, \texttt{proposal\_id} format validation is not signature authentication or replay resistance. A subsequent, separately conducted adversarial review of the archived Gate and Reference code identified a shared input-schema counterexample outside the frozen 72-case cohort: deleting required but nullable \texttt{proposal.target\_member} from the payment baseline left the historical Gate and Reference at \texttt{ADMIT}/\texttt{PERMIT}, although the frozen input contract requires \emph{presence} with a \texttt{null} value for payment. This is a required-field presence defect, \textbf{not} evidence of forged payment approval or an exploited hardware execution path. It shows that agreement between Gate and Reference, including the archived 144/144 replay, must not be generalized to full input-contract conformance.

A fourth issue arises under authorized alternatives. If $\Phi=\operatorname{anyAuthorized}(q,\allowbreak\operatorname{require}(s_A),\allowbreak\operatorname{require}(s_B))$ and the lawful, committed witness $w$ selects $s_A$, it is not valid to authorize the action from a favorable evaluation of $s_B$ when $s_A$ denies. The reference obligation is $\Psi=\Phi\mid_w=\operatorname{require}(s_A)$, not an unconstrained OR over both branches. This illustrates a separate proof obligation from simple slot inclusion. The GSC-1 theoretical track specifies it; the final two-backend amount-limit experiment must not be cited as a general verified implementation of arbitrary authorized-choice compilation.

These constructions do not, by themselves, prove that the proposed Gate is sound. They establish why a correct local evaluator result is logically insufficient without independently justified source/context premises. Demonstrating useful correctness additionally requires an implementation-level specification, negative and positive cases, conformance with a reference interpretation, and an honest accounting of the untested inputs and assumptions.

\subsection{Adversary Capabilities and Explicit Trust Premises}

The principal adversary may influence the source selector, caller-supplied claims of completeness, fetched policy bodies, intermediate caches, untrusted contextual facts, and incorrectly correlated backend outputs. The adversary can omit required sources, substitute an expired version, present a wrong issuer, alter action or target fields, select an unauthorized governance branch, or supply a result produced for another candidate. It may also arrange coordinated inconsistencies across software components. Independence is not established merely because components have different process names or are called separate approval services. The paper does not assume that the caller-controlled set of loaded sources can authenticate its own completeness.

The guarantee boundary is nevertheless conditional. The proposed admission contract relies on at least four classes of externally justified or separately verified premises, each with a distinct failure mode.

\textbf{Constitutional adequacy and independent authority.} The normative authority inventory must be sufficiently complete for the declared boundary. The verifier must possess an appropriate independent authority basis for determining the roster of slots and governing rules. This premise is often called A0 in the broad formal model. It is not a claim about global authority enumeration, and cannot be discharged by hashing the inventory under inspection. If the recognized authority intentionally omits an external rightful governor and no independent source reveals the omission, the admission component cannot manufacture that missing knowledge. The meaningful soundness claim is therefore constitution-relative.

\textbf{Authentication, currentness, and phase identity.} Constitution versions, designated policy heads, status information, and any checkpoint used to establish the applicable epoch must be authentic, properly authorized, and sufficiently current for the phase. A valid signature on a stale constitution does not settle which constitution the present Candidate must obey. Neither a digest nor a monotonic-looking integer proves anti-rollback unless verified against a trustworthy freshness or checkpoint mechanism. Where the phase contract demands currentness and its witness is unavailable, the verifier must refrain from positive admission. The R3 Final experiments install a static manifest as a trusted fixture; they do not establish signatures, checkpoint infrastructure, real signer identities, or revocation publication correctness.

\textbf{Backend semantic and adapter correctness.} For the declared semantic fragment, the backend is assumed to evaluate its business predicate correctly and the adapter is assumed to bind the exact native result to the corresponding candidate and mapped input. Unsupported constructs, evaluator errors, and ambiguous output mappings must not be silently converted to permission. The assumption cannot be omitted simply because two independently implemented backends agree on tested inputs. Nor does disagreement imply which implementation is correct without an independent normative interpretation. The deliberately faulty-backend ablation recorded in the \textbf{original R3} artifact (\texttt{REAL-02}, simulated forced \texttt{ALLOW}) demonstrates the boundary: forcing \texttt{ALLOW} for an over-limit business input can yield a Gate/Reference conflict when the Gate is intentionally not a second business-policy interpreter. That injection is an artificial failure of the backend premise, not a documented native OPA or Cedar semantic bypass.

\textbf{Admission-verifier soundness and trusted invocation.} Source-seat enumeration fidelity, provenance checks, exact context binding, frozen quorum computation, decision/result correlation, and host-floor checks must be correct and actually invoked on the paths for which admission is claimed. They are not proven for all possible inputs because a collection of unit tests and recorded mutations passes. The R3 Final archived cases, independent Reference replay, and evidence-tamper checks provide falsifiable observations for a restricted input family; they do not supply a general refinement proof from the mathematical contract to an arbitrary deployed Gate. Moreover, a malicious downstream actor that ignores the Gate result can cause an external action unless independently mediated by the later execution boundary, which is outside this study's empirical claim.

Several related dependencies follow from these classes. Material applicability facts must be authenticated or otherwise admitted by a declared fact-trust rule; unknown facts cannot conveniently narrow Required. Signer control-domain separation is an organizational and cryptographic property, not implied by distinct strings in a JSON roster. The proposal's stable identifier is necessary for some profiles but insufficient to prove origin. If an execution environment relies on a fixed host safety floor, that floor cannot be weakened merely by substituting or omitting an ordinary policy source. If an experiment tests only a single tenant context, it cannot support a general cross-tenant isolation claim; the R3 Final \texttt{tenant\_not\_bound} case is specifically labeled \texttt{NOT\_CLAIMED}.

The four premises make the study \textbf{falsifiable and bounded}, not vacuous. They identify which obligations an admission implementation can check locally and which must be discharged by a surrounding root-of-trust, policy semantic specification, approved governance process, or protected release mechanism. A security claim that hides those dependencies is stronger in wording but weaker in explanatory value. A system that always rejects might satisfy a one-way no-unsafe-admission condition while failing any useful availability or functional-conformance objective; thus safety and usefulness must be evaluated separately.

\subsection{Frozen Governance Snapshots and Time-of-Check Boundaries}

A governance operation can modify the membership or threshold rules by which governance operations are approved. This creates a self-reference risk if the operation is allowed to use the state that it is attempting to create as its own source of authority. To avoid that circularity, a normal governance-changing proposal must be evaluated against the \textbf{applicable authenticated pre-state} and its governing operation descriptor. Let

\begin{equation}
G_0=(e_0,M_0,h_0,\rho_0),
\quad M_0=\{A,B\},\quad h_0=2,
\label{eq:c3-7}
\end{equation}

where $\rho_0$ identifies the authorized rules for the operation. Suppose principal $A$ proposes to remove principal $B$. The proposed post-state may have membership $\{A\}$ and a single-member threshold, but those prospective values must not be used to authorize the very transition that produces them. Under the assumed normal 2-of-2 rule, an approval from $A$ alone is insufficient for the transition. The fact that $B$ is the target does not, by itself, erase $B$ from the voting basis in the applicable pre-state. Valid emergency or exclusion exceptions, if intended, require independently constituted rules rather than opportunistic recalculation.

This is a rule about \emph{which snapshot governs a proposal}, not a universal impossibility theorem for member removal. Full governance security also depends on signer authentication, the distinction between nominal identities and independent control domains, replay resistance, concurrency, status and credential changes, and a coherent cross-root epoch. A cryptographically signed roster can still be stale; two different signed governance sources can disagree about their current versions. The correct implementation must explicitly determine whether it can establish a coherent applicable snapshot or must report an unresolved conflict. The bounded R3 experiments model the 2-of-2 to 1-of-1 issue using fixed fixture data and test quorum/target/version relationships; they do not authenticate real signers or evaluate the entire membership lifecycle.

Time creates a further separation. Let $t_v$ be the verification time, $t_l$ a later release time, and $t_r$ a possible Redemption time. It is invalid to infer solely from successful source and context checks at $t_v$ that the same policy version, governance membership, Candidate commitment, or authorization evidence is suitable at $t_l$ or $t_r$. A state transition, revocation, new checkpoint, or concurrent governance update may change the relevant facts. This is the time-of-check/time-of-use problem for execution admission: a historical positive result is evidence about the checked phase, not an irrevocable authorization for all later phases.

The appropriate mitigation belongs to a specified multi-phase contract: for example, a later authority may require a fresh comparison against the ERC's exact committed version, an independently protected checkpoint, or one-time Grant/Redemption rules. The earlier EBL-Core work separates those lifecycle concepts and does not authorize using an old Grant merely because a newer, superficially stricter policy also allows the same Candidate \cite{Wu2026EBLCore}. Paper IV does not replace that lifecycle with an admission-only Boolean. Our study ends at pre-execution eligibility and neither constructs nor observes actual protected Redemption or external effects.

A final provenance distinction matters for the proposed architecture: an EffectiveArtifact or compiled program may be reproducibly linked to a set of loaded inputs and may retain a verifiable digest, yet it cannot by self-description prove that the trusted constitution's complete applicable inventory was supplied. Conversely, a perfectly complete source-binding decision may be rendered obsolete before execution unless the downstream phase rechecks the required epoch. Both facts motivate separating source binding, policy evaluation, admission eligibility, and final execution authority.

\subsection{Research Obligations and Empirical Scope}

The problem definition yields three research questions for this paper and one explicit downstream question. \textbf{RQ1} asks whether a required governance-source inventory can be resolved from an independently authorized basis, with SourceSlots retained when their bodies are missing, expired, or revoked. Its scope is a specified constitution, not unknown worldwide authorities. \textbf{RQ2} asks which typed provenance, candidate, policy-result, and frozen-governance checks are necessary for a positive pre-execution admission decision. \textbf{RQ3} asks whether distinct policy evaluators can serve the backend role under a common finite semantic profile while admission-specific checks remain independently specified. \textbf{OQ1}, concerning final non-bypassable execution and protected effects, remains outside the evaluated scope.

Several proof obligations follow, rather than completed theorems. A closure argument must show that the resolver's declared applicable inventory is faithfully enumerated under the accepted constitution and permitted facts. A path-preservation argument must show that a compiled or adapted representation does not allow the Candidate in a state where the fixed normative path $\Psi$ forbids it. A phase-binding argument must establish that an observed native result belongs to the exact Candidate, source commitments, version, and context used by admission. A governance argument must establish that the frozen pre-state and its quorum were used for the operation's actual applicable phase. Finally, a release argument must separately show that an execution path cannot bypass admission and that later-phase checks remain valid. None follows automatically from finite empirical agreement.

The broader GSC-1 research profile expresses one-way finite-domain safety as an implication from a candidate program's \texttt{ALLOW} to a reference permit on a justified nonempty domain. That implication does not guarantee functionality, since a program that always denies can satisfy one-way safety. The distinct source-to-IR derivation artifact $\delta$ remains unimplemented in the narrowed GSC-1 profile, and the broad formal model's T1 is a theorem \emph{target}, not an established mechanized result. The GSC-1 Go prototype and its evidence are a separate research track; the OPA/Cedar study neither implements every GSC-1 construct nor upgrades its conditional lemmas to general proofs.

The completed real-engine experiment offers narrower empirical evidence. Across the archived R3 Final case family, 72 cases were evaluated in two backend configurations, yielding 144 strict records: 120 recorded native OPA/Cedar evaluations and 24 pre-native rejections. These archived native policies did not perform only amount comparison: for supported payment inputs, their positive relation also required the \textbf{current-governance-state quorum}; the local Gate separately checked the \textbf{frozen-governance pre-state}. The two are different predicates and can disagree without constituting a backend bug or an unsafe positive admission. In the Cedar route, the adapter projects and counts approvals before constructing the native policy input. This division of labor is profile-specific, not a general law that all business engines must use current-state quorum. Eight permissive native controls were reported separately, and the experiment's tamper harness detected 18 declared evidence-manipulation operators. Independent offline review replayed the archived Gate and Reference behavior and checked native-output evidence correlations; it \textbf{did not independently reexecute the macOS OPA/Cedar binaries}. Earlier rounds must remain visible: R1 and R2 received independent \texttt{REJECT} audits; R3 received a \texttt{CONDITIONAL} audit; only R3 Final received a \textbf{finite-domain archival} \texttt{ACCEPT} for the frozen cohort. A subsequent manuscript-level assessment was \texttt{CONDITIONAL} and found the required-nullable \texttt{proposal.target\_member} input not covered by that cohort; its analytical reproduction is not an additional native engine execution and does not retroactively change the archived experimental records. Those labels are not substitutes for a formal proof or a production deployment certification.

The recorded tests also delineate important exclusions. The final amount-limit profile is not a proof of arbitrary policy-language equivalence, arbitrary issuer trust, authenticated global source completeness, full alternative-path compilation, cross-tenant isolation, real signing or approval independence, asynchronous race freedom, or protected Hub/Security/Executor mediation. Its backend semantic-correctness premise remains indispensable. Historical attempts at an equal-assumption OPA/Cedar/AuthzForce peer comparison belong to a separate experimental track and cannot be represented as completed by the final two-backend result. The validity of the architectural contribution therefore rests on exposing its contract and the conditions that make it meaningful, not on a fabricated claim of superiority to existing engines.

We consequently position the paper's contribution as a \textbf{typed, constitution-relative admission specification}, a proposed separation of evaluation from admission checks, and a falsifiable bounded two-backend implementation study with an explicit adversarial repair history. Section 4 constructs the admission contract at the chosen level of abstraction, including the treatment of failure, unsupported semantics, context mismatch, and necessary evidence; later sections discuss architecture and the archived evaluation without conflating their trust models. Whether a general, externally certified execution boundary can make these conditions non-bypassable at the physical-effect level remains a substantive open research question.

\section{Binding-Complete Execution Admission Contract}

\subsection{Contract boundary and design requirements}

Section 3 established that a policy decision and an admissible execution candidate are different objects of judgment. We now specify the intervening contract. Its purpose is not to implement yet another business-policy evaluator, but to make explicit the obligations that must be satisfied \emph{around} such an evaluator before its result may contribute to a positive pre-execution decision. This is a compositional specification: one component evaluates a bounded business predicate; another verifies constitution-relative source obligations, exact request identity, governance state, provenance, and the legality of using that result. The components may coexist in one process, or reside in separate services, without changing the logical distinction between their responsibilities.

We call the proposed interface the \textbf{Binding-Complete Execution Admission Contract}, abbreviated \textbf{BCEAC}. ``Binding-complete'' is always qualified by an independently accepted governance constitution, its declared boundary, and the applicable snapshot. It does not assert that an implementation can discover an unreported rightful authority beyond that accepted basis. ``Admission'' means eligibility to approach a later execution-control stage, not issuance of a Grant, Redemption of an earlier Grant, or observation of a protected external effect. The label BCEAC names the \textbf{research contract in this section}; it is not asserted to be an implemented production protocol or an existing wire-format identifier.

A satisfactory contract has five distinct requirements. First, the set of constitutionally applicable obligations must be derived independently of the caller's selected policies and of the backend's decision. Second, the selected or loaded policy objects must faithfully discharge those obligations under the prescribed identity, issuer, revision, digest, scope, validity, and semantic-profile rules. Third, authorization and governance-specific conditions must refer to the \textbf{same} Candidate and the applicable frozen governance state. Fourth, an evaluator's result must be mapped, correlated, and interpreted under an explicitly supported semantic profile; an isolated \texttt{ALLOW} string is insufficient. Fifth, the output must remain a typed \textbf{pre-execution} result, incapable by itself of dispensing with the independent release or execution boundary.

These conditions are not exclusive to a particular EBL policy language. A future EBL Native evaluator may implement a policy backend directly; OPA or Cedar may contribute through profile-specific adapters; a conventional policy decision point combined with a sufficiently trusted enforcement wrapper may also implement the same logical obligations. Engine independence here denotes \textbf{conditional substitutability at a declared interface}, not interchangeability of arbitrary native policy programs. This formulation therefore identifies an auditable composition property rather than attributing intrinsic incompleteness to existing engines merely because they are policy engines.

The proposed EBL architecture places this contract between the external, typed governance/execution interface and the later Hub--Security Module--Executor path. A language or structured external description may help author these inputs, but all third-party backends need not compile from a single EBL textual syntax. Conversely, the existence of an \texttt{EffectiveArtifact} or a backend-accepted program does not establish admission: an artifact committed to the wrong required-source inventory can remain byte-perfect while omitting a mandatory governor. The decisive question is which independent constitutional and contextual commitments are used to justify the artifact and the specific native decision.

\subsection{Profile-indexed inputs, authority classes, and outputs}

A practical admission interface must expose its \emph{scope} before making claims about correctness. Let $\rho$ identify an admission profile containing canonical encodings, supported action and policy schemas, backend input and output mappings, the bounds of the supported fragment, admissible governance operation types, mandatory verification predicates, and error-handling rules. Two evaluators are comparable under this paper only when a common $\rho$ has been declared and the comparison concerns precisely the inputs that it admits. Unsupported constructs are not silently approximated as permissive constructs.

Let $c$ be the fully materialized Candidate, $b$ the named governance boundary, $e$ the relevant constitution epoch, and $f$ the applicability facts. Let $\mathcal{T}$ be the verifier's independently installed trust configuration: authorized roots, issuer rights, operation registry authority, and requirements for version/currentness evidence. The current constitution $C_e$, its slot manifest and policy-version heads are resolved \emph{under} $\mathcal{T}$. This dependency direction is important. A caller may carry references or bytes for those objects, but the objects cannot designate their own previously untrusted roots merely because they parse successfully.

We describe the conceptual input as a typed record rather than one monolithic opaque ``authorization proof'':

\begin{equation*}
\begin{aligned}
X_\rho = \langle\,&c,\ b,\ e,\ f,\ \mathcal{T},\ C_e,\ \kappa,\ M_e,\ H_e,\\
                   &w,\ S,\ G_0,\ q_{\mathrm{op}},\ \pi,\ a,\ d_\pi,\ t_v\,\rangle.
\end{aligned}
\end{equation*}

\textbf{Notation crosswalk.} Section 3 used $q$ for an authorized alternative \emph{rule}, and $\pi$ as a bundled backend-and-adapter profile. In this chapter $q_{\mathrm{op}}$ denotes an operation with approval evidence, $q_{\mathrm{alt}}$ denotes an alternative rule, and the backend, adapter, and shared semantic profile are separately named $\pi$, $a$, and $\rho$. These are refinements of notation, not different authorities.

Here $\kappa$ is the independently authenticated checkpoint or currentness witness \emph{when required by the chosen profile}; $M_e$ is the constitution-committed slot manifest; $H_e$ is the designated source-version status under the relevant authoritative head; $w$ is the authorized choice-path witness; $S$ is the collection of selected and loaded exact policy objects; $G_0$ is the applicable \textbf{frozen} governance pre-state; $q_{\mathrm{op}}$ is the proposed governance operation and its authorization evidence; $\pi$ identifies the policy backend; $a$ is its adapter; $d_\pi$ is the correlated native policy result; and $t_v$ is the verification phase time. A non-governance-changing operation may have no quorum requirement, but its profile must say so rather than inventing a permissive default when the relevant descriptor is unavailable.

The tuple denotes \textbf{logical input groups}, not a claim that any archived R3 Final JSON document has precisely these fields or contains actual cryptographic witnesses for each element. In particular, R3 Final accepts a harness-installed \texttt{trusted\_manifest} fixture and does not implement an independently authenticated $C_e$ or checkpoint protocol. Its local \texttt{proposal\_id} is a typed diagnostic identifier, not a signed authority capability. The general notation is deliberately stronger in its trust premises than that limited experiment.

Inputs fall into three trust classes. \textbf{Constitutive inputs} specify what should govern: verified constitutional rules, authorized alternative paths, operation descriptors, and applicable source seats. \textbf{Observation inputs} report the actual Candidate, selected bodies, versioned snapshots, contextual facts, and policy evaluation. \textbf{Local verifier requirements} define permitted schemas, trusted roots, result-correlation and error semantics, and the host safety floor. The contract must not allow an observation input to reclassify itself as a constitutive input. For example, a selected policy file may contain an applicability field, but that field cannot remove a SourceSlot established by the constitution's independent applicability rule.

The output is a sum type, not a Boolean, and it remains distinct from the backend's result and from an external execution effect:

\begin{equation*}
\begin{aligned}
\mathsf{BindingResult}_\rho
  &= \mathsf{BVerified}(\beta)\mid\mathsf{BRejected}(r)\mid\mathsf{BUndetermined}(r),\\
\mathsf{PolicyDecision}_\rho
  &= \mathsf{Allow}\mid\mathsf{Deny}\mid\mathsf{Indeterminate}\mid\mathsf{Error},\\
\mathsf{AdmissionDecision}_\rho
  &= \mathsf{Eligible}(\alpha)\mid\mathsf{Reject}(r)\mid\mathsf{Unresolved}(r).
\end{aligned}
\end{equation*}

The symbols in this tuple do not assert that all input groups are obtained from a trusted source: $b$ is checked against an authenticated candidate/action-to-boundary registry under $\mathcal{T}$; $\kappa$, $M_e$, $H_e$ and $f$ are accepted only through the profile's designated trust rules. The fixed path $\Psi$ is a \textbf{derived, validated result} of constitutional resolution and witness instantiation, not an extra caller-authorized input; the trace $\tau_\pi$ is derived from the designated backend invocation and checked against committed inputs and outputs. A missing, invalid or ambiguous required derivation cannot default to a positive result. Every invoked verifier obligation must be defined on a bounded supported domain or terminate with a typed non-positive result.

The proposed result $\alpha$ is a phase-specific record of the eligibility decision and its commitments, not an execution bearer token. Its existence is a design choice for the general contract; the concrete R3 Gate returns \texttt{ADMIT}, \texttt{REJECT}, or \texttt{INDETERMINATE} and does not thereby issue a P2 Execution Grant. Likewise, the independent R3 Reference returns \texttt{PERMIT} or \texttt{FORBID} under its local oracle semantics. We preserve those interface differences rather than retroactively renaming archived data.

\subsection{Constitutional closure and body-independent required obligations}

Before defining the checks, fix an authenticated \textbf{verification environment} for one attempt:

\begin{equation*}
\Gamma_\rho := \langle b,e,\mathcal{T},C_e,M_e,H_e,\kappa,f,t_v,\mathsf{IssuerRights}_e,\mathsf{OperationRegistry}_e\rangle.
\end{equation*}

This environment is accepted only after applicable trust checks; its fields may be \textbf{unknown} or \textbf{invalid}, rather than implicitly total. We use $\mathsf{V}$, $\mathsf{F}$, and $\mathsf{U}$ for verified, conclusively contradicted, and unresolved \emph{evidence assessments}. If a Boolean mathematical relation is used below, it denotes the relation \textbf{on an admitted and complete $\Gamma_\rho$}. A concrete verifier must lift that relation to these three assessment values for missing or conflicting evidence. An authenticated contradiction can produce $\mathsf{F}$ without depending on unrelated unavailable facts; an unauthenticated claim of contradiction cannot.

The first precondition concerns the \emph{authority basis}, not policy contents. In particular, the trusted Candidate/action-to-boundary mapping and the operation registry must be checked independently of the caller-provided $b$ and $q_{\mathrm{op}}$; merely finding a matching boundary string in an untrusted request does not establish a constitutional scope. For a constitution $C_e$, Candidate $c$, and trusted applicability facts $f$, the obligation resolver produces $\Phi(c,b,e,f)$ as specified in Section 3. A verifier cannot claim that $\Phi$ is complete merely because every policy supplied in $S$ was evaluated. The required set must originate in the accepted constitutional state, not be reverse-engineered from selection success. This rejects the circular procedure ``the required policies are exactly the policies that happened to be loaded.''

We separate \textbf{normative adequacy} from \textbf{enumeration fidelity}. Normative adequacy states that the authorized constitutional process has correctly designated the relevant policy-source seats for boundary $b$. It is an external premise: the verifier cannot derive the truthfulness of an entire normative roster from that roster's own hash or signature. Enumeration fidelity is narrower and potentially checkable: given an independently anchored constitution, a finite, authenticated, complete declared manifest, and sufficiently trusted facts, has the verifier actually considered the entire declared inventory? A manifest count, canonical ordering, version commitment, publisher authorization, and a currentness witness can help establish the second property, but none proves the first.

For clarity, define a profile-relative closure predicate

\begin{equation*}
\mathsf{Closure}_\rho(c,b,e,f;\mathcal{T},C_e,M_e,\kappa)
\in \{\mathsf{Closed},\mathsf{Invalid},\mathsf{Unknown}\}
\end{equation*}

that returns $\mathsf{Closed}$, $\mathsf{Invalid}$, or $\mathsf{Unknown}$. A \texttt{Closed} result is justified only when the chosen profile's trusted-root, manifest-integrity, applicability-coverage, and currentness requirements are met. An explicit constitutional statement that there are no applicable SourceSlots can justify the empty obligation expression $\top$; failure to fetch any policies cannot. A malformed commitment may be conclusively invalid, while a temporarily unavailable required checkpoint may leave closure unknown. Both outcomes prevent positive admission, although their diagnostic meaning differs.

An applicable SourceSlot persists as an \emph{obligation to obtain a designated, valid policy object}. A source owner may modify or revoke their own policy body according to permitted version rules, but that operation alone does not amend the constitution. A declared slot whose designated body has expired remains a required slot with an unsatisfied discharge; it is not a successful zero-obligation case. A legitimate roster or applicability-rule change is a \textbf{meta-governance transition} whose authorization must be evaluated under the governing prior state. The contract therefore exposes an explicit interface between routine policy updates and constitutional changes, without purporting to settle the entire governance-transition protocol in this paper.

The broader GSC-1 specification gives one example of how a closed, finite declaration could be checked: a separately installed root configuration authenticates an envelope, a complete sorted SlotManifest, committed operation/applicability metadata, and a current designated-version head. An envelope flag such as \texttt{complete=true} has, at most, declaration-relative meaning and cannot independently establish closure. A selected subset's Merkle inclusion proofs, even when individually sound, would not by themselves prove that every mandated page or slot was present. An unavailable current head must not be converted into an assertion that no designated policy exists. These are formal profile requirements and historical prototype concerns; they are \textbf{not} claims that R3 Final verified signature chains, a live signed checkpoint, or a global absence proof.

An adversary might instead supply a coordinated, internally consistent manifest omitting both an operation and its governance snapshot. If the accepted authority inventory is itself inadequate, byte-level consistency and local type checks can still all succeed. The R3 Final \texttt{coordinated\_manifest\_elision\_A0} case is explicitly classified as an \textbf{A0 limitation}, not as a detected and repaired global-completeness attack. This is why our contract's closure judgment cannot be used as evidence that the constitutional sufficiency premise has been discharged.

\subsection{Exact discharge, legal alternatives, and source-to-evaluation separation}

Constitutional closure determines what must be considered, but not whether the appropriate objects were actually selected or whether the backend evaluates them correctly. Once $\Phi$ is known, a legally authorized and candidate-bound witness $w$ chooses a fixed permitted alternative path $\Psi=\Phi\!\mid_w$, if the constitution contains alternative obligations. In the fixed-path profile, $w$ names one choice for each reached \texttt{anyAuthorized} node and must be bound to the Candidate identity, relevant facts, epoch, and authorized choice rule. It cannot be reconstructed after an unfavorable decision by searching for some other branch that returns permission.

For a required slot $s$, let $\mathsf{Head}_{\Gamma_\rho}(s)$ denote the authoritative applicable reference \emph{as of the verified head and verification phase}. It comprises the slot identity, authorized issuer, revision, scope, semantic profile, content digest and status/validity commitments. On an admitted, complete environment $\Gamma_\rho$, the \textbf{normative Boolean relation} for a properly designated discharge is

\begin{equation*}
\begin{aligned}
\mathsf{Discharge}^{\mathrm{norm}}_\rho(s,S;\Gamma_\rho)
\;\Longleftrightarrow\;&
\exists!\,p\in S:\mathsf{ExactDesignated}_\rho(p,\mathsf{Head}_{\Gamma_\rho}(s))\\
&\land\ \mathsf{Active}_\rho(p,t_v;\Gamma_\rho)
\land\ \mathsf{AuthorizedIssuer}_\rho(p,s;\Gamma_\rho)\\
&\land\ \mathsf{ProfileAndScopeOK}_\rho(p,s;\Gamma_\rho)\\
&\land\ \mathsf{NoSlotConflict}_\rho(s,S;\Gamma_\rho).
\end{aligned}
\end{equation*}

The uniqueness sign denotes the \textbf{unique designated active discharge}, not a blanket prohibition against additional independently authorized constraints in $S$. A collection that contains two conflicting references for the same mandatory seat must not satisfy the profile's uniqueness and conflict checks. The corresponding \emph{verifier assessment} $\mathsf{DischargeStatus}_\rho(s,S;\Gamma_\rho)\in\{\mathsf{V},\mathsf{F},\mathsf{U}\}$ is $\mathsf{V}$ only when that normative relation is established with sufficient trusted evidence; it is $\mathsf{F}$ only when a trusted, independent contradiction is established (including a missing required object \textbf{after an authenticated complete selected-set boundary}); otherwise it is $\mathsf{U}$. An unreachable authoritative head, an untrusted assertion of source absence, or an unknown revocation status is not classical $\mathrm{false}$. A confirmed expired or revoked head/body can produce $\mathsf{F}$ while the SourceSlot remains mandatory.

This distinction explicitly avoids reading the failure of $\exists p$ in a partial observation as a proof of absence. Head availability, issuer-rights and status are environment-dependent; no correctness claim is made for arbitrary inputs outside the terminating and error-classified profile domain.

The structured obligation judgment is an \emph{evidence-sensitive} recursion. Only a successfully instantiated fixed path $\Psi=\Phi\mid_w$ is evaluated; $w$ must bind a choice to each reached alternative-node identity (tree address), not merely to a possibly repeated rule identifier. The grammar uses nonempty $\operatorname{all}$ nodes ($n\ge 1$) and authorized alternatives with at least two branches ($m\ge 2$); uninstantiated \texttt{anyAuthorized} nodes never receive a permissive default. With $\mathsf{Conj}_3$ denoting three-valued conjunction (F dominates; otherwise U dominates; V only if all V), define

\begin{equation*}
\begin{aligned}
\mathsf{DischargesStatus}_\rho(\top,S;\Gamma_\rho)
 &= \mathsf{V}\quad\text{iff authenticated empty-path closure holds},\\
\mathsf{DischargesStatus}_\rho(\operatorname{require}(s),S;\Gamma_\rho)
 &= \mathsf{DischargeStatus}_\rho(s,S;\Gamma_\rho),\\
\mathsf{DischargesStatus}_\rho(\operatorname{all}(\Psi_1,\ldots,\Psi_n),S;\Gamma_\rho)
 &= \mathsf{Conj}_3\!\left(\mathsf{DischargesStatus}_\rho(\Psi_i,S;\Gamma_\rho)\right)_{i=1}^{n}.
\end{aligned}
\end{equation*}

For $\top$, absent proof of authenticated emptiness the output is $\mathsf{U}$, or $\mathsf{F}$ if the alleged empty closure is independently contradicted. The mathematical shorthand $\mathsf{Discharges}(\Psi,S)$ used in Section 3 and the positive admission formula means \textbf{the assessment is V in an adequately verified environment}, never that an unchecked Boolean existential returned true. The first line presupposes an authenticated no-required-seat decision; it is \textbf{not} a procedure for assuming emptiness. The selected collection may contain additional constraints only if their source, semantics, and effect on the authorized decision are allowed by the constitution and profile. Additional sources cannot substitute for a missing mandatory seat, and a permissive unauthorized extra source cannot launder an \texttt{ALLOW} that a required source would deny. Conversely, source completeness is not a demand that every branch \emph{inside} every policy body execute. An Auth Key policy that legitimately applies a whitelist-first local rule may avoid a subordinate amount branch; that local evaluation rule does not cancel a separately mandatory Hub or Root SourceSlot.

This distinction also separates \textbf{discharge} from \textbf{semantic preservation}. If all mandatory bodies are correctly present but one denies the Candidate, binding may still be valid and the overall policy result negative. If every necessary body is present but the compiled program permits states forbidden by the chosen reference path, that is a different failure: a loss of decision semantics during translation or combination. We use $\mathsf{Discharges}(\Psi,S)$ for structural source fulfillment and reserve $\mathsf{PreservesPath}(\Pi,\Psi)$ for the relevant semantic non-weakening property. One cannot infer the latter solely from correct hashes of the source objects or from the existence of an \texttt{EffectiveArtifact}.

The broader GSC-1 finite-domain profile defines its semantic safety test over an independently justified, nonempty, bounded state set $D$:

\begin{equation*}
\mathsf{Safe}_D(\Pi,\Psi)
\;\Longleftrightarrow\;
D\neq\varnothing\ \land\
\forall\sigma\in D:\;
\mathsf{Allow}_{\Pi}(\sigma)\Rightarrow\mathsf{Permit}_{\Psi}(\sigma).
\end{equation*}

The chosen path, not an unselected sibling under a global OR, supplies the reference permission. The converse implication is a separate \textbf{functional conformance} question: a program that always denies may satisfy one-way safety while being useless for intended positive actions. Finite checking also depends on a total input guard preventing states outside the declared $D$ from producing positive admission. The stronger source-to-IR structural derivation witness $\delta$ is not implemented in the narrowed GSC-1 profile; its mathematical role must not be inferred from the \texttt{Safe\_D} computation.

Crucially, the actual R3 Final OPA/Cedar experiment is not that GSC-1 compiler-checker profile. R3 Final verifies local required-slot and source-content identities. The native OPA/Cedar policies implement the supported payment \textbf{amount-limit together with a current-governance-state quorum predicate} (and a current-quorum rule for the supported remove operation); the separate Gate checks \textbf{frozen-prestate governance}. Cedar's adapter constructs native policies and projects eligible approval counts from the current-state roster, so not all normalization/counting occurs within Cedar itself. It does not generate an authenticated arbitrary $\Phi$ expression, verify nested $w$, construct a checked $\delta$, or prove extensional $\mathsf{Safe}_D$ for every state of a justified domain. Treating its policy programs as if they were the abstract GSC-1 $\Pi$ would import a property that the archived implementation did not establish.

\subsection{Frozen joint governance and exact contextual alignment}

Binding source bodies is insufficient if the authorization evidence describes another action or another governance state. The Candidate commitment must be consistent with the proposal's operation and parameters, including the target and relevant state-version commitments where required by the profile. A policy result for a payment may not be repurposed as approval to remove a team member merely because both requests happen to produce the same backend Boolean. Likewise, a correctly formed governance snapshot for an unrelated operation cannot serve as the authorization snapshot for the proposed one.

We distinguish the currently observable mutable governance state $G_t$ from the \textbf{authenticated applicable pre-state} $G_0$ for the operation. For a governance-changing operation $q_{\mathrm{op}}$, write

\begin{equation*}
\mathsf{JointGov}_\rho(q_{\mathrm{op}},c,G_0,\mathsf{Approvals})
\end{equation*}

for the profile-specific quorum, eligibility, operation-descriptor, target, and snapshot-binding checks. The operation descriptor must itself be trusted and associated with the correct constitution epoch; a missing or unrecognized descriptor cannot default to a unilateral action merely because no workflow entry was found. The governance state supplied by the caller may be useful as an observation, but it is not permitted to redefine the historically applicable approval threshold.

For example, suppose a normal removal proposal was created under $G_0$ with members $\{A,B\}$ and threshold two. The proposed removal of $B$ does not authorize the system to recompute the applicable requirement as a one-member quorum merely because the post-action roster would contain only $A$. The proposal must bind its own target, pre-state snapshot version, authenticated roster and threshold, approval identities, and constitutional operation descriptor. A valid approval set must satisfy the frozen rule applicable to \emph{that} proposal; a post-proposal membership calculation cannot authorize the transition that produced it. This example characterizes a conditional governance invariant and does not establish real signer independence or every lifecycle and concurrency case.

More generally, the contract must distinguish a \emph{well-formed claimed approval record} from a \emph{verified approval by an eligible and independent actor}. Checking that approval identifiers are unique members of a static list is not equivalent to signature validation, key control-domain independence, or proof that those approvals actually authorize the bound Candidate. A production-level verification predicate would need those dependencies. In the archived R3 Final study, the frozen-member and target cases are fixed structural fixtures; the experiment explicitly does not authenticate human approvers or demonstrate quorum under a live signed constitution.

The distinction between constitutional changes and ordinary source-owner edits is essential here. A source-owner policy update may alter that source's own predicate within allowed rules, including a legitimate local branch that permits a transaction. It cannot, solely by rewriting its body, remove a mandatory SourceSlot, weaken a separate Root floor, or approve a change to the trusted operation registry. Altering required seats, authorized delegates, quorum requirements, or the constitution itself calls for a separately authorized old-to-new transition under the \textbf{prior} authority. We leave the complete semantics of exceptional freezes, root rotation, simultaneous proposals, and recovery transitions to a separate profile; no unmentioned emergency shortcut is silently incorporated into $\mathsf{JointGov}$.

Contextual alignment is thus a conjunction of separately justified relations: the exact Candidate and action, the declared governance boundary, the applicable epoch, the authorized source roster, the fixed governance snapshot, the selected path, and the backend invocation all refer to one coherent admission attempt. Where a required relation cannot be established because trusted data are absent, the result is unresolved rather than a guessed favorable state. Where a verified relation is false---such as an exact target mismatch or a conclusively unauthorized approver---the result is rejection. This separation supports explainable fail-secure behavior without conflating missing evidence and established violation.

\subsection{Backend evaluation, adaptation, and decision provenance}

The policy backend has a narrowly stated responsibility: evaluate a supported business predicate for the correctly mapped request under the designated policy and semantic profile. The admission component does not gain the right to trust a result merely because a binary has the name of a recognized engine, or because a native output was parsed into an \texttt{ALLOW} value. The contract must also establish that the invocation was \textbf{the right invocation} and that its output was interpreted according to the declared adapter semantics.

Let $\mathsf{Encode}_{\rho,\pi}(c,S,f)$ be the adapter's profile-specific mapping from bound source objects and the exact Candidate to a native request $u_\pi$. Let $\mathsf{Eval}_\pi(u_\pi)$ be the corresponding backend execution, and let $\mathsf{Decode}_{\rho,\pi}$ map the native output, error and termination status into a typed policy decision. A valid evaluation trace is a related tuple, not a bare decision bit:

\begin{equation*}
\begin{aligned}
\tau_\pi=\langle &\rho,\pi,a,\operatorname{cid}(c),\mathsf{ContextCommit},
\mathsf{SourceCommit},\\
&\mathsf{NativeInputCommit},\mathsf{InvocationId},
\mathsf{NativeOutput},\mathsf{ExitStatus},d_\pi\rangle.
\end{aligned}
\end{equation*}

The profile states which of these commitments are required and which independent evidence supports them. A successful parser is not a substitute for authenticating the installed policy set; a command log, by itself, does not establish who actually ran a process. The general contract can require stronger runtime provenance than a laboratory fixture. In R3 Final, the independent audit checked archived case hashes, engine-input projections, output-to-adapter mappings, invocation identifiers and Gate/Reference correlations. It did \textbf{not} independently reexecute the macOS binaries or authenticate their historical execution with an external attestation service.

We write $\mathsf{Mapped}_\rho(\tau_\pi,c,S,f)$ for the condition that the adapter's inputs, policy commitments, Candidate identity and output mapping agree with the accepted context. It is distinct from $\mathsf{CorrectBusiness}_\rho(\pi,a)$, the semantic premise that the particular backend and adapter correctly implement the supported business relation on valid inputs. For the \textbf{full declared} supported domain $\mathcal{D}_\rho$ of exact, well-formed contexts and states, the business correctness premise is universally quantified; a claim restricted to isolated tested fixtures cannot silently become this universal statement. An intended refinement property may be written as

\begin{equation*}
\begin{aligned}
\mathsf{CorrectBusiness}_\rho(\pi,a) \Longrightarrow
\forall z=(c,S,\sigma,\Psi)\in\mathcal{D}_\rho:\\
\mathsf{DecodedAllow}_{\rho,\pi}(z)
\iff\mathsf{Permit}^{\rho}_{\mathrm{business}}(z).
\end{aligned}
\end{equation*}

This is a \textbf{required semantic-correctness premise or separately discharged verification obligation}, not a consequence of installing an adapter. The permission relation itself must be independently specified, including legal source combination, extras, selected branch and current governance conditions. \textbf{For any profile that claims fixed-path semantic non-weakening}, BCEAC additionally requires the separate, universally scoped connection

\begin{equation*}
\forall z=(c,S,\sigma,\Psi)\in\mathcal{D}^{\mathrm{path}}_\rho:
\quad \mathsf{Permit}^{\rho}_{\mathrm{business}}(z)
\Longrightarrow\mathsf{Permit}_{\Psi}(c,\sigma),
\end{equation*}

where $\mathcal{D}^{\mathrm{path}}_\rho$ is an independently bounded and guarded subdomain of $\mathcal{D}_\rho$, and $\mathsf{Permit}_{\Psi}$ interprets the \textbf{fixed authenticated} conjunction/authorized alternative rather than an OR reconstructed from favorable results. The profile must define both relations and their combination rules; mere source presence does not prove this implication. Under these two premises, a correctly decoded backend \texttt{ALLOW} entails $\mathsf{Permit}_\Psi$ \emph{on that bounded path domain}. This is only a \textbf{conditional composition argument}, not an assertion that R3 Final verified arbitrary GSC-1 paths, or that the Gate recomputes business semantics. Profiles claiming \textbf{structural binding only} must explicitly forgo path-semantic non-weakening conclusions. Bidirectional business equivalence also asks more about false denials than the one-way path property does. Unsupported, error, or undefined decisions are excluded from a positive branch and must follow the profile's typed failure rules.

A backend returning \texttt{DENY} on a valid, fully bound Candidate provides a normal negative policy decision. It must not be recorded as a failure to bind necessary policy sources. A backend returning \texttt{ERROR}, an unsupported feature result, an ambiguous or unparseable output, or a mismatched invocation cannot provide positive admission. The precise reason may be an invalid invocation that is conclusively rejected or an unresolved capability/evidence condition; the profile must define the classification. Native output conventions are not universal. In particular, an adapter must not normalize a Cedar validation or unsupported-semantics error into a successful \texttt{ALLOW}, and OPA's JSON value types must not be silently coerced across the exact integer contract when that representation is significant.

The failure mode exposed by a deliberately faulty-allow backend makes the semantic premise unavoidable. Consider a correctly bound amount-limit policy requiring $amount\leq limit$. If a malicious or defective backend returns \texttt{ALLOW} when the actual amount is above the limit, a Gate that checks only structural binding, governance, and invocation legality can still report positive admission under its specified design. That is \textbf{not} evidence that source binding was bypassed; it is evidence that the backend-correctness premise is not optional. The \textbf{original R3} faulty-backend ablation artifact records the simulated faulty-ALLOW \texttt{REAL-02} control exhibiting this boundary; the separately archived R3 Final audit describes a distinct additional simulated \texttt{amount\_zero} to \texttt{777} mutation. Neither control is a verified native-engine bypass or part of R3 Final's strict native invocation count. An architecture seeking protection against arbitrary compromised backend semantics would need an additional trusted reference evaluator, proof system, or redundant verification argument, each with its own specification and cost. BCEAC does not claim to supply such a mechanism by renaming a policy decision.

This division also clarifies ``engine-agnostic.'' We can replace $\pi_1$ with $\pi_2$ for a given $\rho$ only if each backend is configured for the same intended relation, its adapter faithfully maps the same committed decision context, its error and unsupported cases cannot become positive, and all admission-specific checks remain applicable. These are \textbf{conditions of substitution}, not a theorem that arbitrary OPA and Cedar policy modules are equivalent. If their native languages disagree on numeric types, missing attributes, combining operators, default-deny conventions, or extension semantics, the disagreement must be resolved in the declared profile or excluded from the comparison domain rather than obscured by an \texttt{ALLOW} label.

\subsection{Typed admission rule, error precedence, and fail-secure behavior}

We now state the necessary positive-admission rule. Define the \textbf{complete, finite, profile-authorized} obligation inventory $\mathcal{O}_\rho(X_\rho)$, resolved independently of which checks happened to execute, as separately checkable components: legitimate and appropriately current governance basis; independently resolved fixed-path obligations; exact discharge of mandatory sources; exact Candidate and contextual alignment; valid frozen-governance authorization where required; valid, correlated and supported backend evaluation; an \texttt{ALLOW} result under the named semantic-correctness premise; and the independent admission host floor. Let $\mathsf{Pos}_\rho(o)$ denote that the relevant obligation has a verified positive result, rather than a missing, unsupported or inconclusive one.

The positive-admission \textbf{necessity} condition is

\begin{equation*}
\begin{aligned}
\mathsf{Eligible}_\rho(X_\rho)\ \Longrightarrow\ &
\mathsf{ClosureOK}_\rho(X_\rho)
\land\mathsf{PathAuthorized}_\rho(X_\rho)\\
&\land\bigl(\mathsf{DischargesStatus}_\rho(\Psi,S;\Gamma_\rho)=\mathsf{V}\bigr)
\land\mathsf{ContextOK}_\rho(X_\rho)\\
&\land\mathsf{JointGovOK}_\rho(X_\rho)
\land\mathsf{Mapped}_\rho(\tau_\pi,c,S,f)\\
&\land(d_\pi=\mathsf{Allow})
\land\mathsf{SemanticProfileOK}_\rho(X_\rho)
\land\mathsf{HostFloorOK}_\rho(X_\rho).
\end{aligned}
\end{equation*}

This is a \textbf{one-way constraint} on a positive result. It must not be promoted to a mathematical sufficiency theorem about release or external execution, and it permits a profile to require additional independently protected conditions. The names $\mathsf{ClosureOK}$, $\mathsf{JointGovOK}$, $\mathsf{Mapped}$ and $\mathsf{HostFloorOK}$ denote obligations whose concrete tests must be specified; writing them in a conjunction does not prove that any implementation correctly computes them. The semantic correctness of the backend, authentic constitutional roots and currentness, and soundness of the relevant verification components are \textbf{external premises or independent proof obligations}. The term $\mathsf{SemanticProfileOK}_\rho$ means that the declared profile has a justified contract for the actual native business relation and, \textbf{if claiming path non-weakening}, a valid domain-scoped $\mathsf{Permit}_{\mathrm{business}}\Rightarrow\mathsf{Permit}_\Psi$ obligation; it is \textbf{not} satisfied merely by a Boolean flag asserted by the caller or by the backend itself. No logical conjunction can turn an unverified premise into an experimental result. The rule also does not require a single function to evaluate its clauses in the displayed order.

The admissibility result must preserve the difference between proven nonconformance and lack of sufficient evidence. Let each mandatory obligation return a three-valued assessment $\mathsf{V}$ (verified), $\mathsf{F}$ (conclusively invalid), or $\mathsf{U}$ (unresolved). For the abstract result aggregator, the contract can require:

\begin{equation*}
\mathsf{AdmissionClass}(O)=
\begin{cases}
\mathsf{Reject}, & \exists o\in O:\mathsf{status}(o)=\mathsf{F},\\
\mathsf{Unresolved}, & (\nexists o:\mathsf{status}(o)=\mathsf{F})
 \land (\exists o:\mathsf{status}(o)=\mathsf{U}),\\
\mathsf{EligibleCandidate}, & \forall o\in O:\mathsf{status}(o)=\mathsf{V},
\end{cases}
\end{equation*}

provided $O=\mathcal{O}_\rho(X_\rho)$ was authentically and completely enumerated and all listed checks terminate with $\mathsf{V}/\mathsf{F}/\mathsf{U}$; if the required-inventory closure itself is unavailable, the result is non-positive even when the executed-check list is empty. Any profile-specific additional condition \textbf{must be included in} $\mathcal{O}_\rho$, or handled by an explicit independent, equally non-positive authorization rule; it cannot be silently omitted from the aggregator. An authenticated standalone failure $\mathsf{F}$ dominates unrelated $\mathsf{U}$, but a putative failure dependent on unauthenticated facts cannot be promoted to $\mathsf{F}$. A verified native \texttt{DENY} is a conclusive \textbf{negative admission} reason, not evidence of malformed provenance. A known mismatch of authorized issuer, required digest, proposal target, or frozen snapshot is likewise a negative result. Unavailable authoritative currentness, unknown required-body status, unsupported profile semantics, or an unobservable native decision may yield \texttt{Unresolved} where the profile cannot establish a proven violation. The model deliberately treats \texttt{Reject} and \texttt{Unresolved} alike with respect to the \textbf{absence of positive execution-admission authority}, while keeping their epistemic meanings distinct for audit and repair.

This presentation is a profile-design target, not a statement that every historical implementation has exactly these return names or categorizes every failure identically. The frozen R3 Final implementation does \textbf{not} fully refine the proposed taxonomy: some known type-invalid inputs, such as a Boolean or non-integral frozen quorum threshold (\texttt{MUT-02} and a separately observed \texttt{2.0} variant), are reported as Gate \texttt{INDETERMINATE}, even though the independent Reference returns \texttt{FORBID} and the normative schema classifies them as invalid. A composite UNKNOWN source-status/digest-conflict input can similarly let an early UNKNOWN mask an independently provable digest mismatch. These are observed legacy classification/precedence discrepancies, not positive admissions, and the frozen outputs must not be relabeled. The R3 Gate uses \texttt{ADMIT} for its local positive decision, \texttt{REJECT} for many known invalid/denied cases, and \texttt{INDETERMINATE} for its defined uncertainty and some malformed cases. Its independent Reference uses a \texttt{PERMIT}/\texttt{FORBID} oracle, in which both R3 Gate rejection and unresolved outcomes may correspond to \texttt{FORBID}. Manuscript notation must retain those differences when interpreting results.

Fail-secure behavior also includes the ordering of authority creation. A caller cannot remove a necessary source by causing a validation timeout; a selected-body fetch failure cannot be converted to an authenticated empty slot set; and a backend unable to interpret a required rule cannot manufacture permission by invoking a permissive fallback. A verified error in a necessary commitment cannot be neutralized by an unrelated successful policy call. In all these cases, the system must refrain from positively admitting the Candidate. There may be operational reasons to offer an explicitly authorized alternative path or a later retry, but any such path must derive from the constitution and produce a \emph{new} validly bound decision; it is not an ad hoc exception to the present attempt.

The model does not assume that aggressive rejection is evidence of correct \emph{functionality}. An always-rejecting admission component trivially avoids positive unauthorized decisions yet does not implement the intended business or governance service. Functional conformance and liveness under declared valid scenarios must therefore be measured separately from the one-way safety rule. The earlier GSC-1 distinction between $\mathsf{Safe}_D$ and $\mathsf{FunctionalEq}_D$ makes this explicit; R3 Final's mixture of valid-positive, valid-negative, malformed and unsupported fixtures provides finite observations about the chosen implementation, not an all-input service-availability proof.

\subsection{Binding evidence, EffectiveArtifact identity, and phase separation}

For accountability, an admission component may produce a non-authority-bearing binding-evidence object $\beta$. This object should commit the independently resolved authority context and the exact objects actually checked: Candidate identity, boundary and epoch, constitution and manifest identities, required path and choice witness, selected policy-version commitments, relevant facts and governance pre-state, semantic profile and adapter identity, backend invocation/output linkage, verification time, and verifier implementation identity. If a current checkpoint is required, its evidence should be part of the bound context rather than supplied by an unauthenticated request field alone. The abstract commitment can be represented as

\begin{equation*}
\beta_\rho=\mathsf{Commit}_\rho(\operatorname{cid}(c),b,e,
\mathsf{GovRef},\Psi,w,\mathsf{Sources},\mathsf{Facts},
\mathsf{Prestate},\rho,\tau_\pi,t_v).
\end{equation*}

A commitment is not a certificate of truth unless its components were checked against independent trusted inputs. A self-reported roster hash proves neither adequate source discovery nor the physical independence of purported signers. Similarly, a compiled policy \texttt{EffectiveArtifact} may preserve the selected rules with perfect byte integrity yet lack a faithful link to an undisclosed required SourceSlot. Such an artifact is not itself a constitution and does not acquire final execution authority from being signed, cached, or labeled effective. The artifact's meaning is conditional on the trusted inventory and precise versioned relationship against which it was admitted.

The role of $\beta$ is primarily traceability, replay resistance \emph{when an appropriate external phase-binding protocol exists}, and prevention of accidental confusion between different decision contexts. Its inclusion in the abstract contract must not be mistaken for a claim that the GSC-1 prototype proves an independently checked per-source compiler derivation witness $\delta$, or that R3 Final emitted a cryptographically authenticated, externally time-stamped version of this record. Local \texttt{SHA256SUMS.txt} checks establish archive-byte consistency, not an external provenance authority. A separately executed audit of raw results can verify internal correlations without independently certifying that the recorded native process took place on its alleged historical host.

Time makes the difference between evidence and execution authority especially important. Let $t_v$ be verification, $t_l$ a later release, and $t_r$ a Redemption phase as in Section 3. A record of \texttt{Eligible} at $t_v$ does not entitle a caller to bypass a later checkpoint, reuse an authorization after an incompatible version transition, or treat a stale policy head as current. The phase relation must be specified by the downstream execution-control design. An exact-version ERC/Grant lifecycle may require revalidation even if an updated policy is stricter but still favorable. Those obligations belong to the EBL-Core release-and-redemption machinery and any actual Hub/Security Module/Executor enforcement argument \cite{Wu2026EBLCore}.

Thus, the proposed output is not a Grant, Ticket, Redemption permission or proof of effect. At most it can be an input to a later protected authorization or arbitration decision, if that stage independently establishes the integrity and currentness required for its own phase. Neither the presence of a positive admission record nor a collection of post-hoc logs proves that all physical or software paths capable of causing the external transition are mediated by that stage. End-to-end non-bypassability must be evaluated against the actual host, recovery, update, and effect paths, which this paper does not test.

\subsection{Conditional properties, falsification obligations, and the limit of the claims}

The contract is designed to be falsifiable. Its key properties can be stated as conditional obligations without asserting completed verification. \textbf{Required-source non-elision} asks whether a source seat independently required under the accepted constitution can disappear from the verifier's positive-admission reasoning merely because a selected body was absent, expired, revoked or mislabeled. \textbf{Exact contextual preservation} asks whether source, Candidate, proposal and governance-version commitments can be substituted without invalidating the positive decision. \textbf{Path non-weakening}, for a profile that admits authorized alternatives, asks whether a program can borrow a permissive unchosen branch after the path was fixed. \textbf{Backend-conformance dependence} asks whether the architecture accurately discloses the consequences of a semantically faulty backend, rather than concealing them within a generic admission guarantee.

For non-elision, suppose closure is sound relative to accepted $C_e$, witness $w$ is lawfully fixed, the applicable slot $s$ occurs on $\Psi$, and independently trusted evidence conclusively establishes that $s$ lacks a valid designated discharge. Under a sound implementation of the discharge check and the positive-admission necessity rule, one can derive the following \textbf{conditional model property}:

\begin{equation*}
\Bigl(s\in\mathsf{RequiredSeats}(\Psi)
\land\bigl(\mathsf{DischargeStatus}_\rho(s,S;\Gamma_\rho)=\mathsf{F}\bigr)\Bigr)
\Longrightarrow\neg\mathsf{Eligible}_\rho(X_\rho).
\end{equation*}

Its argument is direct: positive admission requires a $\mathsf{V}$ discharge status for the fixed path; a required leaf with an independently established $\mathsf{F}$ makes the three-valued conjunction non-positive. This is a consistency consequence of the \textbf{specified relation}. It is not a mechanized proof that a concrete Gate correctly computes $\Psi$, authenticates $s$, or implements all inputs without bugs. For an \emph{unknown} status the conclusion remains ``not positively eligible,'' but its diagnostic result may be unresolved rather than a conclusive rejection. The same argument is constitution-relative; it cannot recover a source omitted by an inadequate sole accepted authority in the first place.

For contextual preservation, let $X$ and $X'$ be two distinct binding contexts whose committed Candidate, target, boundary, frozen snapshot or required-source version differs in a field the chosen profile treats as decision-relevant. Assuming exact canonicalization, collision-resistant commitments and sound cross-object comparisons, a decision trace or binding evidence belonging exclusively to $X$ cannot justify positive admission of $X'$ \emph{without independent evidence binding the changed context}. This is a \textbf{conditional anti-substitution obligation}, not a claim of authenticated signatures or time freshness. The precise set of committed fields is profile-dependent: the local R3 \texttt{base\_state\_version} is type-checked only and the experiment does not claim anti-rollback or equality to a live authoritative current version.

For fixed-path non-weakening, a profile that checks an independently interpreted source program or an equivalent reference relation may require that an admitted backend positive result entail permission under the selected $\Psi$, on the exact domain for which soundness was established. A bounded GSC-1 checker can test this via $\mathsf{Safe}_D(\Pi,\Psi)$, provided its trusted domain $D$ is nonempty, entirely visited and protected by a total input guard. The absence of a verified general source-to-IR derivation certificate and a concrete all-input refinement proof remains explicit. R3 Final has no independent checked compiler witness $\delta$, no arbitrary authorized-path verifier, and no proof of GSC-1 theorem target T1; accordingly, these broader obligations are \textbf{model and separate-profile research targets}, not results inferred from the OPA/Cedar case counts.

A related \textbf{conditional substitutability criterion} can be formulated for two backends $\pi_1$ and $\pi_2$ under one $\rho$. For the intersection of their supported, faithfully mapped input domain, suppose both satisfy the same normative business relation and the admission verifier correctly checks the same non-backend obligations. Then replacing one backend with the other need not alter the intended \emph{admission-specific checks}. If both return the same correctly interpreted \texttt{ALLOW}/\texttt{DENY} on that restricted domain and encounter no distinct profile errors, the resulting abstract admission decision is expected to agree, subject to identical contextual evidence. This is an \textbf{interface-level conditional claim}, not a theorem that the two implementations satisfy the premises or that their unsupported domains and runtime failures coincide. Where those premises fail, divergent or unresolved outcomes are legitimate evidence against the unrestricted claim.

The distinction between a correct evaluator and a trustworthy admission boundary is therefore symmetric in one respect: neither component subsumes the other. The Gate cannot justify missing constitutional obligations by pointing to a correct backend \texttt{ALLOW}; the backend cannot repair a deliberately omitted mandatory source that never reached it. But the Gate also cannot justify a wrong business \texttt{ALLOW} by pointing to correct slot bindings. A complete end-to-end argument would need separate premises for constitutive authority, source selection and exact versions, backend semantics, admission-verifier soundness, and protected downstream mediation. This decomposition supplies the structure of the paper's evaluation without asserting that every premise has been established.

\subsection{Relation to the evaluated profiles and resulting architectural claim}

To prevent scope inflation, we distinguish three levels of artifact that the research workflow has produced. The \textbf{broad constitutional formal model} introduces explicit SourceSlots, authenticated legal choice witnesses, compile/path-preservation obligations, phase-separated results and a theorem target T1 under external authority assumptions. The \textbf{GSC-1 finite checker profile} narrows semantic preservation to an independently enumerated, nonempty finite state domain and reports separate safety and functionality conditions while leaving structural source-to-IR derivation $\delta$ unimplemented. The \textbf{R3 Final real-engine profile} exercises a smaller local amount-limit policy and structured Gate/Reference schema through native OPA and Cedar adapters. These artifacts share a research motivation, but do not constitute one universally verified implementation.

The R3 Final mapping illustrates both the usefulness and the limitations of BCEAC. Its fixed source fixtures do \textbf{not} establish that a general policy program's \texttt{ALLOW} entails $\mathsf{Permit}_\Psi$ for arbitrary authorized alternatives. That implication belongs to the stronger semantics-preserving profile's stated trust or verification obligation, and must not be attributed to R3 Final's bounded structural checks. Its \texttt{trusted\_manifest} and \texttt{required\_slots} fixture stand in for a \textbf{preinstalled, trusted declared source basis}; they do not exercise a genuine signed constitution, externally current checkpoint, or independent universal manifest-discovery protocol. Its \texttt{loaded\_constraints} records test schema and exact policy-body commitments, including issuer, revision, digest, status and semantics checks. Its \texttt{candidate}, \texttt{proposal}, and \texttt{governance\_snapshots} fields support exact action, target and frozen-snapshot checks for the supported test operations. For supported payment operations, the observed backend positive relation includes applicable \texttt{amount <= limit} constraints \textbf{and a current-state quorum}. For member-removal operations, the current-state quorum remains a native evaluation input without a payment amount test. OPA encodes its current-quorum logic in Rego; Cedar's adapter performs source filtering and approval counting and emits the corresponding native policy/request. In contrast, the admission Gate independently checks the \textbf{frozen governance quorum} associated with the proposal. A properly computed native \texttt{DENY} caused by a changed current roster can therefore coexist with an independently \texttt{PERMIT} frozen-prestate Reference; this is a possible false denial, not an unsafe admission. Neither the entire profile nor its backend correctness premise can accurately be reduced to amount comparison alone. The Gate deliberately does \textbf{not} repeat the business amount predicate. These correspondences are \textbf{partial, experiment-specific observations}, not an equivalence between all abstract contract inputs and concrete R3 JSON fields.

A further \textbf{post-freeze, independent audit counterexample} limits what the 144-record agreement demonstrates. The R3 Final input contract marks \texttt{proposal.target\_member} \textbf{required but nullable} for payments (the present value is \texttt{null}). Deleting the key from an otherwise valid payment \texttt{CASE-01} produces historical Gate \texttt{ADMIT} and independent Reference \texttt{PERMIT}: both used Python \texttt{get} in a way that treats missing and explicit \texttt{null} alike. The independently constructed probe did \textbf{not} run the native macOS binaries, and it does not prove a real-world execution bypass. It establishes a shared implementation/schema omission \textbf{outside} the frozen cohort. The archived R3 Final ACCEPT is an accurate statement about its tested cohort, \textbf{not} proof that the implementation satisfies every required field of its frozen input contract. No original input or accepted result is altered to conceal this limitation.

Several local fields show why interface detail matters. R3 requires \texttt{proposal\_id} to have a valid identifier shape, but that check is diagnostic; the identifier is not proof of an eligible real-world approver. \texttt{boundary\_id} and \texttt{epoch} are checked as identities and types, but the fixture does not establish live cross-boundary trust isolation or replay-resistant checkpoint currentness. \texttt{base\_state\_version} is required and checked as an exact nonnegative integer within the declared range, yet it is \textbf{not} compared against a trusted live state-version value. The test case \texttt{tenant\_not\_bound} is explicitly \texttt{NOT\_CLAIMED}; a coordinated manifest omission is explicitly \texttt{A0\_LIMITATION}. Treating these cases as positive evidence of signed constitutional authority or complete execution protection would contradict the frozen input contract.

At the local decision interface, R3 distinguishes native evaluation from rejection \emph{before} invoking a backend. The archived R3 Final result contains 72 cases across two backends, 144 strict records, 120 recorded strict native evaluations and 24 pre-native rejections; additional permissive controls were executed separately. The two-backend study provides instance-level evidence that binding and governance checks can remain independently articulated when the backend is changed under a common supported profile. It does not demonstrate a new independent policy-language equivalence theorem, a generalized arbitrary-backend adapter, or a universally correct compiler. The experiments' own adversarial history is material: R1 and R2 were independently rejected; R3 received conditional audit; only the frozen R3 Final received a bounded offline acceptance. Counts and their precise evidentiary qualifications are reported in the Evaluation section rather than promoted into a formal guarantee here.

The proposed execution-admission architecture thus has a deliberately conditional claim. \textbf{Relative to an independently accepted and sufficiently adequate governance constitution, correct source and snapshot binding, a faithful supported backend, and sound admission verification}, it specifies independently testable conditions under which a backend's policy decision may become \emph{eligible for consideration} by an execution-control layer. It neither appoints that layer as universally non-bypassable nor proves that a protected effect follows. The practical research contribution is the explicit interface of required obligations, provenance, typed failure and bounded conformance---not an assertion that one brand of policy engine is intrinsically more secure than another.

\section{Engine-Agnostic EBL Architecture}

\subsection{Architectural objective: preserve admission obligations across evaluator choices}

The preceding sections distinguish two questions that are frequently compressed into a single authorization response. The first is whether a policy evaluator permits a particular candidate under the semantics and inputs it was given. The second is whether that evaluator's result may be used as a legitimate contribution to the admission of the \emph{same} candidate under an independently established governance constitution. The Binding-Complete Execution Admission Contract (BCEAC) formalizes the latter question. This section develops a system architecture intended to make the separation explicit, inspectable, and compatible with different policy evaluation technologies within declared semantic limits.

The architectural premise is that governance-source obligations are not a private property of the program used to compute a business decision. If a mandatory policy source is excluded before evaluation, a correct evaluator can faithfully compute \texttt{ALLOW} for the reduced input without identifying the exclusion. Conversely, a verifier that checks the identity, provenance, and presence of every required source cannot necessarily identify an incorrectly implemented business predicate. The two responsibilities are complementary, rather than interchangeable. Our architecture therefore keeps the authority for defining \emph{what must participate} distinct from the mechanism computing \emph{what those participating policies decide} and from the later mechanism able to cause an external action.

We use \textbf{EBL} to denote the proposed execution-boundary architecture, not an assertion that all participating policies must be authored in a new domain-specific language. The construction has three layers: (i) an external semantic interface, optionally expressed in EBL Language; (ii) a governance-binding and admission architecture with a policy evaluation interface; and (iii) a separately controlled release-and-execution layer. Layer 2 is the primary subject of this paper. Layer 1 establishes how its inputs are described and interpreted; Layer 3 establishes a critical non-conclusion: admission eligibility does not itself cause, certify, or irrevocably authorize an external effect.

The architecture is \textbf{engine-agnostic only in a conditional and profile-relative sense}. A backend qualifies for a particular profile when its supported domain, policy-combination semantics, adapter projection, typed results, and error behavior are explicitly declared and correctly implement the profile's normative business relation. An arbitrary Rego program, Cedar policy set, or other backend is not automatically substitutable for an arbitrary EBL program. Nor does a general compiler correctness claim follow from observing agreement on a finite set of evaluation cases. This architecture permits multiple implementations to occupy a bounded \emph{evaluation role}; it does not eliminate the need to establish the correctness of whichever implementation actually occupies that role.

This distinction also avoids attributing a capability gap to conventional access-control architectures without justification. A trusted policy decision point coupled to a sufficiently complete, authoritative, and non-bypassable enforcement wrapper could discharge the same admission obligations. BCEAC does not assert that such a construction is impossible or unprecedented. Its proposed contribution is to articulate the obligations as a separately auditable, typed, constitution-relative contract, and to show how a change in evaluator need not silently change the source-of-authority contract. The difference is an explicit interface and assurance decomposition, not an assertion that existing evaluators are intrinsically defective.

\subsection{Three-layer decomposition and typed interfaces}

\begin{figure}[p]
\centering
\begin{tikzpicture}[
  font=\footnotesize,
  >=stealth,
  node distance=4mm,
  layer/.style={draw, thick, rounded corners=3pt, align=center, inner sep=5pt, fill=blue!3},
  box/.style={draw, rounded corners=1pt, align=center, inner sep=3pt, fill=white},
  side/.style={draw, dashed, rounded corners=1pt, align=center, inner sep=3pt, fill=gray!8},
  arr/.style={->, thick},
  darr/.style={->, dashed, thick},
  lab/.style={font=\scriptsize, align=center, text=black!65}
]
\node[layer, text width=9.0cm] (L1) {\textbf{LAYER 1 --- EXTERNAL SEMANTIC INTERFACE}\\[1pt]
Optional EBL Language / other structured authoring format\\
Candidate $\cdot$ boundary $\cdot$ declared semantic profile\\
governance $\cdot$ facts $\cdot$ evidence $\cdot$ execution constraints};

\node[box, text width=9.0cm, below=17mm of L1] (auth) {\textbf{Verifier-installed authority $T$ (independent side input)}\\
Constitution $C_e$ $\cdot$ authenticated manifest $M_e$\\
Heads $H_e$ $\cdot$ checkpoint $\kappa$ $\cdot$ operation rules\\
Accepted phase/state evidence $\cdot$ verification time $t_v$};

\draw[arr] (L1.south) -- node[right, lab, xshift=1mm]{Canonical typing} (auth.north);

\node[box, text width=6.4cm, below=10mm of auth] (phi) {Structured obligations $\Phi$};
\draw[arr] (auth.south) -- node[right, lab, xshift=1mm]{Closure and obligation resolution} (phi.north);

\node[box, text width=6.4cm, below=9mm of phi] (psi) {Fixed authorized path $\Psi$};
\coordinate (inst) at ($(phi.south)!0.5!(psi.north)$);
\node[side, text width=2.5cm, right=8mm of inst] (w) {authorized witness $w$};
\draw[arr] (phi.south) -- node[left, lab, xshift=-1mm]{Instantiate$(\Phi,w)$} (psi.north);
\draw[darr] (w.west) -- (inst);

\node[box, text width=6.4cm, below=5mm of psi] (rpsi) {Required slots $R_\Psi$};
\draw[arr] (psi.south) -- (rpsi.north);

\node[box, text width=7.2cm, below=8mm of rpsi] (dis) {exact discharge / context preflight};
\draw[arr] (rpsi.south) -- (dis.north);
\node[side, text width=2.5cm, left=5mm of dis] (sel) {Selected objects $S$};
\node[side, text width=3.2cm, right=5mm of dis] (bnd) {verifier-bound context $B_\rho$\\ (state + \mbox{attempt} + path)};
\draw[arr] (sel.east) -- (dis.west);
\draw[arr] (bnd.west) -- (dis.east);

\node[box, text width=8.4cm, below=8mm of dis] (eval) {\textbf{Policy Evaluation Interface}\\
{[}EBL Native $\mid$ OPA $\mid$ Cedar $\mid$ qualified other{]}\\
profile $\rho$ + adapter $a$ + backend $\pi$};
\draw[arr] (dis.south) -- (eval.north);

\node[box, text width=7.8cm, below=7mm of eval] (res) {typed result $E$ + invocation trace $\tau_\pi$};
\draw[arr] (eval.south) -- (res.north);

\node[box, text width=8.8cm, below=7mm of res] (adm) {\textbf{Independent admission verification}\\
MatchResolution$(X,R)$ + MatchEvaluation$(X,R,E)$\\
source binding $\cdot$ semantic obligations $\cdot$ frozen\\
governance $\cdot$ evidence $\cdot$ mapping $\cdot$ host floor};
\draw[arr] (res.south) -- (adm.north);

\node[box, text width=6.6cm, below=7mm of adm] (out) {Eligible$(\alpha)$ / Reject / Unresolved\\[2pt] \textbf{ELIGIBILITY ONLY --- NO GRANT}};
\draw[arr] (adm.south) -- (out.north);

\node[layer, dashed, fill=gray!8, text width=10.5cm, below=11mm of out] (L3) {\textbf{LAYER 3 --- EXECUTION CONTROL}\\
\textbf{(NOT VERIFIED HERE)}\\
Hub arbitration $\rightarrow$ Security Module $\rightarrow$ Executor\\
fresh release checks $\cdot$ mediation $\cdot$ effects};
\draw[arr] (out.south) -- (L3.north);

\node[anchor=south east, font=\small\bfseries, inner sep=1pt]
  (L2title) at ([xshift=-4mm,yshift=4mm]auth.north) {\shortstack[r]{LAYER 2 --- EBL ADMISSION\\ ARCHITECTURE}};

\begin{scope}[on background layer]
\node[draw, dashed, rounded corners=4pt, fit=(L2title)(auth)(out)(sel)(bnd), inner sep=8pt] (L2) {};
\end{scope}
\end{tikzpicture}
\caption{Architectural specification of the proposed EBL admission pipeline.
Arrows denote data and control relationships, not evidence that all components
or protections have been implemented. Layer 3 is shown only to mark the
separate execution boundary; it is not verified in this paper.}
\label{fig:architecture}
\end{figure}
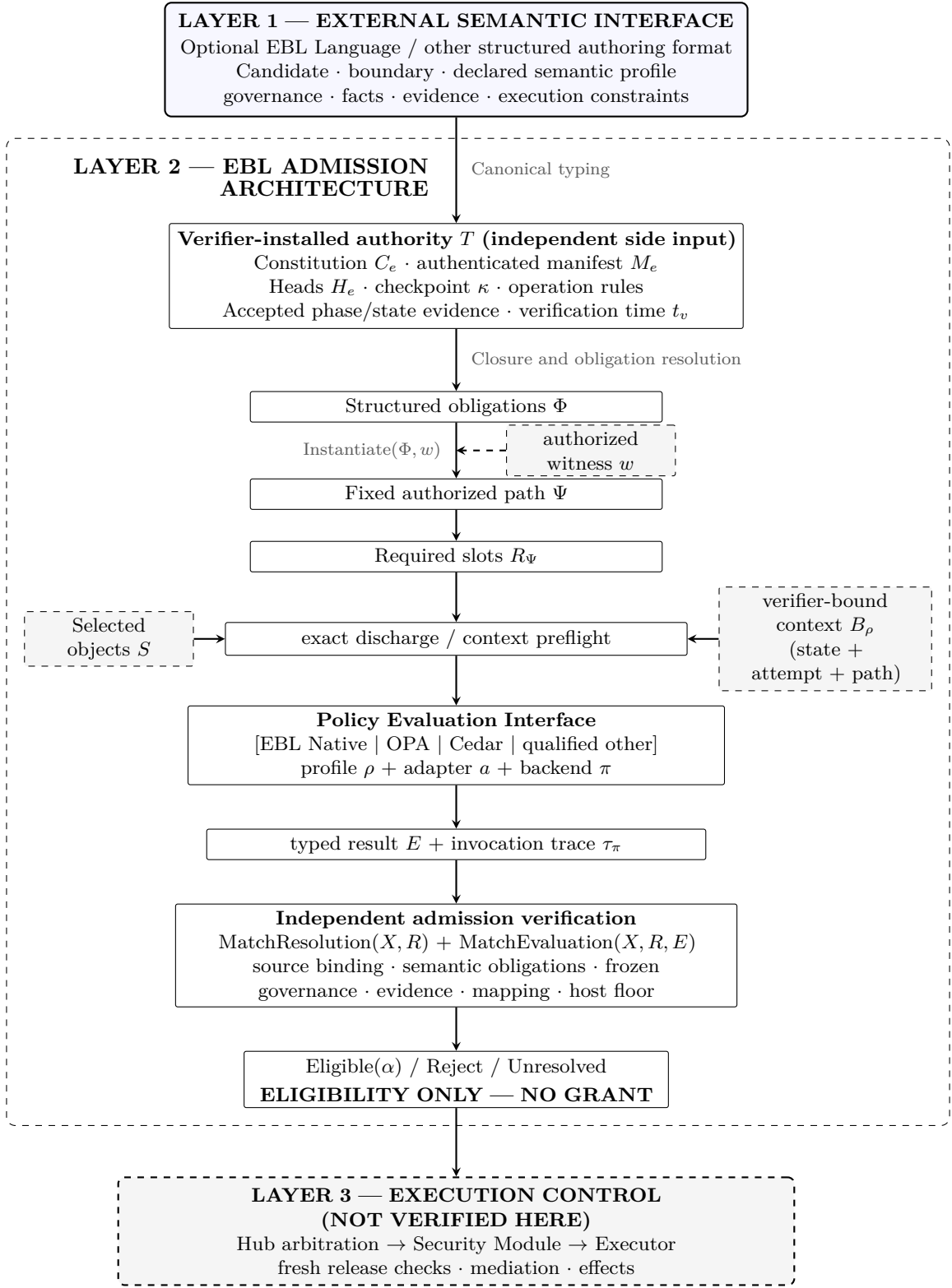

Figure~\ref{fig:architecture} (architectural specification) depicts the intended dependencies. Arrows denote data and control relationships, not evidence that all components or protections have been implemented.

The architectural pipeline is \emph{logical}, not a requirement that every predicate be called exactly once or in the displayed order. For example, an adapter may reject an invalid JSON representation before invoking a native evaluator, and a post-evaluation checker may independently detect a source mismatch. Some checks can occur in parallel; some require earlier authenticated evidence. The security-relevant ordering is the \textbf{dependency ordering}: the selected set or native result must not establish the constitution, the authorized path, or the independent host floor from which its own legitimacy is subsequently inferred.

Let $\rho$ denote an explicitly versioned admission semantic profile. We reuse, rather than redefine, the logical input record $X_\rho$ from Section 4.2: it includes $c,b,e,f,\allowbreak\mathcal{T},C_e,\allowbreak\kappa,M_e,H_e,\allowbreak w,S,G_0,\allowbreak q_{\mathrm{op}},\pi,a,\allowbreak d_\pi,t_v$. Here $d_\pi$ is the correlated evaluation-result projection defined below; it is not an additional vote whose value can override a conflicting result object. The authenticated verification environment is \textbf{exactly} the Section 4.3 environment

\begin{equation*}
\Gamma_\rho=\langle b,e,\mathcal{T},C_e,M_e,H_e,\kappa,f,t_v,
\mathsf{IssuerRights}_e,\mathsf{OperationRegistry}_e\rangle.
\end{equation*}

The two notations express different logical views, not two independently configurable roots of authority. The verifier constructs a candidate environment view from the corresponding $X_\rho$ fields, its \textbf{independently installed} $\mathcal{T}$, and the profile-authorized authentication/fetch mechanisms for committed manifests, heads, rights and currentness evidence. A caller-supplied $\mathcal{T}$, $M_e$, or $\kappa$ is not made trustworthy by occupying a tuple field. Before acceptance, a required component of this view may be available and verified, conclusively invalid, or unresolved; $\Gamma_\rho$ is \emph{not} presumed complete merely because the formal tuple can be written. The verifier must cross-check the declared boundary/epoch against the trusted Candidate/action-to-boundary and operation registries. In particular, a digest or pointer to $M_e$ or $H_e$ is not a substitute for obtaining and verifying the necessary materialized records.

For a particular attempt $u$, let $X^{\mathrm{res}}_\rho=\mathsf{ProjectResolution}_\rho(X_\rho,u)$ include the exact candidate identity and a purported choice-path witness to be validated under the independent constitution, but not a selected-source or evaluator-result-derived authority declaration. Our architectural interface family is then:

\begin{equation}
\begin{aligned}
\mathsf{Resolve}_\rho &: (X^{\mathrm{res}}_\rho;\Gamma_\rho)
        \longrightarrow \mathsf{ResolutionResult}_\rho,\\
\mathsf{Evaluate}_{\rho,\pi,a} &: (c,S,\sigma,B_\rho)
        \longrightarrow \mathsf{PolicyEvaluationResult}_\rho,\\
\mathsf{Admit}_\rho &: (X_\rho,R_\rho,E_\rho)
        \longrightarrow \mathsf{AdmissionDecision}_\rho.
\end{aligned}
\label{eq:a5-1}
\end{equation}

Here $R_\rho$ and $E_\rho$ denote the actual returned resolution and evaluation records, and $B_\rho$ is the \textbf{verifier-accepted BoundContext}, constructed from $X_\rho,R_\rho,\Gamma_\rho$, and authenticated profile-appropriate state $\sigma$. These interfaces are \emph{abstract design obligations}, not deployed APIs or the concrete Python signatures from R3 Final. Semicolons in the resolver signature emphasize the environment dependency, not an effect-free function that can ignore its time, source-fetch or checkpoint inputs. The resolver must either operate on a coherent, explicitly bounded evidence snapshot or acquire evidence through an identified trusted retrieval rule; an unavailable record, expired checkpoint, or nonterminating fetch must end, within the profile's specified bound, in a typed \textbf{non-positive} result. On a fixed accepted evidence snapshot and profile, the semantic result must not change due to an undeclared ambient state. This is an interface-level requirement and not a claim that a general resolver has been implemented or proved terminating.

We further specify the result shapes by \textbf{conceptual sum types} rather than unverified Python field schemas. A successful $\mathsf{ResolutionResult}_\rho$ is $\mathsf{Resolved}(u,\allowbreak\mathsf{AuthorityRef},\allowbreak\Phi,w,\allowbreak\Psi,R_\Psi,\allowbreak\mathsf{ContextRef})$, in which $\Psi$ and $R_\Psi$ were derived and checked from the trusted constitution and legal node-keyed witness $w$. The other branches are $\mathsf{ResolutionInvalid}(r)$ for a demonstrated violation and $\mathsf{ResolutionUnresolved}(r)$ for missing, unauthenticated or otherwise inconclusive necessary evidence. Neither branch supplies an acceptable $R_\Psi=\varnothing$ default. An evaluation record is $\mathsf{Evaluated}(u,\mathsf{Decision},\tau_\pi,\mathsf{ContextRef})$, or a typed non-positive failure such as $\mathsf{NotInvoked}$, $\mathsf{Unsupported}$, $\mathsf{MappingError}$, or $\mathsf{EvaluationError}$, with its phase and provenance preserved. The successful decision retains the profile's typed $\mathsf{Allow}/\mathsf{Deny}/\mathsf{Indeterminate}/\mathsf{Error}$ distinction and its underlying native trace. A pre-native failure must never be fabricated into a native $\mathsf{Deny}$ observation. As in Section 4, admission returns $\mathsf{Eligible}(\alpha)$, $\mathsf{Reject}(r)$, or $\mathsf{Unresolved}(r)$; $\alpha$ is not an execution Grant.

\textbf{Evaluation-state provenance and BoundContext.} The profile defines the type, trust rule, temporal interpretation, and canonical identity of $\sigma$. It can be a current business-state view or a collection of profile-accepted facts with authenticated or explicitly assumed provenance; it must not be inferred solely from a native decision or adapter's unsigned assertion. In particular $\sigma$ is \textbf{not silently identified with} the frozen governance pre-state $G_0$. The R3 profile deliberately distinguishes current-state quorum evaluated by its native business relation from the frozen-prestate quorum checked by the independent Gate. The verifier's expected state commitment $\mathsf{StateRef}_\rho(\sigma)$, including any required state-version, source, freshness and phase markers, must therefore be fixed under the independently defined profile rule for the attempted evaluation.

The conceptual helper $\mathsf{BuildBoundContext}_\rho(X_\rho,R_\rho,\Gamma_\rho,\sigma)$ returns either a verified context $B_\rho$, a conclusively invalid result, or an unresolved result. Its accepted context commits to $u,\operatorname{cid}(c),b,e$, the authenticated constitution/manifest/head/checkpoint references, $w,\Psi,R_\Psi$, exact selected source-head commitments, the relevant frozen $G_0$ and operation $q_{\mathrm{op}}$, the expected $\mathsf{StateRef}_\rho(\sigma)$, $\rho,\pi,a$, and the verification/invocation phase. These are logical equality and authentication obligations; the record need not serialize all values into a universal wire format. An adapter may translate this context for its native backend, but the adapter is not empowered to decide retrospectively what candidate, path or state the verifier meant to authorize.

\textbf{Same-attempt result consistency.} Define $\mathsf{MatchResolution}_\rho(X_\rho,R_\rho;\Gamma_\rho)$ to hold positively only when $R_\rho$ is a successfully authenticated, complete resolution for the current $u$, candidate, boundary, epoch, manifest/head/checkpoint view, choice witness and phase. Define $\mathsf{MatchEvaluation}_\rho(X_\rho,R_\rho,E_\rho;\Gamma_\rho)$ to hold positively only when the evaluation was actually performed (where required), the verified $B_\rho$, state reference and invocation trace $\tau_\pi$ refer to that same attempt and exact candidate, scope, path, source versions, profile, backend, adapter, phase and accepted state, and the result decoding is justified by the declared profile. Equality of $u$, $\mathsf{ContextRef}$, or an invocation ID \textbf{alone} is insufficient: the values are checked against independently accepted commitments and the profile's named invocation-authenticity assumption. A backend-supplied correlation label cannot authenticate itself. Every required relation returns a verified, conclusively false, or unresolved evidence assessment under Section 4's dependency-sensitive rules. Thus the following are \textbf{necessary} interface-level conditions for a positive admission outcome, not a sufficient implementation theorem:

\begin{equation*}
\begin{aligned}
\mathsf{Admit}_\rho(X_\rho,R_\rho,E_\rho)=\mathsf{Eligible}(\alpha)
\Longrightarrow{}&\;\mathsf{MatchResolution}_\rho(X_\rho,R_\rho;\Gamma_\rho)=\mathsf{V}\\
&\land\mathsf{MatchEvaluation}_\rho(X_\rho,R_\rho,E_\rho;\Gamma_\rho)=\mathsf{V}\\
&\land d_\pi(X_\rho)=\mathsf{ProjectDecision}_\rho(E_\rho).
\end{aligned}
\end{equation*}

Where $X_\rho$ carries $d_\pi$ as specified in Section 4, $d_\pi$ must be the \textbf{exact verified decision projection} of this same $E_\rho$, including its typed status. It is not an independent source of policy authority. If the record is duplicated for transport, both copies must match; if a trusted comparison establishes a conflict, positive admission is forbidden with a conclusively invalid result. If provenance is unavailable, it remains unresolved. Implementations may instead pass the authoritative $E_\rho$ once and derive $d_\pi$ from it, but this is a representation refinement, not an authorization to accept contradictory duplicated decisions.

For example, $(X_A,R_B,E_A)$ cannot yield positive admission merely because all three values share profile $\rho$; nor may an $\mathsf{Allow}$ computed for $\sigma_1$ justify a check expecting a different $\sigma_2$ unless the profile independently establishes their relevant equivalence and explicitly authorizes that equivalence. Likewise, a $\mathsf{ResolutionUnresolved}$ record cannot be promoted by a favorable $E_A$. These negative constraints follow from the full matching predicates, not from a claim that ordinary nominal types alone enforce cryptographic origin or dependent equality. Failure to obtain a required matching proof leads to a non-positive admission result, with known invalidity and genuine uncertainty distinguished according to Section 4's tri-state precedence. The verifier does \textbf{not} recompute the backend's full business predicate merely to establish these associations.

The architecture establishes an explicit \emph{non-coercion rule}:

\begin{equation}
\begin{aligned}
&\mathsf{BVerified}(\beta)\not\Rightarrow\mathsf{PolicyAllow},\qquad
\mathsf{PolicyAllow}\not\Rightarrow\mathsf{Eligible},\\
&\mathsf{Eligible}\not\Rightarrow\mathsf{ExecutionGranted},\qquad
\mathsf{ExecutionGranted}\not\Rightarrow\mathsf{ExternalEffect}.
\end{aligned}
\label{eq:a5-2}
\end{equation}

Here each arrow is a forbidden \textbf{unconditional implication}; a later stage may establish an additional relation under its own independent premises. In the earlier execution-boundary formulation, issuance and exercise of an action-scoped Execution Grant require their own conditions and version discipline \cite{Wu2026EBLCore}. The current chapter neither replaces that protocol with BCEAC nor silently introduces a coercion between their result types.

\subsection{Layer 1: optional EBL Language and mandatory canonical semantics}

A policy-independent interface still needs a common vocabulary. For a Candidate, the evaluator must know which actor requests which action against which exact target, under what state, fact and proof context. Governance resolution additionally needs to know the independently applicable source families, operation classification, permitted alternative-path rules and constraints that cannot be altered by the individual policy author. These are typed \textbf{semantic commitments}; an EBL textual syntax may be one way to express them, but the architecture cannot depend on every policy being written in that syntax.

We distinguish an \textbf{authoring surface}, a \textbf{canonical admission contract}, and a \textbf{backend representation}. The authoring surface may be EBL Language or a native policy language. The canonical contract defines which fields are decision-relevant, how source obligations are derived, what counts as a legal path and valid policy object, how errors are typed, and how evaluation is bound to the candidate. A backend representation is what a particular evaluator actually consumes: for example, a Rego input/data context, a Cedar request/entity/policy set, or a bounded future EBL Native program. The three representations need not be byte-identical, and their correspondence is an obligation rather than an assumption from naming alone.

A versioned profile should therefore identify, at minimum, (i) canonical Candidate encoding and relevant equality rules, (ii) supported actions, types and ranges, (iii) required source and object schemas, (iv) policy-combination and alternative-path semantics, (v) backend input projection and result decoding, (vi) unsupported, missing and error cases, and (vii) limits of any permitted additional constraints. The profile is not a synonym for the list of loaded policies: its interpreter, registry and admissible domain must be authorized independently of the caller's selection. If a numeric value is represented differently by two backend APIs, the profile must either supply a faithful conversion or exclude that input from the shared supported domain. Quietly converting a semantically unsupported input into a permissive default is not interoperability.

The first three research papers provide related but separate interfaces. The authority-decomposition perspective identifies why the holder of an approval or policy decision need not control the actual cause of execution \cite{Wu2026Authority}. The execution-boundary profile distinguishes canonical action identity, Root and Operational policies, evidence obligations, release conditions and later Grant/Redemption semantics \cite{Wu2026EBLCore}. Evidence-coverage analysis emphasizes that the inventory of required evidence for a chosen scope cannot safely be inferred solely from whatever records happen to have been collected \cite{Wu2026Evidence}. BCEAC adopts the \textbf{methodological discipline of declared inventories and exact scope}, not an identity between retrospective evidence coverage and pre-execution policy-source binding. Those remain different kinds of obligation.

An EBL Native backend remains a legitimate future component. It could evaluate a purpose-built deterministic fragment under the same profile. However, we do not make the unimplemented claim that a general translator preserves arbitrary EBL semantics in Rego, Cedar, or another policy language. An external evaluator qualifies through its verified or explicitly assumed \emph{profile relation}, not through the existence of an EBL-branded source file. This separation is essential to the paper's conditional engine-agnostic claim.

\subsection{Layer 2a: independent authority, closure and required-source resolution}

The admission architecture begins with authority, not with a policy result. Let $\mathcal{T}$ be the verifier-controlled trust configuration and $C_e$ the constitution authorized for boundary $b$ and epoch $e$. A manifest $M_e$, source-version heads $H_e$, operation descriptors, and checkpoint $\kappa$ may supply verifiable commitments and currentness evidence under that configuration. The caller may transport the bytes, but successful parsing, transport, or local hashing does not confer constitutive authority. Equally, a selected policy object cannot nominate its own trust root for the purpose of validating itself.

The resolver performs two conceptually different operations. \textbf{Closure} establishes that the verifier has an authenticated, complete \emph{declared} inventory for the accepted constitution and profile. \textbf{Applicability resolution} determines which declared source seats govern the exact Candidate under authenticated facts. The distinction matters because a currently inapplicable slot may still be part of the trusted constitutional inventory, while an applicable slot does not disappear because its current policy body was not fetched.

Using the notation established in Sections 3 and 4:

\begin{equation}
\begin{aligned}
\Phi &:= \mathsf{ResolveObligations}(C_e,c,f),\\
\Psi &:= \mathsf{Instantiate}(\Phi,w),\\
R_{\Psi} &:= \{\,s\mid \mathsf{require}(s)
                    \text{ occurs on the authenticated chosen path }\Psi\,\}.
\end{aligned}
\label{eq:a5-3}
\end{equation}

The equations in \eqref{eq:a5-3} denote the \textbf{successful resolution branch} only; an invalid or unresolved resolution produces no authenticated $\Phi$, $\Psi$, or $R_{\Psi}$ that admission may use. The resolver must not compute $R_{\Psi}$ from \texttt{Selected}, \texttt{Loaded}, the native engine's reported dependencies, or a successful \texttt{ALLOW}. A missing constitution, unverified manifest continuation, unavailable required checkpoint, or unknown applicability guard cannot be resolved by treating the necessary set as empty. The latter is a legitimate result only if the independently trusted rules explicitly establish an empty applicable set and all non-optional host requirements remain applicable. The broader GSC-1 formal model describes a bounded closure procedure involving trusted root configuration, epoch/checkpoint validation, committed manifest count and pages, issuer rights, and finite guarded applicability; its cryptographic and complete-refinement obligations are not discharged by R3 Final.

A critical limit is \textbf{normative sufficiency}, denoted A0. Even an internally coherent signed manifest can omit an authority whom the actual organization should have included. Two different real-world normative situations can present indistinguishable evidence if the verifier depends on the same compromised sole source of constitutional publication. Closure is thus constitution-relative; it is not a proof of all legitimate authorities in the world. Introducing an admission verifier does not solve this root-of-authority problem. Instead, the architecture makes the root, scope and expected freshness explicit as deployment or verification premises.

The architectural choice-path mechanism is similarly bounded. A legal alternative structure may allow different approved coalitions or different acceptable classes of constraint. But $w$ must be validated under the constitutive rules and fixed for the attempted Candidate and authenticated context. Once $\Psi$ is fixed, permissions from an unchosen alternative cannot be used to justify the chosen path. This is not a requirement to evaluate every possible branch in every request: it is a requirement to account for every \textbf{chosen and independently mandatory} source, while preserving the declared semantics of legitimate short-circuiting inside a source's own policy body. The distinction prevents a local whitelist exception from silently replacing an independently required Hub or Root constraint.

\subsection{Layer 2b: exact policy-body binding and EffectiveArtifact identity}

The distinction between a \textbf{SourceSlot} and a \textbf{PolicyObject} is the architectural mechanism for preventing body availability from defining constitutional necessity. A SourceSlot is a stable required source seat specified by the accepted constitution; a PolicyObject is a particular versioned policy body authorized to occupy that seat. Replacement, revocation, expiration, unavailability, or invalidation of a policy body may change its ability to discharge the seat, but not the seat's constitutive existence. A body that claims a familiar \texttt{slot\_id} is not sufficient evidence that it is the currently designated body for that slot.

For each required $s\in R_{\Psi}$, an exact discharge check must establish the designated source reference, authorized issuer and rights, scope, policy family, revision, semantics identifier, digest, validity and applicable status, and must reject unapproved substitutions and conflicting duplicates. The relevant comparison is against independently authenticated designated heads and the accepted snapshot, not merely between fields supplied by the same untrusted caller. If designated status cannot be established, the checker returns an unresolved outcome rather than inventing a current version. If absence or conflict is established within an authenticated complete boundary, it can return a demonstrated violation. The same surface symptom, such as ``no valid body loaded,'' can therefore have different epistemic causes that the result type must preserve.

HAVENLON's proposed \textbf{EffectiveArtifact} is a useful point at which these relations may be committed. It denotes a versioned effective or compiled representation associated with a policy scope and source/version context, rather than a new authority that decides which independent sources ought to exist. In a finite pure-language profile, the formal model uses $\Pi$ for an executable decision program and $\delta$ for proposed checkable source-to-program derivation evidence. Neither term licenses assuming that an arbitrary OPA input/policy pair or dynamically constructed Cedar policy is precisely that same IR, nor that $\delta$ has already been implemented. The architecture must support distinct \textbf{artifact representations} with a common binding obligation, rather than requiring a universal artifact byte format.

A possible profile-indexed artifact record has the following conceptual fields:

\begin{equation}
\begin{aligned}
\mathsf{ArtifactBinding}_\rho := \langle
 &\mathsf{cid}, b,e,\mathsf{ConstitutionRef},
 \mathsf{CheckpointRef},\\
 &\mathsf{ChosenPathRef},\mathsf{RequiredSourceCommit},
 \mathsf{SelectedHeadVector},\\
 &\mathsf{SemanticProfileRef},\mathsf{AdapterRef},
 \mathsf{BackendArtifactRef},\mathsf{ContextRef}\rangle.
\end{aligned}
\label{eq:a5-4}
\end{equation}

This is a \textbf{proposed logical commitment set}, not the wire schema of the R3 Final JSON fixture or the published schema of a deployed HAVENLON device. An implementation may factor the commitments among an EffectiveArtifact, an invocation record and verification evidence. What matters is that the verifier can establish their exact, authorized relationship without trusting a self-reported \texttt{complete=true} field. An artifact digest binds bytes relative to a claimed reference; it does not authenticate the publisher, establish source currentness, prove semantic translation, or make two decision programs equivalent. The signature, trust-anchor and checkpoint premises remain separate.

We use $\beta$ for the resulting \textbf{binding evidence}. This record should identify what was checked, against which normative basis and profile, at which verification time, and with which exact commitments. It must not become an authority-bearing execution capability solely because it is serializable or carries a hash. A successful binding check can coexist with a legitimate policy \texttt{DENY}; conversely, an evaluator can produce \texttt{ALLOW} when binding remains unresolved. An architecture that conflates these situations cannot faithfully account for why an action was not admitted.

\subsection{Layer 2c: the Policy Evaluation Interface and conditional substitution}

A policy backend's responsibility is to evaluate the \textbf{declared business relation} on the supported, accurately projected context. The admission architecture supplies the exact Candidate, selected valid objects, relevant state, and profile-bound evaluation arguments; the adapter translates these into the native invocation and decodes the result. This introduces at least three separable correctness questions: whether the original business relation is the one authorized by the profile, whether the adapter faithfully maps the original typed context into the native representation, and whether the backend actually implements the intended relation within its supported domain.

Let $D_\rho$ be a declared guarded domain of supported inputs and states, and let $z=(c,S,\sigma,\Psi)$ include the fixed authorized path relevant to an attempt. Section 4 defines a profile-specific $\mathsf{Permit}^{\rho}_{\mathrm{business}}(z)$ and, for profiles claiming semantic non-weakening, a separately interpreted $\mathsf{Permit}_{\Psi}(c,\sigma)$. A qualified backend/adapter pair must satisfy, as a verified property or an explicit premise, a relation of the form:

\begin{equation}
\forall z\in D_\rho:\quad
 \mathsf{DecodedAllow}_{\rho,\pi,a}(z)
 \Longleftrightarrow
 \mathsf{Permit}^{\rho}_{\mathrm{business}}(z).
\label{eq:a5-5}
\end{equation}

For a \textbf{path-preserving} semantic profile, the additional obligation is not optional:

\begin{equation}
\forall z\in D^{\mathrm{path}}_\rho:\quad
 \mathsf{Permit}^{\rho}_{\mathrm{business}}(z)
 \Longrightarrow
 \mathsf{Permit}_{\Psi}(c,\sigma).
\label{eq:a5-6}
\end{equation}

These obligations have different purposes. Equation~\eqref{eq:a5-5} specifies faithful evaluation of the profile's entire business predicate; Equation~\eqref{eq:a5-6} prevents that predicate from permitting an action forbidden by the fixed lawful governance path on the \emph{separately declared} path-preservation domain. The second is not entailed by exact provenance of a selected source, because the native policy-combination semantics may still be wrong. An \texttt{all(A,B)} requirement cannot be implemented as \texttt{A OR B} merely because both policy bodies were loaded. Nor can an unknown or malformed input simply be removed from the checked domain after an unfavorable native result. The admitted domain and its guard must be fixed independently of the backend outputs.

These equations do not announce universal equivalence between EBL Native, Rego, Cedar or any third-party language. They constrain a \textbf{qualified pair for one profile}. In an actual deployment, sufficient evidence of adapter/backend correctness may require independent specification, focused tests, verified translation, proof-carrying artifacts or an explicitly bounded trust assumption. Finite agreement alone supplies observations, not a theorem over all future program inputs. Likewise, an evaluator that is malicious or incorrectly returns \texttt{ALLOW} can violate \eqref{eq:a5-5}; a Gate that deliberately avoids recomputing the business predicate cannot generally repair that fault. The architecture's claim requires backend correctness and must disclose its trust in the adapter as part of the trusted computing base.

A native result is never interpreted independently of invocation. In particular, the admissible result is the one accepted by $\mathsf{MatchEvaluation}_\rho(X_\rho,R_\rho,E_\rho;\Gamma_\rho)$, not any other \texttt{ALLOW} associated with the same profile. The verifier fixes the expected business-state reference before trusting the mapped decision; a native response cannot choose its own expected state, and a repeated $d_\pi$ in $X_\rho$ must equal the verified result projection. The evaluation interface must distinguish \texttt{ALLOW}, \texttt{DENY}, an indeterminate native result, a mapping error, failure to invoke the native engine, and an unsupported policy construct. The invocation record should commit to the profile and adapter versions, the exact Candidate/context projection, the source and artifact identities, and the relevant raw result and diagnostic information. Simply displaying a native \texttt{ALLOW} string does not show that the evaluator received the same Candidate that will later be admitted. Conversely, an adapter's statement that native evaluation occurred is not cryptographic proof of execution unless the profile supplies an independent means of attesting it.

A useful abstraction for two qualifying backends is \textbf{conditional interface substitutability}. Suppose two backend/adapter pairs operate on the same trusted input, share the same supported profile and domain, preserve the complete business relation, and expose compatible error/result conventions to the same admission checker. Then an intended change in evaluator should not remove an independent constitutional SourceSlot, rewrite the applicable governance pre-state, or grant execution authority on the evaluator's behalf. If, additionally, the pairs agree on the correct typed result for a given supported input, an identical admission checker can reach the same abstract outcome given identical other evidence. This is conditional compositional reasoning, not an unconditional \texttt{EngineA $\equiv$ EngineB} theorem. It permits legitimate divergence outside the common domain, under different failure behaviors, or when one implementation does not meet its profile obligations.

\subsection{Layer 2d: preflight, independent admission verification and evidence dependency}

The independent admission stage does not merely apply a final Boolean AND to whatever results are readily available. It needs to know \textbf{which obligations must have been checked} and whether the evidence supporting each conclusion has itself been authenticated. We distinguish a complete, finite profile-derived inventory of necessary verification obligations from the subset of functions a particular implementation happened to run. Let $O_\rho(X_\rho)$ denote the former; its correctness is subject to a profile-specific soundness obligation. A prematurely terminated or empty performed-check list is not evidence that $O_\rho$ has been satisfied.

As established in Section 4, each necessary check can yield a known-valid result, an independently demonstrated violation, or unresolved evidence. We denote these values by $\mathbf{V}$, $\mathbf{F}$, and $\mathbf{U}$, respectively. The aggregation priority is evidence-aware: an independently established false condition precludes eligibility even if unrelated inputs are unknown, but a purported contradiction that depends on unauthenticated data is not yet an independently established false condition. In the absence of conclusive invalidity, any unresolved necessary condition prevents positive admission. Only an authenticated, complete inventory whose necessary checks all succeed can support a positive result, and then only in conjunction with the admitted native \texttt{ALLOW} and the remaining profile-specific conditions.

One way to describe the admitted dependency is:

\begin{equation}
\begin{aligned}
\mathsf{Eligible}_\rho(X_\rho)\ \Longrightarrow\ &
 \mathsf{TrustedAuthorityAndClosure}_\rho(X_\rho)\\
 &\land\ \mathsf{FixedLegalPath}_\rho(X_\rho)\\
 &\land\ \mathsf{ExactSourceAndArtifactBinding}_\rho(X_\rho)\\
 &\land\ \mathsf{FrozenJointGovOK}_\rho(X_\rho)\\
 &\land\ \mathsf{SemanticProfileOK}_\rho(X_\rho)\\
 &\land\ \mathsf{BoundInvocation}_\rho(X_\rho)\\
 &\land\ \mathsf{DecodedDecision}_\rho(X_\rho)=\mathsf{Allow}\\
 &\land\ \mathsf{AdmissionHostFloorOK}_\rho(X_\rho).
\end{aligned}
\label{eq:a5-7}
\end{equation}

In \eqref{eq:a5-7}, \texttt{BoundInvocation} specifically requires verified $\mathsf{MatchResolution}$ and $\mathsf{MatchEvaluation}$, including accepted state provenance and consistent duplicated decision projections. These obligations must belong to the complete finite verification inventory $O_\rho(X_\rho)$; their absence from an executed-check list cannot make them true. A conclusive mismatch produces a negative classification, while unavailable authenticating evidence blocks eligibility as unresolved unless another independently established violation already applies. An inapplicable or incomplete resolver result cannot be repaired by a backend $\mathsf{Allow}$. Equation~\eqref{eq:a5-7} is an \textbf{architectural necessary-condition specification} inherited from BCEAC, not a sufficient theorem for an external effect. In particular, \texttt{SemanticProfileOK} is not made true by merely naming a profile: for path-preserving claims it must depend on an independently justified instance of \eqref{eq:a5-6}, and the evaluator's correctness premise must have a defined scope. Similarly, listing \texttt{TrustedAuthorityAndClosure} does not construct A0, and listing a host floor does not prove that a deployment enforces it. The precise dependency order and result precedence may differ in a concrete implementation, but the source of each normative obligation may not be inverted.

Preflight serves a distinct role. It can reject a malformed Candidate, unsupported numeric representation, invalid source-body schema or inadmissible target binding \emph{before} native evaluation. Such a rejection is not a policy engine \texttt{DENY} and must not be counted as a successful native evaluation. The policy backend's true \texttt{DENY} is also not equivalent to failed source binding: both mandatory policies could be correctly loaded and faithfully evaluated, but one may legitimately reject the Candidate. The architecture therefore preserves at least three provenance dimensions in an audit record: (1) normative input and binding assessment, (2) native evaluation and its raw/mapped status, and (3) the resulting independent admission judgment.

The rule for \texttt{Unresolved} is intentionally strict about authority but careful about meaning. Missing checkpoint evidence, unknown required source status, unsupported semantics or a failed native invocation cannot lead to positive eligibility. However, they do not all justify the factual statement that a signature was false or a governor explicitly denied the action. A typed unresolved result is often the more precise diagnosis. This distinction remains important when reconciling a proposed normative three-state contract with earlier implementation labels, which may not classify every invalid or unknown field identically. The architecture does not retroactively change archived verdicts to fit its later abstract model.

\subsection{Joint Governance as constitutional input, not a backend-local vote}

Joint Governance is not merely a second policy function added after business evaluation. It affects \textbf{who may alter governing conditions}, which sources must participate, and which snapshot governs an authorization or transition. For ordinary independent constraints, governance can be intrinsic: every mandatory source contributes to the effective boundary through a correctly specified combination rule. For discrete state or authority transitions, the governance design also permits a workflow-based proposal and approval process. The governance architecture distinguishes \texttt{UNILATERAL}, \texttt{INTRINSIC}, \texttt{PROPOSAL\_AUTO\_COMMIT}, \texttt{PROPOSAL\_GRANT\_EXECUTION}, and \texttt{CONSTITUTIONAL\_BLOCKED} operation classifications. These are interface-specific rules, not a claim that one global approval page determines every form of authority.

For an operation $o$, an authorized operation descriptor must say whether a frozen quorum is required, which membership and threshold apply, and which subsequent transition semantics are contemplated. An untrusted application cannot quietly relabel an operation that changes another party's authority as ordinary \texttt{UNILATERAL} work, nor can a proposed new constitution authorize the transition by which it installs itself. The relevant pre-state rule must be valid \textbf{before} the proposed change. This is why the proposed removal of member B from an authenticated $\{A,B\}$ roster with threshold two cannot silently be authorized by a threshold of one computed \emph{after} removing B. In the described scenario, the candidate's governance proof must satisfy the applicable frozen two-of-two rule or fail eligibility. The argument is about a specified fixed pre-state, not a completed proof of all member lifecycle, concurrency, credential recovery or emergency-transition semantics.

The precise classification also governs what happens \emph{after} a proposal passes. In \texttt{PROPOSAL\_AUTO\_COMMIT}, the governance design contemplates a deterministic, revalidated commit under a final Hub veto. In \texttt{PROPOSAL\_GRANT\_EXECUTION}, passing the proposal instead contemplates issuing a constrained Execution Ticket to a designated actor, followed by later presentation, freshness checks and a final decision at the execution boundary. Neither path is equivalent to an admission proof immediately mutating external state. Ticket issuance, consumption, replay resistance and actual executor mediation are not established by the present paper's OPA/Cedar experiment.

The separation between intrinsic and workflow governance prevents two opposite mistakes. One is making every independent policy-body constraint depend on a human voting workflow even when a pure and deterministic combination rule suffices. The other is permitting the policy-body owner to use a routine constraint update to change the constitution, remove other mandatory source seats, or lower the governance threshold. An independent source may update a constraint within its authorized seat under a specified policy update contract, but the right to \emph{change the rules defining necessary seats or their authority} is a different and higher-level operation. A change to the required-source inventory must itself satisfy the old authorized governance transition semantics. Otherwise, ``changing the policy'' becomes an uncontrolled channel for changing who is permitted to govern.

This design places constitutional invariants outside ordinary governable policy objects. Approval, even unanimous approval, need not disable fixed requirements such as the independent execution boundary, evidence integrity, or non-optional host safety conditions. Their deployment enforcement remains an assumption or a separate proof obligation, not a conclusion derived from expressing them as descriptors. A well-typed descriptor is necessary for classifying governance; it is not by itself evidence that protected hardware, authentic signers or physically independent control domains exist.

\subsection{Layer 3: admission-to-execution handoff without authority inflation}

Layer 3 is the boundary at which consequences become possible, and precisely for that reason the paper separates it from the Layer 2 evaluation. In the proposed HAVENLON design, Hub arbitration, a Security Module and an Executor occupy a later control path capable, when properly implemented and mediated, of applying an action-scoped final decision. They may recheck versions, currentness, fixed invariants, secure state, approval provenance, device health or local veto conditions. The previous execution-boundary research distinguishes an Execution Release Contract, subsequent Grant issuance and exact-version Redemption \cite{Wu2026EBLCore}. BCEAC can supply an upstream admission obligation to such a system; it does not unilaterally revise those existing grant and redemption semantics.

An architectural handoff may convey a correlation record $\alpha$ describing the admitted Candidate, governing epoch, chosen-path and policy-source commitments, decision profile and checking time. It must not be mistaken for a bearer token authorizing an arbitrary downstream call. A safe realization would need an independently specified rule establishing which component may consume that record, how it verifies its provenance, whether a newer state invalidates it, how multiple presentations are handled, and how the execution component prevents bypass through a different API, recovery path, privileged administrator or software update. Such properties are not consequences of source-binding completeness at $t_v$.

Time is therefore a first-class architectural boundary. Let $t_v$, $t_l$ and $t_r$ denote binding verification, release and Redemption time. Even when an admission conclusion was valid at $t_v$, policy heads, membership, facts and credentials may change before $t_l$ or $t_r$. The later authority must either re-establish required currentness and exact version conditions or rely on a separately specified, sound locking/lease mechanism that covers the interval. The R3 Final profile did not verify authenticated checkpoint acquisition, live anti-rollback, signed approval provenance, one-time ticket consumption, full executor mediation or a protected external effect. The same limitations apply to a logically correct architecture diagram: depicting a Security Module in the final column does not demonstrate that all alternative execution routes actually pass through it.

A complete system-level security claim would need to revisit the authority-decomposition problem: whether any principal or coalition can achieve the protected effect through a pathway that avoids the purported final veto, including administration, recovery, updates or direct use of underlying credentials \cite{Wu2026Authority}. This paper does not solve that problem by redefining eligibility as execution. Instead, it preserves the opportunity for a later, separately verifiable end-to-end argument. The status distinction is substantive:

\begin{equation}
\mathsf{AdmissionEligible}(c,t_v)
\neq \mathsf{ReleaseAuthorized}(c,t_l)
\neq \mathsf{GrantRedeemed}(c,t_r)
\neq \mathsf{EffectObserved}(c).
\label{eq:a5-8}
\end{equation}

The symbols in \eqref{eq:a5-8} designate non-identical judgments, not proof that any two outcomes must always differ. The last step also raises independent evidence-coverage and cutoff obligations: a record that an attempt was admitted or dispatched does not by itself establish the complete account of a consequential external outcome \cite{Wu2026Evidence}.

\subsection{Concrete realization boundary: R3 Final and the separate GSC-1 track}

The architectural decomposition is illustrated, but not completely implemented, by the archived real-engine R3 Final study. That profile uses a harness-installed \texttt{trusted\_manifest} and a set of declared required slots as the experiment's accepted source basis. The Gate checks source/body provenance, schema and contextual bindings, supported versions and semantics, governance snapshot relations, proposal identity and its applicable frozen quorum. Adapters invoke actual OPA or Cedar evaluations for the supported business context, and the Gate independently interprets the correlated result. The independent Reference is a separately coded oracle for the local fixture contract, not a cryptographic trust service or a formal proof of general checker soundness.

The backend predicate is more precise than the shorthand ``amount limit.'' For supported payments, the native path includes applicable amount constraints \textbf{and the current-state governance quorum}. A member-removal path uses current quorum without a payment amount comparison. OPA expresses its corresponding logic in Rego, whereas the Cedar adapter performs input projection, source filtering and current-member approval counting and constructs native Cedar policies and requests. Separately, the Gate checks the \textbf{frozen applicable governance snapshot}, so a correctly computed native rejection under a changed current roster can coexist with permission under a reference interpretation of the frozen pre-state. This can be a functional false denial, not an unsafe admission. Backends must preserve the full declared profile relation, including this current-state condition, rather than being exchanged on the basis of a fictional amount-only semantics.

R3 Final archives 72 distinct cases evaluated across two backend configurations, yielding 144 strict records. It records 120 strict native evaluations and 24 pre-native rejections, plus eight separately classified permissive native controls. The archived finite cohort and its offline review supply evidence that a separate admission function can be exercised with more than one actual policy engine on the shared, narrowly supported profile. They do not establish general semantic equivalence, universal input validation, a verified translation, independent signer identity, current checkpoints or final execution security. R1 and R2's prior rejection histories are preserved; the acceptance of the later frozen R3 Final corpus is explicitly bounded to its archived tested cases and audit method.

The distinction between cohort conformance and complete input-contract conformance is material. A later independent adversarial review constructed a required-but-nullable field omission: deleting \texttt{proposal.target\_member} from an otherwise valid payment input still resulted in Gate \texttt{ADMIT} and Reference \texttt{PERMIT}, despite the frozen R3 input contract requiring the field's presence, with \texttt{null} as its payment value. The two implementations shared a Python missing-versus-null pattern. The construction did not rerun native binaries and is not evidence of a deployed protected execution bypass. It is nevertheless a valid counterexample to inferring full schema soundness merely from Gate--Reference agreement over the original cohort. This limitation is disclosed rather than resolved; the corresponding implementation defect remains open. Legacy \texttt{INDETERMINATE} classifications for certain known invalid fields and status-first diagnostic precedence also remain implementation limitations rather than being retroactively relabeled to resemble the new normative three-state model.

The \textbf{GSC-1} line of work addresses a different, broader part of the specification. It describes a finite constitution/manifest closure profile, lawful path witnesses, a bounded source-to-program checking target, and a formal distinction between structural binding and a program's business decision. Its \texttt{Safe\_D} condition expresses a one-way non-weakening property on a nonempty, independently fixed admissible domain, and functionality must be separately assessed to avoid calling an always-denying evaluator practically equivalent. The general source-to-IR derivation $\delta$, theorem target T1 and complete checker-refinement argument remain unproved or unimplemented. The GSC-1 material cannot simply be appended to the OPA/Cedar fixture and described as a single system that has experimentally verified all three architectural layers.

Consequently, this paper uses two complementary but non-identical sources of support. The formal-model track supplies typed definitions and conditions under which a stronger source/path preservation result could be established. R3 Final supplies \textbf{bounded empirical observations} about specific backend invocation and admission separation under installed test fixtures. Neither supplies evidence that the proposed Layer 3 is non-bypassable. The Evaluation section will report these tracks with distinct denominators, assumptions, proof statuses and provenance; the present section uses them only to clarify which portions of the architecture have actually been exercised.

\subsection{Architectural obligations and falsifiable boundaries}

The value of the three-layer proposal lies partly in making its failure points explicit. An implementation can be challenged independently on (a) \textbf{constitutional completeness relative to its admitted roots}, (b) \textbf{required-source and exact-policy-object binding}, (c) \textbf{lawful chosen-path and governance snapshot preservation}, (d) \textbf{typed, correlated backend evaluation}, (e) \textbf{admission-verifier correctness and diagnostic behavior}, and (f) \textbf{downstream mediation and fresh release checks}. These are not six names for a single assertion of security; they are different premises or testable interfaces, often owned by different trust domains and possessing different failure consequences.

For example, if an authenticated constitution says both A and B must participate, a run admitting only A's body despite verified knowledge that B is required falsifies the source-binding obligation under the named premises. If both required bodies are present but a path-preserving backend composes \texttt{A OR B} where the fixed path requires \texttt{A AND B}, its evaluation violates the independent semantic-preservation obligation even though source presence is satisfied. If the correct backend denies because current quorum is insufficient while the frozen pre-state independently satisfies a different quorum, the difference requires a functional and profile analysis, not an automatic judgment that binding security was bypassed. If a native backend deliberately reports an incorrect \texttt{ALLOW} for an over-limit Candidate, the example violates backend semantic correctness and illustrates exactly why the Gate cannot substitute for a business-evaluation proof. If a real Executor can be invoked through an unmediated channel after a correct rejection, that is a Layer 3 complete-mediation failure, not necessarily a defect in the earlier policy evaluation.

Each boundary requires its own evidence standard. Finite tests can falsify universal-sounding descriptions and provide bounded conformance observations, but cannot enumerate every unknown governing authority or prove arbitrary-program equivalence. Digests and signatures can commit to records, assuming trustworthy keys and validation, but do not independently prove that the accepted constitution names every rightful stakeholder. A \texttt{BVerified} result can describe exact checked relationships at one time, but does not establish the safety of a later, potentially stale Redemption. Engine substitution requires semantic equivalence or appropriate one-way refinement on a declared common domain, not mere agreement of human-readable \texttt{ALLOW} labels.

Finally, the architecture is not offered as an exclusive way to organize a secure PDP/PEP system. An already trusted wrapper could realize equivalent constitutional source checking and admission mediation. Nor does the proposal yet establish comparative performance, lower implementation complexity, hardware resistance or deployment readiness. Its contribution is a typed decomposition that preserves the independence of governance authority from evaluator choice and keeps eligibility separate from causal execution authority. Under explicitly accepted constitutive premises and correctly instantiated profile semantics, the resulting BCEAC describes which evidence an evaluator's \texttt{ALLOW} must accompany before an action may advance toward the independent execution boundary. \textbf{Whether the eventual executor can be bypassed, whether the governance roots are genuinely sufficient, and whether all implementations soundly realize the interface remain separate questions.}

\section{Implementation and Experimental Evaluation}

\subsection{Evaluation questions and evidence boundaries}

The proposed Binding-Complete Execution Admission Contract (BCEAC) distinguishes the completeness and validity of required governance sources, the semantic decision of a policy evaluator, and the eligibility of the same candidate to advance toward a separately governed execution boundary. The preceding sections specify these obligations conditionally. Their implementability and their interaction with actual policy engines nevertheless require examination. In particular, a correct policy decision may be computed over an incomplete selection of policy bodies; a complete source selection may still be evaluated by an incorrect backend; and agreement between an admission verifier and a reference implementation may conceal a common input-validation error. An evaluation that reports only the fraction of cases in which two programs agree would not distinguish these possibilities.

We therefore organize the available evidence into two \textbf{non-equivalent} experimental tracks. The \textbf{GSC-1 track} is an isolated Go research prototype for finite, constitution-relative closure, source-slot binding, chosen-path resolution, and bounded extensional checking of an independently constructed intermediate program. The \textbf{real-engine track} exercises a narrower, fixed experimental profile with archived native Open Policy Agent (OPA) and Cedar evaluations, independent admission and reference implementations, and adversarial mutations. The first track studies a richer source/path model but does not attach it to the latter track's native engines. The second uses real engine invocations in a smaller policy fragment but does not implement GSC-1's general path semantics. No combined end-to-end executable system, proof of cross-track refinement, or protected external execution is evaluated here.

The evidence addresses four questions. \textbf{EQ1} asks whether a finite prototype can distinguish required source seats from currently available policy bodies and preserve legal choices, versions, and typed uncertainty. \textbf{EQ2} asks whether the bounded business-evaluation role can be attached to two actual engines while an independent gate enforces separate governance and binding requirements. \textbf{EQ3} asks whether targeted adversarial cases and audit replays expose defects that ordinary expected-case agreement would miss. \textbf{EQ4} asks what these results do \emph{not} establish about constitution adequacy, semantic equivalence, authenticated governance, or the downstream execution chain. These are empirical and methodological questions, not formulations of a previously proved universal theorem.

Throughout this section, \textbf{recorded native execution} means that an archived invocation, engine identity, input and raw output purport to record a process run in the original macOS environment. \textbf{Independent offline verification} means source inspection, parsing or replay of archived data without rerunning those native executables. \textbf{Pre-native rejection} means the gate/harness rejected an input before the policy engine was invoked; it is not a native DENY. \textbf{Simulated backend response} means a deliberately injected or analytical decision, not an OPA or Cedar evaluation. We preserve these labels even where all recorded checks report success. In the formal terminology of Chapter 4, a positive policy decision, exact binding, and successful pre-execution admission remain distinct judgments, and no positive admission result itself constitutes an Execution Grant.

Reviewer independence in this study refers to a separately conducted examination of implementation and archived evidence, not institutional independence. Parts of the adversarial review were AI-assisted: large-language-model tooling was used within the project, under author direction, to inspect the implementation and archived evidence. These reviews do not constitute external peer review, third-party security certification, or independent re-execution of the historical native OPA/Cedar binaries. Responsibility for interpreting and reporting the findings remains with the manuscript authors.

\subsection{Experimental design, frozen inputs, and units of observation}

The principal units are different in each track and must not be merged. A GSC-1 \emph{named case} is one selected fixture and its predeclared oracle; a \emph{variant} is another fixture-derived observation, possibly directed at a boundary case; a \emph{timing trial} is a repeated measurement under a stated configuration. In the real-engine track, a \emph{case} is a fixed scenario, whereas a \emph{strict record} is its evaluation under a particular backend configuration. A \emph{native record} has an archived actual engine invocation; a \emph{pre-native record} has no native invocation. A negative control or tamper-operator outcome belongs to a further denominator and cannot be counted as another strict case merely because it was processed by an auditing tool.

Table~\ref{tab:6-1} defines the two evidence cohorts. The versioned source trees, fixture sets, scripts, reference implementations, decision logs and audit reports were supplied as distinct archives. We inspected the underlying JSON/JSONL and implementation source files rather than relying solely on prose summaries. The counts in the table describe the supplied archived runs; the present analysis relies on those archived execution records, and no additional native-engine rerun is claimed.

\begin{table}[htbp]
\centering
\small
\caption{Separate experimental cohorts and their denominators.}
\label{tab:6-1}
\begin{tabularx}{\textwidth}{@{}>{\raggedright\arraybackslash}p{2.55cm} >{\raggedright\arraybackslash}p{2.75cm} Y Y@{}}
\toprule
Track & Primary units & Archived scope & Interpretation \\
\midrule
GSC-1 Go prototype & Named test fixtures & 44 named cases: 21 P0 regressions and 23 P1 additions & Oracle conformance within one bounded experimental specification \\
GSC-1 variant cases & Derived fixture records & 48 variants, including 12 E12 performance-corpus variants & Additional inputs; not 48 independent deployments \\
GSC-1 falsification controls & Additional analytical/runtime probes & 4 recorded controls & Separate from the 44 named cases \\
GSC-1 timing & Recorded repetitions & 360 trials over 12 configurations $\times$ 30 trials & Local prototype diagnostics, not a cross-engine benchmark \\
R3 Final strict cohort & Case $\times$ engine & 72 cases $\times$ 2 engines = 144 strict records & Fixed, narrower OPA/Cedar experimental profile \\
R3 Final strict native & Engine invocation records & 120 native: 60 OPA and 60 Cedar & Native execution recorded in the original archive \\
R3 Final pre-native & Early-stop records & 24: 12 per engine & Not native evaluator decisions \\
R3 Final permissive controls & Separate engine records & 8 native controls: 4 per engine & Deliberately permissive control arm, outside 144 strict records \\
\bottomrule
\end{tabularx}
\end{table}

The GSC-1 archive additionally contains program tests, race tests and vet/suite logs. Their pass events include ordinary test hierarchies, so the number of test pass events cannot be added to the number of independent named fixtures. Likewise, the R3 Final audit recorded 18 adversarial evidence-tampering operators, while its other \texttt{negative\_controls.json} categories include twelve normative controls and nineteen evidence controls. Those are distinct classifications in the archived harness, not interchangeable estimates of the number of native executions.

We report an outcome as \emph{observed} only for the specified input, time and semantic profile. A case designed to show the limit of a claim, such as a hidden-but-rightful source absent from the trusted constitution, is not converted into a passing security experiment. Unsupported inputs, explicitly unclaimed tenant properties, and A0 limitations are likewise kept outside any count of positive security validations. The general trust and completeness assumptions of Sections 3--5 continue to govern interpretation of every experiment.

\subsection{GSC-1: isolated implementation of finite closure and binding}

GSC-1 implements an experimental finite specification using a separate Go module, with a custom \texttt{GSC1-ENC1} typed encoding and deterministic artificial signing fixtures. Its verifier-installed root configuration specifies scoped authority and a fixed domain; signed envelopes, manifests, source heads, checkpoints, issuer rights, and policy bodies carry distinguishable commitments. It uses Ed25519 verification in a synthetic experimental trust environment but neither an external identity or currentness service nor production hardware. The repository's test keys and installed time/checkpoint values are laboratory inputs, not evidence of independent real-world governing principals.

The implementation divides its work across a strict codec, authentication and issuer-right checks, closure checking, authorized path resolution, exact policy-body binding, a direct source-policy reference evaluator, and a separately constructed tree compiler/interpreter. An essential ordering requirement is that closure be determined from independently installed constitutive metadata and authenticated materials: selected policy bodies and an evaluator's favorable decision cannot define the required inventory. A source seat persists if its currently designated body becomes unavailable, expires, or is revoked. If a needed head or checkpoint cannot be authenticated or obtained, its absence is not automatically a proven negative fact or a reason to remove the seat.

For a successfully resolved fixed legal path $\Psi$, the finite binding track checks the bodies selected for its \texttt{SourceSlot} leaves, including slot and issuer identity, designated revision and digest, supported semantics, authorization and applicable validity. A chosen alternative is a candidate- and context-bound witness, not a license to borrow a permit from an unchosen sibling. The prototype intentionally supports only restricted guard locations and bounded expression resources. Inputs beyond its grammar or limits cannot be certified by default. This makes it a tractable test of selected BCEAC obligations rather than a general-purpose policy-language equivalence checker.

The checker also separates structural source binding from \textbf{extensional preservation} on an independently installed finite evaluation domain $D$. The relevant one-way relation is:

\begin{equation}
\operatorname{Safe}_{D}(\Pi,\Psi)
\;\equiv\;
(D\neq\varnothing)\;\land\;
\forall \sigma\in D:\
\operatorname{Allow}_{\Pi}(\sigma)
\Rightarrow
\operatorname{Permit}_{\Psi}(\sigma).
\label{eq:e6-1}
\end{equation}

The separate functional-equivalence and false-denial measures are:

\begin{equation}
\begin{aligned}
\operatorname{Exact}_{D}(\Pi,\Psi)
&\equiv \forall\sigma\in D:\
\operatorname{Allow}_{\Pi}(\sigma)
\Leftrightarrow\operatorname{Permit}_{\Psi}(\sigma),\\
\operatorname{UnsafeAllow}_{D}
&=\left|\{\sigma\in D:\operatorname{Allow}_{\Pi}(\sigma)
\land\neg\operatorname{Permit}_{\Psi}(\sigma)\}\right|,\\
\operatorname{FalseDenial}_{D}
&=\left|\{\sigma\in D:\operatorname{Permit}_{\Psi}(\sigma)
\land\neg\operatorname{Allow}_{\Pi}(\sigma)\}\right|.
\end{aligned}
\label{eq:e6-2}
\end{equation}

These equations characterize a finite research profile, not a generic interpretation of all native OPA/Cedar programs. The reference relation and test domain are independently stipulated; the compiler cannot redefine $D$ to exclude states on which it behaves incorrectly. An all-denying program can satisfy the one-way safety condition while producing false denials, and an all-denying reference domain can make a program trivially exact without showing useful positive service. Consequently, the prototype records \texttt{EXACT}, \texttt{SAFETY\_ONLY}, \texttt{BROKEN} and \texttt{NOT\_TESTED} separately, and a verified binding cannot be identified with business \texttt{ALLOW} or a Release Grant. The stronger source-to-program derivation claim $\delta$, including a certified account of how a particular compiler constructed $\Pi$ from all required sources, is explicitly \texttt{NOT\_IMPLEMENTED}.

The \texttt{Beta} record in this prototype commits to the checked finite tuple and is subject to full revalidation if reused. It is not a bearer execution capability. The synthetic simulated boundary retains \texttt{ReleaseEligibility=INELIGIBLE} and \texttt{ExecutionResult=NOT\_ATTEMPTED} for the reported named/variant cases. The experimental checker therefore supplies bounded observations concerning admissible inputs to future execution control, not observations of a completed execution-control path.

\subsection{GSC-1 results, counterexamples, and archival integrity}

The archived final P1 run contains 44 named records, 48 fixture-variant records, four additional falsification records and 360 performance-trial records. Its named-case result summary reports \textbf{21/21} P0 regression cases and \textbf{23/23} P1 additions matching their predeclared oracles. The 44 named binding outputs comprise \textbf{16 \texttt{B\_VERIFIED}}, \textbf{21 \texttt{B\_REJECTED}} and \textbf{7 \texttt{B\_UNDETERMINED}} judgments. This distribution matters: a test matching its expected rejection or unresolved status is a test of discrimination, not evidence that the associated external candidate is permissible. The 48 variants also matched their frozen expected labels in the subsequent archived evidence review. The \texttt{measurements.csv} data portion contains 452 rows: 44 named cases, 48 variants and 360 timing trials. These counts are mutually traceable but should not be presented as 452 independent security attacks.

The fixtures exercise, among other cases, unavailable or authenticated-missing bodies, issuer-right or designated-revision mismatch, changed source heads, malformed closure inventories, illegal alternative witnesses, deliberately weakened compiled programs, strict domain bounds, and reuse of a \texttt{Beta} record with altered candidate, facts, path or program. They include positive controls to prevent a purportedly secure always-rejecting checker from being mistaken for a useful evaluator. One deliberately incorrect program dropping a required denial clause must fail the extensional check on a domain containing the discriminating state. In contrast, legitimate local short-circuiting or an all-denying program may satisfy certain one-way checks without being functionally identical. These examples support the need to distinguish exact binding, one-way non-weakening and application utility.

Two implementation-level classification regressions were found during the P1 work. In one, a missing envelope caused authority evaluation to borrow an unjustified default epoch; in another, an unauthenticated head-status string was treated as though an authenticated conflict had been established. Both were first preserved as failing developmental cases and then corrected in the later source snapshot. The revision retained independently provable negative evidence while allowing genuinely missing authority to remain \texttt{UNKNOWN}/\texttt{B\_UNDETERMINED}. Another early failure arose from a test comparison between nil and empty chosen-path representations rather than a changed policy decision. The final recorded isolated Go tests, race tests, vet invocation, and suite commands exited successfully. The test ledger retains those prior failures instead of presenting an uninterrupted success history.

Archival integrity qualifies the run's acceptance status. An independent C5 review found \textbf{four historical P0 CSV files with byte-level line-ending drift} relative to their previously recorded digests. The discrepancy was explainable exactly by CRLF-to-LF normalization without observed numerical-field changes, but that observation did not identify its cause or retroactively restore the prior bytes. Thus C5 recorded \texttt{NO-GO/BLOCKED} even while the later P0/P1 functional observations were technically consistent with the saved oracles. A subsequent authorized forward process preserved the observed LF files and C0-hash-matched CRLF reconstructions as separate provenance evidence; later read-only checks of the archival checkpoint did not erase the historical C5 verdict. This distinction is necessary when using the results as archived empirical evidence rather than as an unqualified end-to-end validation certificate.

The later E0-A evidence review rechecked the 44 named records, 48 variants and 360 timings against saved inputs, oracles, JSONL outputs and the local evidence chain, and gave \textbf{conditional permission to use the bounded observations}. That work did not rerun the Go program, reproduce all signature operations independently, or certify the original oracle semantics as universal. Of the planned E01--E48 named investigations, E45--E48 remained explicitly deferred; the archived timing series did not establish a cross-engine performance advantage. The stronger result, that a general GSC-1 checker or the intended production admission architecture is sound for all valid inputs, remains a proof/refinement obligation.

\subsection{Real-engine testbed and implementation responsibilities}

The second track evaluates the E1-Lite real-engine lineage. The R3 Final profile accepts two declared operation families, \texttt{payment.transfer} and \texttt{team.member.remove}, with a fixed synthetic input contract. For each strict case the harness records the candidate, an installed \texttt{trusted\_manifest} declaring required policy slots and governance information, selected policy bodies, a proposal, a current governance state, a backend request, the expected normative outcomes, and independent evidence identifiers. The Gate and Reference are separate Python implementations. OPA evaluates a Rego policy through its adapter; Cedar receives a generated policy/request assembled by its Python adapter. The profile uses canonical digests and explicit versions but \textbf{does not authenticate real governance signers or obtain a live trusted checkpoint}. Its manifest is a harness-installed trust assumption, not a demonstration that all rightful policy sources in an actual institution have been discovered.

The recorded native environment used \textbf{OPA 1.21.2 (Rego v1)} and \textbf{Cedar CLI 4.13.0 (Cedar language version 4.5)} on macOS arm64. The relevant binaries' identities, versions, hashes and invocation commands are saved in the experimental provenance and the early independent real-engine audits. The later transferred R3 Final evaluation archive excluded the executable bytes; its independent reviewer parsed archived native outputs and re-executed the Gate/Reference \textbf{offline}, but did not independently launch the same native binaries. A recorded version command, checksum and raw output improve traceability without proving historical processes in a different execution environment. The formal business-correctness assumption is therefore separate from the fact of offline trace agreement.

The native business predicate is \textbf{not} correctly described by an amount-only comparison. On the supported payment path, applicable policy amount limits are combined with a \textbf{current-state governance quorum}; a supported member-removal path uses the current quorum without the payment amount comparison. In the Cedar arm, the Python adapter additionally performs validity/source filtering, projects source attributes and computes current approval intersection before constructing the native policy. The correct evaluation unit is consequently the \emph{adapter plus the native evaluator}, not the Cedar runtime alone. These implementation facts are important when describing engine substitution: matching the amount comparison while omitting the current quorum would be a change of business relation, not merely a change of execution technology.

The admission Gate independently checks typed inputs, exact source/issuer/revision/digest relationships, proposal-to-candidate bindings, designated governance snapshots, and the \textbf{frozen pre-state quorum} relevant to the proposal. The independent Reference specifies expected Gate-level permissibility for this experimental contract and produces \texttt{PERMIT} or \texttt{FORBID}. The backend returns the native business decision; the Gate's admission result is a further interpretation of the backend response together with independently checked non-business requirements. It does not reimplement the amount-limit decision, and its agreement with the Reference depends on the premise that the backend correctly computes the declared business relation. A \texttt{DENY}, an unsupported or invalid mapping, an engine failure, and a pre-native stop are separately typed rather than being folded into an undifferentiated negative test result.

\subsection{Why the real-engine lineage includes rejected intermediate results}

The historical development sequence supplies a more informative security evaluation than the final pass count in isolation. The first independently examined real-engine round, \textbf{R1}, recorded a narrow original cohort that appeared consistent under its own expected inputs. The extended adversarial inspection nevertheless found \textbf{15 admission/reference conflicts under additional inputs}, alongside shared validation and audit-chain weaknesses, and returned \texttt{REJECT} with four P1 and four P2 findings. Real policy engines had been invoked: their actual operation did not make an incomplete surrounding admission contract correct.

In \textbf{R2}, fixes closed earlier counterexamples in the specified cohort, but independent inspection found new positive-admission violations and shared input-contract defects. The R2 audit reported two P1 and four P2 findings and again returned \texttt{REJECT}. Among the conceptual problems exposed was the danger of allowing the Gate to recheck the same business amount predicate as the native engine: such duplication could make apparent admission safety rely on Gate-side business logic rather than demonstrate separation between the evaluation and admission roles. The subsequent design narrowed the Gate's role to schema, governance, binding and legal backend-output checks, leaving the policy evaluator responsible for the bounded business predicate.

The initial \textbf{R3} was audited offline as \texttt{CONDITIONAL}: it had 67 case IDs and 134 strict case-engine records, but still omitted enforcement of the required diagnostic \texttt{proposal\_id} and lacked complete correlations in the evidence self-auditor. Its archive reported 110 strict native records and 24 pre-native records, with eight separate permissive native controls. The final R3 closure introduced five additional malformed \texttt{proposal\_id} cases per engine, strengthened both Gate and Reference validation, and expanded the audit linkage among exact case inputs, invocation IDs, native outputs, comparison rows, log entries and negative controls. The initial R3's original 134 strict records were retained in the later archive with their recorded decisions, mappings and status classifications unchanged.

The \textbf{R3 Final} offline audit returned \texttt{ACCEPT} only for that named finite and frozen cohort. The earlier R1/R2 \texttt{REJECT} and R3 \texttt{CONDITIONAL} verdicts are not overwritten by this later verdict. They document why input-contract exactness, independently expected authority and an audit chain that can resist deliberate record tampering are not optional reporting details. A frozen archive can be internally consistent while a new, untested adversarial input still exposes a common implementation defect. Thus the correct proposition is that the revised cohort and specified checks were accepted by a bounded offline reviewer, not that an arbitrary input to an arbitrary backend has been proved safe.

\subsection{R3 Final quantitative observations and native/early-stop separation}

Recounting the supplied \texttt{results.json} produces \textbf{72 distinct case IDs} and \textbf{144 strict case-engine records}, split evenly between OPA and Cedar. In that cohort, \textbf{120 records} carry native-invoked flags, comprising 60 per engine; \textbf{24 records} are pre-native, comprising 12 per engine. The separately stored eight permissive controls are recorded as native evaluations, bringing the archive's native record count, when controls are included, to \textbf{128}. The corresponding raw-output directory contains \textbf{152} JSON records: 144 strict plus eight controls. This arithmetic prevents two frequent reporting errors: treating early stops as native DENY decisions, and counting the eight permissive controls inside the 144 strict records.

\begin{table}[htbp]
\centering
\small
\caption{R3 Final strict status distribution, independently recounted from the archived result table.}
\label{tab:6-2}
\begin{tabularx}{\textwidth}{@{}>{\raggedright\arraybackslash}p{3.3cm} r Y@{}}
\toprule
Status & Count & Interpretation \\
\midrule
\texttt{PASS} & 111 & Fixture-specific recorded expectations satisfied \\
\texttt{PASS\_PRE\_NATIVE} & 24 & Expected early non-native handling; no engine invocation \\
\texttt{UNSUPPORTED} & 5 & Explicit out-of-support or unsupported mapping; not a demonstrated general property \\
\texttt{NOT\_CLAIMED} & 2 & Intentionally excluded property, including tenant-boundary semantics \\
\texttt{A0\_LIMITATION} & 2 & Constitution-adequacy limitation, not an identified hidden source \\
\texttt{FAIL} / \texttt{BLOCKED} & 0 / 0 & No such status in this particular frozen strict table \\
\textbf{Total} & \textbf{144} & One status per strict case-engine record \\
\bottomrule
\end{tabularx}
\end{table}

Across all 144 strict records, the Gate returned ADMIT in ten records: six classified as PASS, two as NOT\_CLAIMED, and two as A0\_LIMITATION. The latter four concern properties outside the evaluated assurance claims.

The native decision, Gate result and Reference result cannot be treated as one outcome. Their archived 72-case comparison reports no between-engine differences in Gate or Reference results on this cohort, while identifying a backend difference for \texttt{REAL-04} and native-decision differences for \texttt{MUT-01}, \texttt{REAL-04} and \texttt{unsupported\_semantics}. Differences at the engine-result or support level are therefore not evidence that the engines are universally equivalent. Conversely, equality of Gate and Reference within the observed cohort is not by itself a proof that both implementations validate the entire input contract. The R3 Final historical independent audit parsed the 120 strict native raw-output records and replayed Gate and Reference against all 144 archived backend results without finding a discrepancy in the tested comparisons; these are \textbf{offline verifications of recorded evidence}, not new native engine executions.

The separate controls add diagnostic support. The final negative-control file records \textbf{8/8} permissive native controls detected, alongside \textbf{12/12} normative controls and \textbf{19/19} evidence controls with their respective meanings. The self-audit results record \textbf{18/18} predeclared tamper operators detected, classified as seventeen \texttt{FAIL} and one \texttt{BLOCKED} under the archival auditor. Such checks demonstrate sensitivity to the specified mutations of the stored evidence chain. They do not prove that every falsification of a real execution trace would be detected, and the controls should not be described as 18 additional business-policy native invocations.

The archive also reports that none of the fifteen previously identified R1 unsafe admissions reappeared as an unsafe admission in the named R3 Final tested cases. This is a regression statement concerning the tested transformations, not a universal absence-of-bypass theorem. In particular, the offline audit's \texttt{ACCEPT} cannot be read as establishing semantic correctness of arbitrary future Rego or Cedar programs, full schema completeness, cryptographic signer verification, or resistance to a deliberately incorrect backend's favorable decision.

\subsection{Adversarial analysis: separating specification failures from violated premises}

Three adversarial patterns demonstrate why the admission contract must expose its premises. First, \textbf{a missing required policy body} can create a favorable evaluator result because the evaluator has no intrinsic means of identifying an absent governing source. Under a constitution that independently declares both source seats as mandatory, the gate must fail the positive source-binding obligation when one designated source is missing or invalid. An unavailable body cannot be interpreted as deletion of the seat. This pattern motivates source closure independent of selected policies; it does not justify claiming that the gate can discover a rightful source omitted from the trusted constitution itself.

Second, \textbf{governance and state are distinct references}. The real-engine payment/remove profile uses the current-state quorum for the native business predicate but evaluates proposal authorization against the frozen pre-state in the Gate and Reference. For example, a frozen roster $\{A,B\}$ with threshold two and approvals from both members may authorize the pre-state proposal, while a changed current roster $\{A,C\}$ with threshold two makes the native current-state quorum fail. Under those deliberately different relations, Gate rejection and Reference permission can coexist as a \emph{false-denial or functional divergence relative to the frozen reference}. This is not evidence that a correctly implemented native backend was insecure, and the two quorums cannot be exchanged without altering the declared profile. A member-removal proposal likewise cannot bootstrap itself from the smaller post-removal roster to satisfy an older two-member quorum.

Third, \textbf{business correctness is not implied by binding correctness}. The original R3 backend-fault ablation includes an over-limit candidate for which a deliberately incorrect or forced \texttt{ALLOW} makes Gate \texttt{ADMIT} while Reference returns \texttt{FORBID}. The relevant ablation source is the \textbf{original R3} faulty-backend ablation artifact, not the R3 Final adversarial evidence record. This construction is a dependency counterexample: it violates the assumed correctness of the business evaluator and therefore does not refute a theorem explicitly conditioned on a correct backend. It is, however, decisive against any interpretation that the admission Gate by itself checks or repairs all semantic bugs in the evaluation engine.

A separate, later adversarial review found a \textbf{common-mode schema defect in R3 Final's frozen implementation}. The contract requires \texttt{proposal.target\_member} to be \emph{present}, while allowing its value to be \texttt{null} for payment. Deleting the field entirely from an otherwise valid payment fixture caused both Gate and Reference to return \texttt{ADMIT} and \texttt{PERMIT}, respectively. Their use of the Python \texttt{.get} pattern conflates an absent field with an explicitly present null value. The probe employed a marked simulated backend response, \textbf{not a new native OPA/Cedar execution}. It shows that agreement between two independently coded programs can coexist with a shared validation blind spot; the recorded 144 strict comparisons do not certify full schema conformance. The implementation defect remains open. Similarly, the historical Gate's handling of exact-integer threshold type errors and diagnostic precedence does not fully refine the newer normative \texttt{Reject}/\texttt{Unresolved} distinction. Neither issue was repaired by revision of the manuscript.

Finally, the trusted constitution's adequacy assumption, A0, is an epistemic limit rather than a field that can be verified merely by hashing a manifest. If two possible worlds provide the same accepted constitution but only one contains a hidden rightful source, no verifier restricted to that constitution can distinguish the worlds. The \texttt{A0\_LIMITATION} records preserve this qualification. The observed system detects omissions \textbf{relative to a trusted declared inventory}; it does not prove the inventory is globally complete. These distinctions keep actual in-profile violations, violated trust assumptions and explicitly unsupported future scenarios in different logical categories.

\subsection{Evidence provenance, replay and reproducibility}

The supplied evidence chain includes versioned Python sources, Rego, Cedar adapter construction logic, declared case fixtures, canonical input projections, recorded output blobs, run and invocation IDs, logs, comparison tables, negative controls, checksums and independent review reports. R3 Final's checksum manifest fixes individual archived artifacts while its native engine identities are reported from the original environment. The independent offline reviewer checked available bytes and parsed native output records, recomputed Gate and Reference decisions from stored backend responses, and verified record-to-input correlations. This provides stronger support than a top-line \texttt{PASS} count, while remaining weaker than an independent live rerun of the original native processes under observed, reproducible execution conditions.

The GSC-1 recorder likewise retains source/fixture snapshots, command metadata, stdout/stderr and exit status, including failed development runs, as well as a separately versioned performance corpus. Its follow-up E0-A audit compared case records to independently stipulated oracle files and reviewed the precise line-ending discrepancy noted above. The artificial signatures are real operations on synthetic fixture keys, but neither the laboratory clock nor the checkpoint's installed high-water state is backed by an authenticated external service or secure monotonic hardware. In both tracks, byte-for-byte reproduction of an archived computation and independent normative justification of its trust roots are distinct standards of evidence.

A further peer-comparison workstream evaluated possible common profiles and candidate versions for OPA, Cedar and an XACML implementation, but its build/fetch and equal-assumption implementation gates did \textbf{not} complete a three-engine comparable evaluation. Some later official artifact acquisition attempts remained \texttt{BLOCKED}. Consequently, there is \textbf{no} defensible result here showing a three-engine performance or correctness ranking, universal substitution across different policy languages, or lower overhead than a conventional enforcement wrapper. The 360 GSC-1 timings measure isolated prototype configurations; without a completed common protocol, independently validated native mappings, and matched cost accounting, they cannot be used to imply a production or competitive performance advantage.

The study's reproducibility is therefore conditional on the availability of the archived sources and outputs and on the stated trust assumptions. External reviewers can reproduce the finite reference checks and audit logic where the supported runtime is available, and can independently examine the event lineage; they cannot infer that omitted macOS binary bytes were included in the transferred archive, that a particular historical process is cryptographically witnessed merely because its command was logged, or that the paper has supplied machine-checkable proofs of all interface refinements. The published experimental claim must retain these distinctions even if a later implementation resolves some outstanding problems.

\subsection{Interpretation relative to the BCEAC and engine-agnostic architecture}

Taken together, the two experimental tracks support a \textbf{bounded compositional engineering claim}. The GSC-1 prototype demonstrates that a finite constitution-relative source-closure and path-binding procedure can be implemented with typed uncertainty and independently stipulated source/program comparisons within its declared limits. The R3 Final archive demonstrates that, on a narrower well-specified fixture profile, actual OPA and Cedar evaluations can occupy a backend role while a separate Gate evaluates non-business source/governance requirements and a separate Reference evaluates the intended local contract. The latter track further demonstrates, through rejected intermediate rounds and later controlled replays, that the integrity of the input schema and the evidence chain is part of the scientific result rather than an administrative appendix.

This evidence does not establish that the GSC-1 verifier is an implementation of Chapter 5's entire BCEAC interface, or that the R3 Final Gate implements the general \texttt{Resolve}, \texttt{BoundContext}, \texttt{MatchResolution} and \texttt{MatchEvaluation} relations. Those are newly specified \textbf{normative interfaces} subject to future profile-specific refinement. Likewise, the existence of policy snapshots, hashes and recorded decisions does not discharge the conditional backend-correctness relation or the path-preservation implication over every possible supported state. In the real-engine cohort, the chosen business fragment and its adapter projections must be understood as part of the exact tested profile; an arbitrary evaluator cannot be swapped in on the strength of a compatible output label alone.

The two complementary observations are best expressed as non-equivalences:

\begin{equation}
\begin{aligned}
&\text{Bounded fixture conformance}\not\Rightarrow
  \text{universal checker soundness},\\
&\text{archived native evaluation}\not\Rightarrow
  \text{independent live replay},\\
&\text{source binding verified}\not\Rightarrow
  \text{business correctness},\\
&\text{business ALLOW}\not\Rightarrow
  \text{eligible execution admission},\\
&\text{admission eligibility}\not\Rightarrow
  \text{protected external effect}.
\end{aligned}
\label{eq:e6-3}
\end{equation}

Here $\not\Rightarrow$ rejects \emph{unconditional} implication; additional independently demonstrated premises may establish narrower relationships. In particular, a qualified backend and correctly implemented admission verifier may allow the same candidate to advance when the necessary contract is fully and soundly discharged. That conditional progression is the subject of BCEAC. The experiments do not prove it for arbitrary implementations, and they do not prohibit it for a correctly protected system.

\subsection{Threats to validity and unverified obligations}

\textbf{Construct validity.} The installed \texttt{trusted\_manifest}, synthetic signers, chosen test fixtures and reference implementations stand in for real governance authority. A0 adequacy, signer independence, issuer-right authenticity across administrative domains, current checkpoint acquisition and protection from rollback are assumed or simulated, not independently established in the deployment sense. A valid digest does not authenticate a rightful source roster. An apparent negative result may derive from a correctly unsupported profile rather than a violation of the stated semantics, while a passing normal-input cohort may omit a required-field edge case.

\textbf{Internal validity.} Gate and Reference are separately implemented, but they can share assumptions or programming mistakes, as the required-nullable example demonstrates. A native engine and its adapter together determine what business relation was actually evaluated. Evaluation and admission may be correlated through harness projections and output mappings; incorrect correlation can yield misleading agreement unless checked independently. The recorded independent offline audit improves traceability but does not establish the historical native execution events with external attestation. The GSC-1 test author and prototype implementer are not institutionally independent, and a test suite that compares to predeclared oracles remains limited by oracle correctness.

\textbf{External validity and coverage.} GSC-1 admits finite bounded domains with a fixed grammar, synthetic time and controlled heads. R3 Final covers a restricted payment/member-removal profile with particular adapter implementations. Neither track covers arbitrary third-party business-policy constructs, concurrent constitutional changes, unexpected engine failures outside the tested mappings, state-dependent adversarial input beyond its fixed contracts, or general alternative-path semantics across native engines. No cross-engine fair performance comparison has been completed; no real asset movement or physical effect has been measured.

\textbf{Unproved formal obligations.} A general theorem establishing universal binding-complete admission, source-to-IR derivation correctness $\delta$, soundness of every concrete \texttt{MatchResolution}/\texttt{MatchEvaluation} refinement, policy semantics equivalence, and the full non-bypassability of the execution chain remain outstanding. The newer normative distinction between known-invalid and unknown inputs is not uniformly realized by the frozen older Gate implementation. A future study would require independently fixed profiles and truth relations, stronger input-schema oracles, authenticated source/currentness services, adversarial implementation testing, protected and fresh release/redemption checks, and a controlled observation of real effects. Closing such obligations requires new versioned evidence, not revision of old verdicts or reinterpretation of past \texttt{PASS} labels.

\subsection{Evaluation conclusion}

The empirical record supports a constrained but meaningful separation: governance-source completeness and exact input binding can be examined independently of a policy evaluator's business verdict, and the resulting eligibility judgment can be treated as another typed stage rather than a synonym for policy \texttt{ALLOW}. A finite Go prototype exercised richer constitutional and path checks; a distinct real-engine experiment recorded OPA and Cedar participating in a narrower profile with separately tested admission criteria. Independent challenges rejected two earlier real-engine versions, conditionally accepted a third, and accepted the final \textbf{archived finite cohort} only after additional schema and evidence-correlation checks. Subsequent inspection still found an uncorrected common-mode required-field omission outside that cohort.

Accordingly, the evidence supports \textbf{the feasibility of a bounded architecture and the utility of explicit falsifiable contracts}, not a universal correctness theorem, arbitrary backend interchangeability, or a deployed secure execution system. The remainder of the paper treats authenticated constitutional adequacy, trusted state/currentness, backend and verifier refinement, and non-bypassable release-to-effect mediation as explicit limitations and future verification targets. The distinction between an accepted admission \emph{judgment} and an independently controlled real-world \emph{action} remains fundamental.

\section{Discussion and Limitations}

\subsection{What the evaluated evidence permits us to conclude}

The central result of this study is best understood as a \textbf{separation of obligations} rather than a claim that one tested implementation has solved the entire execution-security problem. Chapters 3 and 4 identify the conditions under which an evaluator's decision may contribute to a typed execution-admission judgment: an independently accepted constitutional basis, exact candidate and scope identity, a justified set of required source obligations, valid discharge of those obligations, a supported business-evaluation relation, and a sound correlation between the evaluation and the admitted context. Chapter 5 places those conditions in a proposed architecture that permits different policy-evaluation backends only under explicitly bounded semantic and trust assumptions. Chapter 6 tests selected fragments of this decomposition. None of these steps establishes that a protected external effect must, or cannot, occur in a deployed system.

The two experimental tracks illuminate different parts of this argument. The isolated GSC-1 implementation examines finite constitutional closure, chosen paths, source-to-body binding, and bounded source/program relations. Its 44 named cases, 48 variants, four additional falsification controls, and 360 timing observations constitute distinct experimental units; the 452 CSV data records are not 452 independent security attacks. The separate R3 Final real-engine archive contains 72 scenarios tested in two backend configurations, yielding 144 strict case-engine records, of which 120 record native execution and 24 are pre-native. Its eight permissive native controls lie outside that strict cohort. The records support the proposition that some necessary source/governance checks can be exercised separately from policy evaluation in two real engines on a declared narrow profile. They do not show that the GSC-1 source/path checker has been integrated with R3 Final, or that either implementation realizes every interface specified in Chapter 5.

The evaluation history is itself a substantial part of the result. R1 and R2 received independent \texttt{REJECT} verdicts; R3 was \texttt{CONDITIONAL}; and R3 Final received a bounded archival/offline \texttt{ACCEPT}. These are findings about different versions and evidence states, not successive votes that can be averaged into a single confidence score. The later cohort addresses identified faults in the earlier ones, but does not retroactively make an unsafe historical admission safe. Conversely, a later adversarial counterexample outside the final cohort---the required-but-nullable \texttt{proposal.target\_member} omission---does not mean that the original 144 recorded cases were miscounted. It instead disproves the stronger, unjustified inference from cohort agreement to \textbf{complete input-schema soundness}.

It is therefore useful to distinguish three statements that are sometimes conflated. First, the BCEAC is a \emph{normative contract}: it specifies necessary obligations and external premises for a positive pre-execution judgment. Second, the testbeds are \emph{bounded implementations and observations}: they demonstrate selected behavior on specified fixtures with recorded assumptions. Third, an end-to-end system safety claim would be a \emph{compositional theorem about actual authority and effects}: it would require both those earlier components and separately protected release, consumption, and execution paths. Only the first two, in their qualified senses, are advanced in this paper. An experimental \texttt{PASS} status, a \texttt{B\_VERIFIED} structural result, a native \texttt{ALLOW}, and an \texttt{Eligible} judgment are not interchangeable descriptions of success.

This distinction affects the interpretation of negative evidence. A tested nonpositive judgment may establish that a specified malformed, conflicted, unsupported, or incompletely justified candidate was not positively admitted in that case. It is not proof that every unsafe request would be rejected. Similarly, an \texttt{A0\_LIMITATION} status identifies a boundary of the accepted authority inventory, and \texttt{NOT\_CLAIMED} excludes an undeclared property from the paper's assurance claim. Treating either status as another ordinary security pass would conceal precisely the assumptions that BCEAC makes inspectable. The appropriate contribution is thus \textbf{a falsifiable and bounded admission architecture with explicit failure obligations}, not a numerical estimate of general attack resistance.

\subsection{Constitution-relative completeness and the A0 epistemic boundary}

The most consequential assumption is not that a policy object's hash can be checked, but that an independent authority has identified the \emph{right normative universe to check}. A constitution, manifest, designated-head relation, and applicability rules can determine which \texttt{SourceSlot} obligations a candidate must discharge only relative to an accepted boundary and authority basis. A fetched set of policy bodies is not entitled to define that universe. This asymmetry is essential: a missing, revoked, expired, or unreachable policy body cannot erase its corresponding required source seat. Nor may a favorable backend result justify narrowing the inventory after the outcome is known.

The A0 constitutional-adequacy premise has an epistemic limit. Consider two possible worlds in which the verifier receives the same authenticated constitution and all the same visible evidence. In the first, that constitution enumerates every rightful governor for the declared operation. In the second, an additional rightful governor is omitted from the accepted constitution, and no independent mechanism exposes the omission. An admission procedure restricted to the identical accepted view cannot distinguish these worlds simply by rehashing the view or checking it against itself. Schematically, the interpretation is:

\begin{equation}
\begin{gathered}
\mathsf{View}_{T}(W_1)=\mathsf{View}_{T}(W_2),\\
\mathsf{RightfulRequired}(W_1)\ne\mathsf{RightfulRequired}(W_2)\\
\Longrightarrow\quad\text{the accepted view alone does not identify the rightful difference}.
\end{gathered}
\label{eq:d7-1}
\end{equation}

Equation~\eqref{eq:d7-1} states a conditional indistinguishability argument, not a newly proved universal impossibility theorem over every conceivable external information source. Its force lies in fixing what information the verifier is assumed to receive. Authenticating the wrong roster can establish that a particular publisher signed it, but cannot establish that the roster is normatively complete. Similarly, a well-formed \texttt{EffectiveArtifact}, a signed \texttt{Beta} record, or an audit trail of source retrieval cannot become a second constitutional authority merely by referring to its own successful validation.

A realistic deployment would therefore have to justify the origin, jurisdiction, membership rules, and update procedure of its trust roots independently of the evaluator that consumes them. It would also need to account for multiple governing domains, cross-root conflicts, status updates, emergency authority, and the possibility that a previously authentic roster is no longer current. The experiments use installed or synthetic authority fixtures and do not evaluate those organizational and cryptographic trust questions. They establish no general test for discovering an unlisted rightful principal.

This limitation does not make constitution-relative binding vacuous. Once a declared authority basis is accepted under explicit premises, one can ask precise, falsifiable questions: whether the resulting \texttt{Required} obligations were derived without consulting the favorable selected set; whether a needed head was independently authenticated; whether a chosen path is legal; and whether the exact designated body was bound to the candidate. A missing declared source can be detected even though an undeclared rightful source cannot be conjured from absent evidence. The correct claim is \textbf{relative completeness under an adequately justified external authority basis}, not global authority discovery.

Joint Governance makes this dependence especially visible. For an ordinary proposal to change its own members or threshold, the applicable authenticated \emph{pre-state}---not the proposed post-state---determines the approval rule. A roster $M_0=\{A,B\}$ with threshold two cannot become a one-member approving authority merely because principal A proposes to remove B. That property is conditional on the applicable pre-state having been correctly established, the operation having been classified under legitimate constitutive rules, and the approvals being authentic and properly bound to the candidate. Neither two distinct fixture IDs nor two successful signature-verification calls, without a justified control-domain model, prove genuinely independent governing principals.

\subsection{Why complete binding does not establish semantic permission}

A second limitation follows from keeping the objects of verification distinct. The \texttt{SourceSlot} inventory says which governing sources are required; a \texttt{PolicyObject} provides a versioned body intended to discharge one of those requirements; an \texttt{EffectiveArtifact} or $\beta$ can commit to particular bodies, versions, paths, and contexts; and a backend computes some declared business relation. Exact object identity is necessary for source-binding claims, but it says nothing by itself about the correctness of the evaluator's interpretation. A complete set of correctly identified inputs can be combined under the wrong policy operator, evaluated against the wrong business state, or passed through an adapter that changes its meaning.

The distinction is sharpest for authorized alternatives. An obligation expression $\Phi$ is structured: \texttt{all(require(A),require(B))} is not equivalent to a permitted choice between A and B. A lawful, reached-node-keyed witness $w$ instantiates a fixed path $\Psi$, whose required seats $R_\Psi$ must be discharged without importing a favorable result from an unchosen sibling. Merely reporting the leaf set $\{A,B\}$ loses the distinction between conjunction and authorized choice. The R3 Final profile exercises only a narrower set of policy and governance relationships; it does not establish general support for arbitrary nested alternatives across OPA and Cedar.

The corresponding semantic issue is separate from structural discharge. Suppose the accepted fixed path requires both A and B to permit, but a correctly functioning evaluator computes a business predicate that permits if \emph{either} A or B permits. Even a perfectly verified pair of loaded bodies does not prevent a false positive relative to the intended path. That evaluator may be correct with respect to its own \texttt{OR} program and still fail the required path-preservation implication. Thus a claimed path-preserving profile must independently justify both the business evaluation relation and its relationship to the authorized path:

\begin{equation}
\begin{gathered}
\mathsf{DecodedAllow}(z)\ \Longleftrightarrow\
\mathsf{Permit}_{\mathrm{business}}^{\rho}(z)\\
\text{for successful evaluations in its declared supported domain},\\
\mathsf{Permit}_{\mathrm{business}}^{\rho}(z)
\ \Longrightarrow\
\mathsf{Permit}_{\Psi}(c,\sigma)\\
\text{on the independently guarded path-preserving subdomain}.
\end{gathered}
\label{eq:d7-2}
\end{equation}

Equation~\eqref{eq:d7-2} \textbf{restates obligations already established by Chapters 4 and 5}. It is not an additional theorem proved by the experiments. The two lines use the same exact candidate, state and fixed path where the profile claims that property; the path domain is not allowed to be restricted after seeing an unfavorable result. A structural-only profile may validly establish less, but it may not borrow the stronger path-preservation conclusion. An adapter that returns the correct decision for a different normalized input has not fulfilled the exact-context contract merely because the resulting Boolean matches a reference fixture.

The original R3 forced-\texttt{ALLOW} ablation demonstrates the complementary limit. When an over-limit business action is deliberately returned as \texttt{ALLOW}, the admission Gate can admit a structurally well-bound input while an independent business-aware Reference forbids it. This does not contradict BCEAC under its explicitly stated \textbf{backend-correctness premise}: the premise was violated by construction. It does disprove an overextended interpretation in which a binding Gate repairs all errors in arbitrary policy code. Independent source verification is not a replacement business evaluator. The R3 forced response was simulated, not an observed vulnerability in the native OPA or Cedar engines.

\subsection{Engine agnosticism is a qualified interface property, not a universal substitution result}

The phrase \emph{engine-agnostic} can obscure two different propositions. One is an architectural proposition: the obligation to validate governing sources and contextual admission need not be expressed in the syntax of any single policy language. The other is a much stronger implementation claim: arbitrary evaluators can be substituted without changing decisions, availability, errors, timing or security. This paper advances the first proposition as a design and specifies \textbf{conditions} under which a bounded form of the second could hold. It does not prove unrestricted substitutability.

Under the BCEAC, a qualified backend includes more than the core policy VM. Its adapter, input projection, generated policy text where applicable, supported operation and profile domain, result decoding, invocation trace, and error conventions determine the effective business relation exposed to the admission verifier. In the archived real-engine profile, payments are subject to applicable amount constraints and a \emph{current-state quorum}. Member-removal decisions rely on current quorum without an amount comparison. The Gate and Reference also apply a proposal-bound \emph{frozen-prestate quorum}. The Cedar adapter performs filtering and approval counting before generating native policy material; treating every such computation as an operation of the Cedar runtime would misdescribe what was actually tested.

This split allows a revealing functional divergence. Let the frozen voting members be $\{A,B\}$ with threshold two and approvals from both. If the current voting members change to $\{A,C\}$ with threshold two, a correctly computed current-quorum native decision may deny while a frozen-prestate Reference permits the already bound proposal. A separate Gate may reject on that native denial. This is a \textbf{false denial relative to the frozen reference} under two deliberately distinct predicates. It is not, on its own, evidence of an unsafe admission or of an incorrectly implemented native predicate. The appropriate comparison must fix which state relationship is normative, whether the profile intentionally imposes both constraints, and which failures are acceptable. A backend substituted only for an amount-limit calculation would not preserve the actual tested business relation.

Two practical consequences follow. First, a meaningful cross-engine correctness experiment needs an independently defined common domain, the same candidate/context projection, equivalent or explicitly refined business semantics, an agreed treatment of unsupported inputs and errors, and a verifier that subjects all configurations to the same host-side admission obligations. Second, native \texttt{ALLOW} agreement alone is insufficient to certify substitution. Two implementations may share a wrong assumption about a required field, accept the same inaccurate context, or omit a governing source that was never represented in the test fixture. Conversely, two qualified implementations may differ in native \texttt{DENY}, \texttt{ERROR}, or \texttt{UNSUPPORTED} on inputs outside their common domain without violating a claim explicitly limited to supported normal completion.

The R3 Final archive supports a narrower assertion: two named engine configurations participated in a fixed, audited cohort while a separately implemented Gate made the final \emph{admission} classification for that experiment. It supplies no universal translation proof $\delta$, and no evidence that the newer Chapter 5 \texttt{Resolve}, \texttt{BuildBoundContext}, \texttt{MatchResolution}, and \texttt{MatchEvaluation} abstract interfaces have been implemented end to end. The unfinished E1B2 comparison work likewise cannot establish a fair OPA/Cedar/XACML ranking. A failed local artifact-acquisition or preflight gate does not show that an XACML PDP is intrinsically unsafe or less efficient. A conventional enforcement wrapper could, in principle, satisfy the same BCEAC requirements; the paper does not claim exclusive architectural novelty from the mere existence of a separate Gate.

\subsection{Fail-secure decisions, functional usefulness, and diagnostic precision}

A fail-secure admission contract cannot be assessed solely by counting favorable or unfavorable decisions. Its central rule is asymmetric: independently established disqualifying facts prevent positive admission, and unresolved necessary facts also prevent positive admission, but for different evidentiary reasons. Chapter 4's \texttt{Reject} and \texttt{Unresolved} represent different epistemic states. A known mismatch in an authenticated designated digest may support a conclusive negative judgment, while an inaccessible checkpoint or untrusted head status should remain unresolved until a trusted basis supplies the missing fact. Returning no \texttt{ALLOW} is not, by itself, evidence that the verifier has classified the cause correctly.

This difference has engineering consequences. A legitimate operator needs to distinguish a request that violates an authenticated boundary from one that may become admissible after independent information is supplied. Monitoring, recovery, retries, human intervention and audit explanation may depend on the classification. The older R3 Final implementation sometimes uses \texttt{INDETERMINATE} for a malformed exact-integer threshold and can allow a status-first path to mask a separately provable digest conflict. Those known F-04 behaviors do not automatically produce an unsafe positive admission in the observed probes, but they do \textbf{not} fully refine the newer normative diagnostic precedence. Changing a paper's taxonomy cannot retroactively fix a frozen program.

Nor is one-way safety equivalent to useful policy service. The GSC-1 condition \texttt{Safe\_D} checks that a compiled program does not allow a state that the fixed reference would deny, on an independently established nonempty finite domain. An evaluator that always denies can satisfy this one-way implication while suppressing legitimate actions. The associated \texttt{EXACT}, \texttt{SAFETY\_ONLY}, \texttt{BROKEN}, \texttt{NOT\_TESTED}, unsafe-allow and false-denial classifications therefore measure different properties. GSC-1's deliberately weakened and all-denying controls illustrate why a security-oriented specification must record both unauthorized positives and functional negatives rather than labeling all nonpositive outcomes as indistinguishable successes.

At the architecture level, the same tension appears in conservative error handling. A timeout, unsupported policy form, or unavailable governance source should not silently become permission; nevertheless, excessive unresolved outcomes can render a system operationally unusable. The present experiments are not an availability study: the 360 timing observations are local prototype trials without a matched cross-engine comparator, and there is no liveness theorem for asynchronous retrieval or admission under failures. A future implementation must separately establish bounded completion of necessary checks, the soundness of their diagnostic classifications, and measurable positive service within its supported profile. A proposed rule that rejects all requests is not thereby validated as a practical execution-admission architecture.

\subsection{Admission cannot certify the causal execution boundary}

The sharpest system-level limitation is the \textbf{phase and causal boundary}. A positive \texttt{Eligible} judgment is a statement about the evidence accepted for a particular candidate, governance context, semantic profile and verification phase. It is not an Execution Grant, an exact-version Redemption, a successful Hub or Security Module veto, or evidence of an external effect. Even a correct admission process is ineffective against a path that reaches the Executor without traversing the intended protected mediator. Conversely, an independently protected executor might deny a request after an earlier positive admission because relevant currentness, host state or release obligations have changed.

The times $t_v$, $t_l$ and $t_r$ mark distinct verification, release and Redemption contexts. A designated source head or checkpoint acceptable at $t_v$ may have been superseded at $t_l$; a governance epoch may have changed by $t_r$; a state-dependent business condition may no longer hold. This makes replay resistance, monotonic version evidence, exact candidate/authorization commitments, bounded leases or revalidation, and sound consumption rules additional downstream obligations. Merely retaining a signed $\beta$, an \texttt{EffectiveArtifact} digest or a favorable policy result cannot establish their continued suitability. The earlier execution-boundary model separates ERC, Execution Grant and exact-version Redemption; the present admission contract neither overrides that lifecycle nor proves its implementation \cite{Wu2026EBLCore}.

{\sloppy In particular, proposed Layer 3 components---\allowbreak Hub arbitration, Security Module and\linebreak Executor---do not become independent control domains by being drawn as different boxes. A complete-mediation argument must examine alternative invocation paths, maintenance and recovery interfaces, credential export, administrator bypasses, local update mechanisms, and behavior under partial compromise. Proving that a malicious evaluator cannot fabricate an accepted source inventory would still leave open whether an attacker could directly cause the protected effect through another route. This is the distinction between workflow permission and causal authority that motivated the earlier authority-decomposition work \cite{Wu2026Authority}.\par}

The laboratory evidence leaves this step entirely open. For the GSC-1 named/variant records, the stored \texttt{ReleaseEligibility} is \texttt{INELIGIBLE} and the stored \texttt{ExecutionResult} is \texttt{NOT\_ATTEMPTED}; those fields do not constitute a tested release mechanism. R3 Final contains Gate classifications, recorded backend calls and audit artifacts, not a protected external transfer or physical act. Neither track observes Grant construction, one-time Redemption, hardware-enforced non-bypassability, or post-effect evidence completeness. In particular, a log of a successful admission does not prove that an action happened, and a log of a requested action does not certify the completeness of its external consequences \cite{Wu2026Evidence}.

For this reason, the phase separation is an assurance \textbf{boundary}, not a negative statement that correct systems must never connect admission to execution. A future system may soundly advance from one judgment to another after discharging the independent obligations at each phase. What the present paper rules out is an \emph{unconditional coercion} from policy \texttt{ALLOW} or admission eligibility to actual execution authority. The right architectural question is which independently trusted component can prevent the effect when a later obligation fails, and how that prevention is tested. That question is not answered by an admission-only corpus.

\subsection{Threats to construct and internal validity}

\textbf{Construct validity.} The experiments operationalize difficult real-world notions through controlled surrogates. GSC-1 synthetic keys, fixed clocks, predeclared manifests and checker-installed domains stand in for authenticated governing principals, real revocation publication, authorized epochs and trustworthy observations of external state. R3 Final's \texttt{trusted\_manifest}, static policy bodies, and structured proposal fields stand in for a legitimate constitutional inventory and qualified requests. Passing those fixtures is informative about the declared test contract, not a measurement of whether a real organization has identified every relevant authority. A SourceSlot may be required by an independently adequate constitutional inventory while its designated policy body is omitted from \texttt{Selected} or unavailable in \texttt{Loaded}. This is a binding or materialization deficiency assessable against the trusted required-slot set. It differs from a world in which a rightful SourceSlot is omitted from the constitutional inventory itself: the latter is an A0 adequacy failure that cannot be identified from the unchanged accepted view alone.

A second construct-validity issue concerns the meaning of the measured outcomes. A native \texttt{DENY} may arise from a current-state business relation rather than a binding failure. An \texttt{INDETERMINATE} may reflect a transport or diagnostic limitation rather than a conclusive policy violation. A pre-native error is not the native engine's decision, and a simulated forced \texttt{ALLOW} is not an observation of a real policy VM misbehaving. The final archive's \texttt{PASS} status denotes success against that run's specific expected behavior, including negative cases; it must not be interpreted as a successful positive authorization or an independently measured security property. The distinction between 120 strict native and 24 pre-native records is consequently substantive, not just bookkeeping.

\textbf{Internal validity.} The nominal separation between Gate and Reference lowers some risks of copying exactly the same program logic, but does not ensure independent error models. The required-but-nullable \texttt{proposal.target\_member} example (F-01) is a direct counterexample: both implementations treat an absent field like an explicitly present \texttt{null}, although the input contract requires field presence. The common-mode error survives successful agreement on the recorded cohort. An independent reference implementation is therefore useful but insufficient unless its oracle specification, serialization boundary and malformed-input semantics are separately challenged. Independent test code and an independently justified \emph{normative oracle} are different things.

Backend adapters are another internal-validity dependency. An engine cannot be evaluated separately from the preprocessing that selects valid constraints, computes current-state quorum, constructs requests, decodes outputs and classifies unsupported behavior. An unchanged native policy may appear to behave differently if the adapter changed an input projection or if the harness accidentally reused stale output from another invocation. R1--R3 repairs concerning input typing, proposal identity, and evidence correlation show why tracking actual requests and raw results matters. Chapter 5 adds normative same-attempt obligations, but R3 Final does not constitute an implementation or refinement proof of those newer interfaces.

\textbf{Audit dependence and experimental iteration.} The cohorts were constructed and extended in response to observed weaknesses, rather than sampled independently from a predeclared distribution of all possible adversarial inputs. The corrected final set is therefore a regression artifact as well as an experiment. Its passing labels cannot estimate an unknown real-world bypass probability or prove that the next mutation is unlikely to expose another common error. The four historical verdict stages and the later F-01 finding make selection and survivorship visible, but visibility is not a statistical correction. Likewise, GSC-1's reference oracles and prototype were developed in the same research effort; source separation and frozen fixtures do not supply institutionally independent oracle validation or external preregistration.

\subsection{Archival provenance and reproducibility as part of the claim}

An evidence chain is a subject of security research in its own right. Hashes, inputs, outputs, decisions and audit reports make a finite claim inspectable; their presence does not automatically prove when and where an original native process ran, or that an authenticated authority approved the tested state. R3 Final preserves source files, native-output records, invocation identifiers and a checksum manifest, permitting offline cross-checking of the stored run relationships. However, some original macOS engine binary artifacts and optional independent-audit supplementary machine files are absent from the transferred research archive. The independent audit of Chapter 6 confirmed internal archival consistency and performed read-only source/data replays; it did \textbf{not} conduct a fresh witnessed native engine campaign. Reproducibility must be stated at that exact level.

GSC-1 provides a different illustration. A historical C5 integrity review identified four P0 CSV artifacts whose byte sequences differed from earlier committed digests because of line-ending changes. The field contents could be reconciled under a specific LF/CRLF transformation, and later forward-only repair work preserved the observed artifacts together with explicitly labeled CRLF reconstructions. That does not retroactively make the reconstructed files original captured bytes, identify the person or tool that caused the drift, or change the historical \texttt{NO-GO/BLOCKED} finding. The later conditional allowance to use bounded observations applies to a reviewed subsequent evidence state; it is not a declaration that the earlier state had always been intact. Separate later P1 CSV recovery records likewise do not change the historical four-P0 denominator.

These episodes motivate a distinction among \textbf{computational reproducibility}, \textbf{archive provenance}, \textbf{normative validity}, and \textbf{external execution attestation}. A stored computation may be reproducible even if an earlier transfer changed file line endings. A checksum-matched archive may be byte-consistent without proving the oracle's normative authority. A logged native invocation may be parsable without constituting an independent cryptographic witness of a historical process. And an authenticated admission record may still lack a corresponding protected execution or post-effect observation. No one of these evidence standards subsumes the others. The paper's finite empirical claim relies on the standards actually achieved and explicitly identifies those that remain absent.

The original R3 faulty-backend ablation, the later R3 Final audit's separately described simulated input mutation, and the subsequently constructed missing-field probe must also remain distinct in provenance and denominator. The forced business \texttt{ALLOW} recorded in original R3 \texttt{ablation\_results.json} was simulated and did not represent an additional native OPA/Cedar call. The later offline-audit mutation summary is evidence at the historical report's level where supplementary machine files are unavailable. The F-01 \texttt{target\_member} probe was a later analytical Gate/Reference check outside the 72-case final cohort. Combining the three into one undifferentiated set of ``native attacks detected'' would invent experiments that were not performed.

\subsection{External validity, alternative implementations, and comparative claims}

Both tracks cover specific models and deployment assumptions. GSC-1 uses a finite, typed and deliberately restricted grammar, bounded source heads, installed evaluation domains and synthetic signing context. R3 Final supports payment transfers and governance member removal with particular Rego, Cedar-adapter, Gate and Reference versions. Neither track supports an inference to arbitrary policy languages, nested alternative-path expressiveness in actual native engines, multi-tenant isolation beyond what is explicitly claimed, live concurrent governance, malicious modifications of a genuinely independent trust root, or authenticated production checkpoint distribution. Scaling the number of fixtures without broadening the semantic and authority assumptions would not by itself remove these gaps.

The most defensible comparative conclusion is consequently architectural, not competitive: it is possible to specify admission obligations that are \emph{not owned by the selected evaluator language}. This does not imply that OPA, Cedar, or any future EBL Native runtime is preferable on security, expressiveness, throughput, latency or operational complexity. The incomplete E1B2 common-profile and artifact-acquisition work cannot support a third-engine comparison. The GSC-1 timing series provides local within-prototype observations but no matched cross-engine cost accounting; even a completed native comparison would need to include the cost of trusted retrieval, source normalization, adapter processing, the admission verifier, failure handling, and any protected downstream release checks.

An existing trusted enforcement wrapper might implement many of these obligations without using the name EBL. The issue is not whether the same safety checks are packaged as a policy language, a wrapper, a broker, a dedicated verifier or a hardware-adjacent module. It is whether its authority basis is independent, its profile and input mappings are checkable, its actual business semantics satisfy the claimed relation, and the paths that can cause effects are separately mediated. The conceptual value of BCEAC is to articulate those obligations and their evidence dependencies in a way that enables scrutiny across implementations. The paper does not claim that ordinary PDP/PEP separation is absent from previous systems or that the mere introduction of another component constitutes a demonstrated security advantage.

\subsection{Remaining implementation defects and assurance limits}

The absence of a blocking objection to the material in Chapters 3--6 should not be mistaken for zero open engineering defects. Three tracked issues require separate, versioned follow-up. \textbf{F-01} is a valid schema-presence flaw in the archived R3 Final Gate and Reference: a field declared required but nullable can be omitted and still be processed as though its value were explicitly \texttt{null}. A future repair would need field-presence validation independent of value type and an oracle that is not subject to the same \texttt{.get} assumption. It must be exercised in a new implementation/evidence version rather than rewriting frozen results.

\textbf{F-04} concerns the divergence between certain historical Gate classifications and the newer normative known-invalid/unknown distinction. A Boolean or floating-point threshold outside the exact required integer type, and an unknown status that masks a separately proved mismatch, can be classified less precisely than the formal contract requires. A future implementation should validate the complete necessary obligations, respect independent-false evidence over unresolved dependent checks, and justify the error precedence it exposes to users and auditors. These observations do not, by themselves, establish an unsafe positive result for the corresponding tested probes, but they prohibit claiming exact refinement to the proposed taxonomy.

\textbf{F-06} is a provenance/documentation follow-up rather than a new evaluator bypass: the existing claim ledger's C11 cross-reference to the faulty-\texttt{ALLOW} ablation points to the wrong historical artifact. The manuscript correctly identifies the original R3 \texttt{ablation\_results.json}, while the broader evidence ledger remains to be updated under separate authorization. Frozen archival specifications that contain earlier dead or inaccurate local paths should be preserved and cross-referenced in a versioned correction, not silently rewritten to present a cleaner history. None of these three open matters is resolved by editorial revision of the \emph{wording and bounded evidence use} in a manuscript chapter.

Beyond the named defects, a complete BCEAC implementation still requires concrete, verified instantiation of Chapter 5's trusted resolution environment $\Gamma_\rho$, its \texttt{BoundContext} and business-state provenance, and the mandatory \texttt{MatchResolution}/\texttt{MatchEvaluation} checks. The admission verifier must have an independently accepted expected context, not reconstruct an expected value from the result it is verifying. Any duplicate policy decision in $X_\rho$ and the evaluator record must have a single authoritative projection or exact typed equality. Source/currentness retrieval must complete under profile-defined resource bounds or yield an appropriately typed nonpositive outcome. These are architecture-level requirements with no claim of general implementation correctness in the existing R3 Final codebase.

\subsection{Prioritized verification agenda}

Future work is not one undifferentiated request for more testing. The unresolved tasks belong to different proof and measurement categories and should be gated in an order that respects their dependencies.

\textbf{First, repair and independently test the narrow implementation contracts.} A successor to the frozen experimental Gate should check every required field for presence, distinguish missing from explicit null, define typed and ordered error evidence, and use an independently constructed malformed-input oracle. The unchanged R3 Final cohort should remain available as a historical comparator, while the new version adds the previously omitted cases. These repairs are necessary to support stronger schema and diagnostic claims; they do not validate trust roots or real effects.

\textbf{Second, instantiate a complete profile-level BCEAC and discharge its local refinement obligations.} The trusted resolution environment must specify exact authority retrieval and phase time, accepted source/head coherence, legal witness selection, state provenance, native invocation correlation, and the complete finite inventory of mandatory checks. An implementation study should test mixing resolution results, candidate identities, source versions, business-state snapshots, and policy outputs across attempts. It should also verify that a successful \texttt{MatchEvaluation} reflects an actual correctly bound adapter/native trace rather than equality of self-reported labels. The general source-to-intermediate-program derivation $\delta$, the broader theorem target T1, and the concrete checker/refinement proof remain separate formalization tasks; none is silently completed by adding interface names.

\textbf{Third, justify the authority and freshness foundations.} Independent constitutional adequacy for real governance domains, authenticated checkpoint and head publication, revocation and epoch changes, issuer/publisher rights, and genuine signer-control separation require their own threat models and verification evidence. A future protocol must consider simultaneous updates and stale but validly signed records. Fail-secure unknown handling should be measured for functional cost and recovery behavior as well as for avoidance of unsupported positive judgments. This work determines which external premises can become proved properties under a specific deployed environment.

\textbf{Fourth, evaluate engine substitution fairly and on a fixed supported profile.} A meaningful multi-engine comparison must predeclare its common semantic relation and operation set, hold candidate/source/context normalization constant, include adapter and host-side verifier overhead, distinguish native, pre-native and unsupported traces, and preserve error classifications. It should report false denials and positive service, not only absence of unsafe \texttt{ALLOW}. A three-engine result cannot be manufactured from one completed two-engine archive and an unrelated blocked artifact fetch. Equal-assumption comparisons should be conducted only when the relevant artifacts and audit conditions are available, and without interpreting absence of an implementation as failure of the other systems.

\textbf{Fifth, compose admission with a protected execution path.} To support end-to-end assurance, a concrete Hub/Security Module/Executor integration would need protected release-state checks, exact-version commitments, defined Grant issuance and consumption, anti-replay and recovery semantics, and a complete-mediation argument that includes direct, administrator and maintenance paths. Its evaluation must observe or securely attest real external effects and correlate those effects with the same candidate and independently governed authorization context. A positive admission result cannot substitute for this experiment or proof. Execution control thereby remains a \textbf{downstream research and engineering target}, not a property established retroactively by the present data.

These priorities have different acceptance criteria. Correcting a schema parser is an implementation repair; proving a finite checker property is a formal claim; witnessing a native process is an empirical provenance claim; ensuring that a specific action cannot bypass a final veto is a system-level security claim. A useful research program should make progress in each category visible without allowing the completion of one to certify the others by implication.

\subsection{Discussion synthesis}

The evaluated evidence and the preceding analysis support a limited but important architectural interpretation. It is possible to describe, implement in bounded fragments, and audit \textbf{pre-execution governance-source binding as an obligation independent of the particular policy evaluator}. A favorable policy decision is informative only after the evaluator's semantics and invocation are related to the same accepted candidate, required sources, authorized governance path and trusted evaluation context. The exact form of those relations is profile-dependent; their governing authority does not arise from the evaluator's output.

The limitations define the perimeter of that conclusion. Constitutional adequacy cannot be established from a self-authenticating inventory; exact source identity does not prove business-semantic fidelity; a one-way no-unsafe-allow relation is not functional equivalence; agreement on an evolving finite corpus does not imply full input-schema or implementation soundness; and phase-specific admission does not confer final execution authority. The historical \texttt{REJECT}, \texttt{CONDITIONAL}, \texttt{BLOCKED} and bounded \texttt{ACCEPT} records are compatible once their versions, scopes and evidence standards are held separate. Their coexistence is evidence of the need for falsifiable contracts and traceable revision, not a defect to be hidden from the publication narrative.

Under its declared external trust premises, BCEAC supplies a vocabulary and a set of necessary checks for the decision to \emph{consider} a policy evaluator's output in execution admission. The present experiments provide bounded demonstrations and counterexamples for selected elements of that program. They leave general refinement, authority adequacy, arbitrary engine substitution, protected release, and real-world causal effects unproved. That boundary is intentional: a policy engine can be replaced only \textbf{conditionally} as an evaluator; the independent obligations for governing-source completeness, binding, and execution admission cannot be discarded merely because an engine returns \texttt{ALLOW}.

\section{Conclusion}

\subsection{From a policy decision to an execution-admission question}

A policy evaluator answers a question about a policy and the input it was given. That answer does not, by itself, establish that the input represents the entire independently required governance basis for an action, that the evaluated source objects are the exact objects authorized for the relevant candidate, or that a downstream component has authority to cause an external effect. The distinction becomes consequential when governance is distributed across independently designated participants, when policy bodies and membership change over time, and when different components may disagree about the candidate, governing state, or admissible execution path. An \texttt{ALLOW} computed over a correct but incomplete input can remain a correct answer to the evaluator's local question while being insufficient evidence for execution admission.

This paper has addressed the narrower, checkable problem that precedes the final exercise of execution authority: \textbf{under an independently accepted and explicitly bounded governance basis, what must be established before a policy evaluator's decision may contribute to admitting a particular action?} The answer is not to treat the policy engine as a constitutional authority, nor to require every deployment to use a new domain-specific policy language. It is to make the obligations linking governance, candidate identity, selected and loaded policies, evaluation context, and the resulting decision explicit and independently assessable. We call this specification the \textbf{Binding-Complete Execution Admission Contract (BCEAC)}.

The qualification in this research question is essential. The contract is \emph{relative to} a declared boundary, trusted constitutional basis, applicable semantic profile, and independently established context. It does not infer the full universe of rightful governors from whichever policy objects happen to have been loaded. Nor does it claim that the trusted constitution is adequate merely because its bytes, signatures, or derivations are internally consistent. These external assumptions must be stated before local completeness claims can be interpreted as meaningful security claims.

\subsection{What the BCEAC specifies}

The first contribution is an explicit separation between \textbf{what governance requires} and \textbf{what a policy evaluator happened to receive}. An independently accepted constitution and applicability context determine a finite set of required \texttt{SourceSlot} obligations for the exact candidate and authorized path. Selected or loaded policy bodies can discharge these obligations only when their designated source identity, issuer authority, revision, digest, validity, and applicable semantics match the expected requirements. A required slot does not disappear when its body is omitted, expired, revoked, unavailable, or replaced by another body carrying a similar label. A legal alternative path must be justified and fixed under the governing rules; a favorable decision cannot retrospectively choose a weaker path or use an unselected alternative as permission for the selected one.

The second contribution is a profile-indexed \textbf{pre-execution admission obligation} separate from the evaluator's business judgment. The contract requires exact correspondence among the candidate and authorized action, frozen joint-governance pre-state where applicable, trusted source and checkpoint relations, materialized evaluation inputs, supported backend semantics, and independently expected invocation context. A native \texttt{ALLOW} is a necessary component only where the declared profile requires such a business result. It is never a substitute for these other conditions. Conversely, checking source provenance is not a substitute for the correctness of the evaluator's business predicate: a deliberately faulty evaluator can return \texttt{ALLOW} for a forbidden action, and an admission checker that intentionally does not recompute the complete predicate need not detect that semantic error. Backend and adapter correctness therefore remain named, falsifiable premises rather than hidden consequences of a successful binding check.

The third contribution is a \textbf{typed fail-secure interpretation} of the admission boundary. Independently established disqualifying evidence, unresolved necessary evidence, unsupported evaluator semantics, and successfully verified mandatory conditions must not be coerced into one Boolean success indicator. In the proposed contract, a proved independent failure can justify a rejecting judgment; necessary evidence whose truth cannot be established requires a nonpositive unresolved judgment in the absence of such a failure. A positive judgment requires the complete finite profile-defined obligation inventory to be satisfied. These distinctions preserve diagnostic meaning without allowing absent or ambiguous material to create permission by default. They are normative obligations of the proposed architecture, not a claim that every archived implementation already refines their precise classification rules.

Taken together, these components identify what an evaluator's decision is \emph{about} and what independently governed evidence must accompany it. They do not establish the adequacy of the governing constitution, prove the semantic correctness of an arbitrary backend, or make the admission interface itself an execution authority. Their value is that such assumptions and possible failure points become objects of separate analysis, implementation testing, and adversarial challenge.

\subsection{The architectural meaning of engine agnosticism}

The architecture developed in this paper places the BCEAC at a boundary that is intentionally distinct from any particular policy language or evaluator implementation. Its optional EBL Language is one possible representation of the governing semantics, not a prerequisite for expressing the admission obligations. Its proposed EBL Architecture separates trusted resolution and required-source derivation, exact body and \texttt{EffectiveArtifact} binding, a policy-evaluation interface, and independent post-evaluation admission verification. The latter must check the actual response against an expected context established independently of that response, including the applicable candidate, source set, authorized path, state, and invocation identity. Merely comparing two self-reported labels or hashes cannot make that context trustworthy.

Within a \textbf{declared supported domain}, a different evaluator may occupy the policy-evaluation role only if its adapter, complete business relation, context projection, result mapping, and error behavior meet the profile's explicit obligations. Where path-preserving admission is claimed, business permission must also respect the independently fixed authorized path on the qualified subdomain. Structural source binding alone neither proves that implication nor provides functional equivalence. In particular, one-way absence of unsafe positives is not equivalent to preserving every legitimate positive decision, and current-state business evaluation may produce functional denials relative to an independently frozen governance reference. These distinctions are reasons to specify substitution conditions, not reasons to assume that all engines are behaviorally interchangeable.

This contribution should not be read as an exclusive architectural claim. A conventional policy decision point coupled to a correctly implemented trusted wrapper and a separately protected enforcement boundary may discharge comparable scoped obligations under the same trust and semantic assumptions. BCEAC's proposed contribution is the \textbf{explicit, compositional and falsifiable contract} by which those obligations can be specified and scrutinized, rather than a claim to have invented policy enforcement, to have proved an intrinsic superiority over existing engines, or to have established a general compiler-equivalence theorem. The conceptual freedom to choose a policy evaluator does not remove the independent responsibility to establish who must govern the action, what was actually evaluated, and why the particular admission judgment is justified.

\subsection{What the archived evaluation establishes}

The experimental evidence supports the feasibility of exercising selected portions of this separation, but its units and trust assumptions must remain distinct. The GSC-1 Go prototype provides a bounded source-closure and path-binding study with \textbf{44 named cases}, \textbf{48 recorded variants}, \textbf{four additional falsification controls}, and \textbf{360 prototype timing observations}. Its \textbf{452 measurement CSV data rows} combine named cases, variants and timings; they do not represent 452 independent attacks or external executions. The prototype's finite oracle comparisons illustrate that verified binding, policy outcomes, one-way safety, functional equivalence, and false denial are separate measurements. They do not complete the proposed source-to-program derivation proof or authorize an execution release. Historical archival checks also documented a four-file C5 line-ending drift and a \texttt{BLOCKED} integrity finding; subsequent derived reconstructions and conditional evidence use do not turn those reconstructed bytes into independently preserved original acquisitions.

The separate OPA/Cedar study addresses a narrower, declared real-engine profile. Its R3 Final archive contains \textbf{72 scenarios} and \textbf{144 strict case-engine records}, with \textbf{120 recorded native} and \textbf{24 pre-native} records, plus \textbf{eight separate permissive native controls}. The strict record statuses comprise \textbf{111 \texttt{PASS}}, \textbf{24 \texttt{PASS\_PRE\_NATIVE}}, \textbf{five \texttt{UNSUPPORTED}}, \textbf{two \texttt{NOT\_CLAIMED}}, and \textbf{two \texttt{A0\_LIMITATION}}. These are case classifications, not a measured probability of resisting attack; the pre-native records must not be counted as native executions. The native process outputs and provenance are archived evidence, while the later independent review of the frozen results was an offline inspection and replay, not an independent rerun of the original Mac binaries. The two experimental tracks do not constitute one integrated implementation of the entire Chapter 5 architecture.

The history of this evaluation is as important as its final bounded cohort. Independent reviews rejected R1 and R2, the R3 review was conditional, and the final archived cohort received a \textbf{bounded offline \texttt{ACCEPT}} after further input and evidence-correlation checks. Earlier defects and counterexamples are evidence about their own versions, not observations to be overwritten by later success. A deliberate simulated faulty-\texttt{ALLOW} ablation additionally demonstrates the dependence on backend semantic correctness; it does not establish an attack on a correctly operating native evaluator. Subsequent analysis identified an uncorrected required-but-nullable field omission shared by the archived Gate and Reference outside the final cohort. These results support the usefulness of a specification that makes omissions and cross-context substitutions testable; they do not support a claim of comprehensive schema correctness, arbitrary hostile-backend tolerance, authenticated governance adequacy, or an independently attested external effect.

Thus, the empirical result is neither a universal theorem nor a security certification. It is a \textbf{bounded demonstration that governance-source and contextual admission checks can be articulated and evaluated separately from business policy evaluation}, accompanied by adversarial evidence of the consequences when that separation or its associated premises are incomplete. The completed two-backend study does not establish a fair, completed three-engine comparison or any general performance advantage. In this setting, transparent negative findings and explicit exclusion labels are part of the result rather than material to be hidden from it.

\subsection{The remaining proof and execution boundaries}

Several obligations remain outside the established result. The adequacy of the accepted constitution---\textbf{A0}---requires independent justification against the real governance domain. Authenticated source, manifest, head and checkpoint publication; issuer rights; currentness; anti-rollback; real control-domain independence; and trustworthy business-state provenance require separate evidence. The proposed \texttt{Resolve}, \texttt{BuildBoundContext}, \texttt{MatchResolution}, and \texttt{MatchEvaluation} interfaces still require concrete construction and refinement against actual implementations. The broad theorem target \textbf{T1} has not been proved, and the general source-to-intermediate-program derivation \textbf{$\delta$} has neither been implemented nor proved correct. Nor has arbitrary native backend substitution or general agreement on semantic outcomes been established.

The archived implementation also retains explicitly identified open defects. \textbf{F-01} is the required-field presence ambiguity involving a required-but-nullable field; \textbf{F-04} is a discrepancy between the legacy Gate's diagnostic precedence and the newer dependency-sensitive failed/unresolved distinction; \textbf{F-06/C11} is a documentary misreference to the original faulty-\texttt{ALLOW} ablation. Each requires its own versioned repair or evidence correction. Acceptance of the manuscript's description does not repair the code, strengthen the archived oracle, or change historical experiment results. Future evaluations should test these repairs with independently justified inputs and retain the original cohorts as provenance rather than editing the past to fit a newer implementation.

Finally, \textbf{execution admission is not final execution authority}. A positive pre-execution judgment is not an Execution Grant, a successful Redemption, or proof that an external effect occurred. A later Hub arbitration decision, Security Module enforcement, Grant lifecycle, freshness transition, or Executor action may fail independently or be bypassed if the eventual system lacks complete mediation. Establishing a protected release-to-effect path would require exact-version and anti-replay commitments, a non-bypassability analysis covering direct and administrative paths, verified temporal and recovery semantics, and evidence of actual effects. The paper presents that downstream execution boundary as an architectural requirement and future research direction, not as an implementation already validated by the present datasets.

These are distinct classes of unfinished work. A schema fix is an implementation correction; a checker refinement is a formal and empirical obligation; authenticating a current external authority is a governance and provenance obligation; proving non-bypassable effect mediation is a system-level security obligation. Satisfying one does not silently discharge the others. A meaningful progression is therefore to preserve the bounded evidence, repair and independently test local defects, establish concrete profile refinements and authority foundations, evaluate substitution under equal assumptions, and only then advance stronger claims about protected real-world execution.

\subsection{Closing perspective}

The central conclusion of this study is deliberately narrower than the ambition of a complete execution-security system. Under explicit external authority and semantic assumptions, a policy evaluator's result can be made \textbf{eligible for execution admission only through a separate, checkable relation to the governance sources, exact candidate, authorized path, trusted state, evaluation provenance, and mandatory host constraints}. This is the conditional claim made by BCEAC. Its value lies in refusing to promote a local evaluator's decision into a statement about the completeness of governance or the exercise of causal authority without the evidence those stronger claims require.

The archived prototypes, rejected intermediate revisions, bounded final observations and acknowledged counterexamples show both why that separation matters and why it is not sufficient on its own. They support an explicit, auditable admission contract and a conditional engine-agnostic architectural direction; they leave a truthful boundary around what has not been established. \textbf{A policy evaluator can be substituted only under a justified semantic and trust contract. The obligations of governance completeness, exact binding, independent admission and downstream execution control do not disappear when the evaluator changes or returns \texttt{ALLOW}.}

\printbibliography[heading=bibintoc,title={References}]
\end{document}